\documentclass[11pt]{article}

\usepackage{color} 
\usepackage{graphicx}
\usepackage{amsmath}
\usepackage{bbm}
\usepackage{amsfonts}
\usepackage{amssymb}
\usepackage{amsbsy}
\usepackage[T1]{fontenc}
\usepackage{theorem}
\usepackage{stmaryrd}
\usepackage{tipa}
\usepackage{textcomp}
\usepackage{subfigure}
\usepackage{epsfig}
\usepackage{lmodern}
\usepackage{framed}
\usepackage[subfigure]{tocloft}
\usepackage{subeqnarray}
\usepackage{alphalph}
\usepackage[refpage,cfg,prefix]{nomencl}
\usepackage{multirow}
\usepackage{xifthen}
\usepackage{enumerate}
\usepackage{color}
\usepackage{wasysym}
\usepackage{bibentry}
\usepackage{todonotes}
\nobibliography*
\usepackage{verbatim}
\usepackage{booktabs}
\usepackage{breakurl}
\usepackage[linesnumbered,algoruled,boxed,lined]{algorithm2e}
\usepackage{url}
\usepackage{setspace}
\usepackage{xr}
\makenomenclature
\usetikzlibrary{decorations.pathreplacing}

\usepackage[labelfont=bf,labelsep=period,justification=raggedright]{caption}

\makeatletter
\renewcommand{\@biblabel}[1]{\quad#1.}
\makeatother

\date{}

\newcommand{\D}[2]{\frac{\partial #1}{\partial #2}}

\newcommand{\ud}{\mbox{d}}

\newcommand{\bs}[1]{\boldsymbol{#1}}

\usepackage{tikz}

\begin{document}

\begin{flushleft}
{\Large
\textbf{The blending region hybrid framework for the simulation of stochastic reaction-diffusion processes}
}
\\
Christian A. Yates$^{1,\ast}$, 
Adam George$^{1}$,
Armand Jordana$^{4}$,
Cameron A. Smith$^{1}$,
Andrew B. Duncan$^{3}$,
Konstantinos C. Zygalakis$^{2 \ast}$
\\
{\bf 1 Department of Mathematical Sciences, University of Bath, Claverton Down, Bath, BA2 7AY, United Kingdom} \\
{\bf 2  School of Mathematics, University of Edinburgh, James Clerk Maxwell Building, The King's Buildings, Peter Guthrie Tait Road, Edinburgh, EH9 3FD, United Kingdom}\\
{\bf 3 Department of Mathematics, Imperial College London, London SW7 2AZ United Kingdom} 
\\
{\bf 4  Centre de math\'ematiques et de leurs applications, CNRS, ENS Paris-Saclay, Universit\'e Paris-Saclay, 94235, Cachan cedex, France}
\\
$\ast$ E-mail: c.yates@bath.ac.uk; K.Zygalakis@ed.ac.uk
\end{flushleft}
\section*{Abstract}
The simulation of stochastic reaction-diffusion systems using fine-grained representations can become computationally prohibitive when particle numbers become large. If particle numbers are sufficiently high then it may be possible to ignore stochastic fluctuations and use a more efficient coarse-grained simulation approach. 
Nevertheless, for multiscale systems which exhibit significant spatial variation in concentration, a coarse-grained approach may not be appropriate throughout  the simulation domain. Such scenarios suggest a hybrid paradigm in which a computationally cheap, coarse-grained model is coupled to a more expensive, but more detailed fine-grained model enabling the accurate simulation of the fine-scale dynamics at a reasonable computational cost. 

In this paper, in order to couple two representations of reaction-diffusion at distinct spatial scales, we allow them to overlap in a ``blending region''. Both modelling paradigms provide a valid representation of the particle density in this region. From one end of the blending region to the other, control of the implementation of diffusion is passed from one modelling paradigm to another through the use of complementary ``blending functions'' which scale up or down the contribution of each model to the overall diffusion. We establish the reliability of our novel hybrid paradigm by demonstrating its simulation on four exemplar reaction-diffusion scenarios.

Key index words: hybrid modelling, stochastic reaction-diffusion, multiscale modelling, partial differential equation, hybrid modelling framework
\section{Introduction}

Many biological and physical systems are inherently multiscale in nature 
\cite{sherratt2005avs,volpert2009rdw,mort2016rdm,khan2011scd, 
dobramysl2015pbd,flegg2013dsn}. The modelling of such systems therefore 
requires multiscale representations which, by their nature, are not well 
captured using a single modelling paradigm. There is a trade-off between, on the 
one hand, ensuring that models are sufficiently detailed that they accurately 
capture known biological and physical phenomena of interest and, on the other, 
achieving model outputs in a timely manner.

The appropriate representation of travelling waves of cells in developmental or 
maintenance contexts is a classic example of a multiscale phenomenon for which 
the trade-off between cheap-but-coarse and expensive-but-accurate modelling 
paradigms is evident. For a pulled wave-front the wave speed is determined by 
the low-density dynamics at the front of the wave \cite{king2017ppf}. It is therefore important 
to represent cell movement and proliferation dynamics at the front using an 
appropriately detailed model. A model that is too coarse may neglect 
important features of the real process. Behind the wave, cell density is higher 
making a fine-grained representation more computationally expensive. Since the fine details are less important in this region we can substitute the more detailed model for a cheaper, coarser 
representation. Coupling modelling regimes at different scales is an open 
question to which a variety of solutions have previously been proposed 
\cite{yates2015pcm,moro2004hms,spill2015ham,schulze2003ckm, harrison2016hac, 
flekkoy2001cpf,rossinelli2008ash,lo2016hcd,chiam2006hss,flegg2015cmc,
flegg2012trm,robinson2014atr,flegg2014atr,dobramysl2015pbd,hellander2012cmm,
klann2012hsg, smith2018arm, franz2012mrd, geyer2004ibd, gorba2004bds, 
alexander2002ars, alexander2005ars,plapp2000mrw,lo2019hsm,ferm2010aas,strehl2015hss,winkelmann2017hmc}. For more details on the different types of hybrid methods available we direct the interested reader to \cite{smith2018seh}.

In this paper we focus on the three main modelling paradigms used for 
representing  reaction-diffusion systems. 
At the coarsest scale (which we refer to as the  \textit{macroscopic} scale) 
we represent the \textit{concentration} of reactant species by 
partial differential equations (PDEs) 
\cite{keller1970ism,keller1971tbc,keller1971mfc,turing1952cbm,painter1999sfj, 
painter2002vfq,hillen2008ugp}. For validity, these models typically require high 
concentrations since assumptions underlying the use of PDEs break down for low 
copy numbers.
Continuum models such as these can usually be simulated extremely efficiently using a wide variety of well-established numerical methods, however, they lack the realism of finer-scale models.

At the next level down, the \textit{mesoscopic} scale, reactant species are 
represented as individual particles and are compartmentalised into contiguous, 
non-overlapping subdivisions of space 
\cite{engblom2009ssr,elf2004ssb,isaacson2013crd,erban2014msr,baker2009fmm,
yates2013ivd,yates2012gfm,mort2016rdm}. Particles are assumed to be well-mixed 
within a compartment and can interact with others in their compartment. These 
models can capture stochasticity in the behaviour of the particles and can be 
simulated efficiently when copy numbers are low. However, when particle numbers 
become large, simulations can become prohibitively slow in comparison to 
macroscale representations. They also lack the accuracy of more fine-grained 
models since the individual particle identities and positions are not retained.

The finest representation we consider is Brownian-dynamics 
models at the \textit{microscopic} scale 
\cite{andrews2004ssc,lipkova2011abd,erban2009smr,Erban2007rbc}. In these models 
the trajectories of all particles are simulated (typically using a discrete 
fixed time-step paradigm) in continuous space 
\cite{smoluchowski1917vem,andrews2004ssc,van2005gfr,sokolowski2019ead}. For a system of $N$ particles, an 
appropriate simulation algorithm must generate $Z N$ Gaussian random 
variables (where $Z$ is the dimension of the system) in order to update the 
particle positions. For simulations incorporating pairwise interactions, $N^2$ pairwise distances must 
also be updated at each time-step\footnote{Note that by a careful partitioning of space the number of comparisons can be reduced dramatically to almost $O(N)$ when particles are only compared with others in their local neighbourhood \cite{robinson2017pbm}.}. Consequently, these methods can be extremely 
computationally intensive. They do, however, provide a comprehensive and 
accurate individual representation capable of incorporating stochasticity into 
particle positions and interaction times. More details on the specific 
implementation of each of these three modelling paradigms will be given in the 
next section.

In general, the aim of a hybrid method is to exploit the complementary advantages and negate the complementary weaknesses of models at different scales. Using a coarse, cheap representation in a region of space in which particle density is high allows for significant computational savings in comparison to the purely fine-scale simulation. Conversely, implementing a fine-scale individual-based representation in regions in which low-copy number effects are of paramount importance can give significant improvements in accuracy in comparison to coarser models. Consequently, one way to achieve accurate simulations that are also computationally tractable is to combine the models' strengths in a hybrid representation.

In this paper we propose a novel hybrid method for coupling PDEs at the 
macroscale to compartment-based models at the mesoscale and a related novel hybrid 
method for coupling compartment-based models at the mesoscale to 
Brownian-dynamics models at the microscale. In each case, the coarser regime 
is coupled to the finer regime through an overlap region. In this overlap 
region, which from now on we will refer to as the \textit{blending region}, 
both representations of the reaction-diffusion dynamics are valid. In the 
blending region the strength of diffusion for each model is determined by a 
spatially-varying \textit{blending function} which is prescribed to be unity on 
one end of the overlap region and zero on the other. The blending 
functions for the two models are complementary so that the sum of the  two 
blending functions at any point in the domain is equal to unity. These 
functions control the relative contribution of each model to the diffusion 
dynamics. This approach is reminiscent of that taken by 
\cite{duncan2016hfs} in a non-spatial context. In \cite{duncan2016hfs} two 
different non-spatial models for stochastic chemical kinetics were coupled in 
copy-number space through a blending region in which both models co-existed.


The remainder of the paper is organised as follows. In Section 
\ref{section:modelling_at_different_scales} we describe the individual 
reaction-diffusion models that we couple together and provide a brief 
justification for why the models can be considered ``equivalent'' and hence are 
suitable candidates for coupling. In Section 
\ref{section:hybrid_blending_algorithms} we present the mechanics of the two 
hybrid blending methods and prove their effectiveness, in Section 
\ref{section:results}, by simulating a number of test scenarios and 
determining whether any bias is introduced by the blending methods. We conclude 
in Section \ref{section:discussion} with a short summary of our findings and 
suggestions for extensions to this work.

\section{Modelling at different scales}
\label{section:modelling_at_different_scales}
Within this section, we describe the three different modelling scales that we will couple in order to create 
two distinct spatially-coupled hybrid methods.  In Section \ref{sect:Modelling_macro} we describe a 
general macroscale PDE for reaction-diffusion systems with a single species, as 
well as different numerical approaches for its solution. Section 
\ref{sect:Modelling_meso} contains a discussion of mesoscale compartment-based 
models and their simulation, while in 
Section \ref{sect:Modelling_micro} we introduce the microscale individual-based 
dynamics.  In Section 
\ref{section:Modelling_Equivalence} we briefly discuss how each of these 
representations of reaction-diffusion processes at different scales might be 
considered to be equivalent in an appropriate limit.

\subsection{Macroscopic representation} \label{sect:Modelling_macro}

Partial differential equations, the macroscale models we employ in this 
paper, can be considered to be appropriate representations of the mean 
behaviour of 
particles at high concentrations. The primary advantage of the PDE 
representation is that there exists a wide range of well-established and 
well-understood tools for their numerical simulation. In rare, simple cases, 
PDEs are amenable to mathematical analysis. However, they typically fail to 
model low 
copy number behaviour.

A generic PDE which describes the spatio-temporal evolution of the concentration of a single species, $c(\bs{x},t)$, at position $\bs{x}$ and time $t$ takes the form:

\begin{equation}
\D{c}{t}(\bs{x},t) = \nabla \cdot ( D(\bs{x}) \nabla c(\bs{x},t)) + \mathcal{R}(c(\bs{x},t),\bs{x},t), \quad \bs{x}\in\mathbb{R}^Z,\quad t\in[0,T],
\label{eqn:Reaction_Diffusion_Equation}
\end{equation}
where consistent initial and boundary conditions need also to be specified. Here 
reactions are represented by the function $\mathcal{R}$, $Z$ is the 
dimension of space and $T$ is the final time to which we wish to evolve the 
solution. Note that the spatially varying diffusion 
coefficient, represented by $D(\bs{x})$, sits inside the first derivative, but 
not the second.
As noted by \cite{van1988dim}, there is no canonical choice of operator 
describing spatially dependent diffusion. In physical applications the form of 
the macroscopic diffusion equation should be dictated by the underlying 
microscopic or mesoscopic process. Since the spatial dependence of the diffusion 
coefficient in our hybrid methods is introduced purely as a modelling 
convenience and does not correspond to any microscopic or mesoscopic ground 
truth, we are effectively free to choose the form of the diffusion operator. 
We adopt the form considered by \cite{benson1993apf} (see equation 
\eqref{eqn:Reaction_Diffusion_Equation}). We 
choose the transition rates in the corresponding compartment-based 
representation (see Section \ref{sect:Modelling_meso}) and the drift and 
diffusion coefficients of the corresponding microscopic position evolution 
equation (see Section \ref{sect:Modelling_micro}) so that diffusion in the 
overlap regions of the hybrid methods satisfies the same form of non-constant 
coefficient diffusion equation.

For the majority of this paper we focus on the following one-dimensional PDE in 
the region $\Omega=[a,b]$:
\begin{equation} 
\label{eq:model_pde}
\frac{\partial c}{ \partial t}=\frac{\partial}{\partial 
x}\left(D(x)\frac{\partial c}{\partial x} \right) + \mathcal{R}(c(x,t)),
\end{equation}
 with constant flux boundary conditions
 \begin{equation}
  D(a)\frac{\partial c}{\partial x} \Bigr|_{x=a}=J_a, \quad D(b)\frac{\partial 
c}{\partial x} \Bigr|_{x=b}=J_b.\label{equation:PDE_boundary_conditions}
 \end{equation}
For a discussion of the implementation of the 
numerical solution of the PDEs employed in this paper please refer to 
Appendix \ref{section:appendixA}. Note that there is no explicit spatial dependence in the reaction term in equation \eqref{eq:model_pde}.

\subsection{Compartment-based representation} \label{sect:Modelling_meso}
Compartment-based methods are coarse-grained stochastic representations. The 
spatial domain is typically divided into compartments, each of size $h$, in 
which particles are assumed to be well-mixed. The reaction-diffusion dynamics 
are characterised by a set of possible events.
Events are either reactions, in which particles can interact with others within 
their own compartment according to some prespecified reaction rates, or jumps 
to  
adjacent compartments with rates which depend on the macroscopic diffusion 
coefficient, $D(x)$, and the compartment size, $h$. Specifically, in order to 
capture diffusion which corresponds to the macroscopic equation 
\eqref{eqn:Reaction_Diffusion_Equation} we must choose the rates of jumping to 
be different depending on the direction of the jump (see equations 
\eqref{equation:left_jump_rate} and \eqref{equation:right_jump_rate} for more 
detail).

Throughout this paper we refer to models at this scale as  \textit{mesoscopic} 
or \textit{compartment-based}. For a discussion of the implementation of the 
numerical simulation of the compartment-based models employed in this paper 
please refer to 
Appendix \ref{section:appendixB}.

\subsection{Brownian-based representation} \label{sect:Modelling_micro}
Individual-based  methods require the recording and updating of large numbers of 
particles' positions. Relative positions for each pair of particles must also be 
maintained at every step if higher-order reactions (higher than first-order) or 
volume-exclusion are to be modelled. For large particle numbers, $N$, the 
$O(N^2)$ computational complexity means that individual-based 
simulation algorithms can become extremely expensive\footnote{As previously noted some of this complexity can be offset by a careful partitioning of space allowing particles to be compared only with others in their local neighbourhood \cite{robinson2017pbm}.}.

In what follows we employ a fixed-time-step algorithm, although we note that 
continuous-time algorithms for Brownian reaction-diffusion dynamics are also 
available \cite{van2005gfr}. The evolution of particle $i$'s position, 
$y_i(t)$, between times $t$ and $t+\Delta t_b$ in the case of space-dependent 
diffusion (corresponding to PDE \eqref{eq:model_pde} and compartment-based 
jump-rates given by equations \eqref{equation:left_jump_rate} and 
\eqref{equation:right_jump_rate}) can be simulated according to the following 
discrete-time update equation
\begin{equation}
y_i(t+\Delta t_b) = y_i(t) +\Delta t_b\frac{\ud D(x)}{\ud x}\Bigg|_{x=y_i(t)} +  \sqrt{2D(y_{i}(t))\Delta t_b}~\xi_i,\label{equation:edge_SDE}
\end{equation}
where $\xi_i\sim N(0,1)$ is a Gaussian random variable with mean $0$ and variance $1$. If required, reactions can be implemented according to a variety of different 
algorithms \cite{van2005gfr,smoluchowski1917vem}. In this paper, we employ the 
$\lambda$-$\rho$ method \cite{erban2009smr}. If two eligible particles come 
within a reaction radius, $\rho$, of  each other they interact with a given 
rate, $\lambda$, according to the appropriate reaction pathway. 

We refer to these models at this scale as  \textit{off-lattice}, 
\textit{microscopic} or  \textit{individual-based} models in what follows.

\subsection{Connections between models at different scales} \label{section:Modelling_Equivalence} 
In attempting to couple together representations of the same phenomenon at 
different scales we need to ensure that, under certain assumptions, they are 
representations of the same process. Pioneering work in establishing the connection between stochastic and deterministic models was undertaken by \cite{gillespie2009dls}, \cite{van2007spp} and \cite{kurtz1972rbs}. In this section we concisely summarise the ways in 
which the models outlined above can be considered to be equivalent and direct 
the interested reader to resources which contain more detailed arguments. 

In order to transition from the mesoscale to the macroscale, we can first use the reaction-diffusion master equation to derive the deterministic mean-field representation of the compartment-based particle numbers \cite{erban2009smr,baker2009fmm,othmer1988mod}. 
It should be noted that for second- and higher-order reactions, the mean-field equations are only approximations of the true mean behaviour of the stochastic system  \cite{erban2007pgs}. Taking the diffusive limit of the mean-field equations gives a corresponding reaction-diffusion PDE. 

The Fokker-Planck equation can be used to connect a microscale stochastic 
differential equation (SDE) model of diffusion to a macroscale model describing 
the evolution of the probability density of a particle's position 
\cite{erban2007pgs,risken1989fokker}. For example, the canonical diffusion 
equation is the macroscopic Fokker-Planck equation corresponding to 
non-interacting particles undergoing simple Brownian motion. 

Although we do not use this macroscopic-microscopic coupling directly in this 
work, we employ it indirectly in order to link the microscopic and mesoscopic 
descriptions together through their connection to the same PDE. Alternatively, 
first-passage time theory can be applied to a particle which moves subject to a 
given SDE in order to derive jump rates between neighbouring compartments in a 
compartment-based representation \cite{redner2001gfp,yates2013ivd}. Connections 
between the models at microscale and mesoscale are stated more rigorously by 
\cite{isaacson2008rbr}.

\section{Hybrid blending algorithms}
\label{section:hybrid_blending_algorithms}
In this section we discuss the two main algorithms of this  paper. In particular, in Section \ref{subsec:hybrid_split} we present the central unifying idea 
behind both of our hybrid methods. The methods can both be understood 
as operator-splitting algorithms in which, in a central overlap region between the two regimes, diffusion is dealt with by both regimes using spatially varying diffusion coefficients. We discuss how to couple the methods discussed in Section \ref{section:modelling_at_different_scales} in order to accommodate this split-diffusion paradigm. In Section \ref{subsection:conversion_rules} we give the specific details of how to convert mass from one modelling regime to another to ensure both models are synchronised and valid representations of the particle density in the blending region. We then present, in Section \ref{subsec:algo}, a generic algorithm for coupling the PDE with the compartment-based approach, as well as  a similarly general algorithm for coupling the compartment-based approach with Brownian dynamics. We emphasise that the generic methods we present for coupling two regimes are independent of the numerical implementations chosen to simulate each regime. However, for ease of use and reproducibility we have provided details of the numerical implementations we chose in Appendices \ref{section:appendixA}-\ref{section:appendixC}.

\subsection{Hybrid modelling interpreted as a splitting algorithm}
\label{subsec:hybrid_split}
In order to illustrate the conceptual framework behind our algorithms we 
consider the following constant coefficient diffusion PDE in $\Omega=[a,b]$:
\begin{equation} \label{eq:main}
\frac{\partial c}{ \partial t}= \frac{\partial}{\partial x} \left( D \frac{\partial c}{\partial x}\right),
\end{equation}
with the following zero-flux boundary conditions: 
\begin{equation} 
 \quad D\frac{\partial c}{\partial x} \Bigr|_{x=a}=D\frac{\partial c}{\partial x} \Bigr|_{x=b}=0.
 \end{equation}
Divide the domain, $\Omega$, into three subdomains $\Omega_{1}=[a,I_{1}], \ 
\Omega_{2}=[I_{1},I_{2}], \ \Omega_{3}=[I_{2},b]$ and write the constant 
diffusion coefficient $D=D_{1}(x)+D_{2}(x) $ where

\begin{equation}
 {\displaystyle D_1(x)={\begin{cases}D,&a\leq x<I_1,\\
 {f_1(x)},&I_1\leq x<I_2,\\
 0,&I_2\leq x\leq b,\end{cases}}}
 \label{equation:diffusion_coefficient_1}
\end{equation}
and 
\begin{equation}
 {\displaystyle D_2(x)={\begin{cases}0,&a\leq x<I_1,\\
 {f_2(x)},&I_1\leq x<I_2,\\
 D,&I_2\leq x \leq b,\end{cases}}}
  \label{equation:diffusion_coefficient_2}
\end{equation}
where $f_1$ and $f_2$ are monotonically decreasing/increasing functions, respectively, with $f_1(x)=D-f_2(x)$ and $f_1(I_1)=f_2(I_2)=D$ and $f_1(I_2)=f_2(I_1)=0$  in order to ensure continuity of $D_1$ and $D_2$ across $\Omega$.

Equation \eqref{eq:main} can now be written as 
\begin{equation} \label{eq:main_split}
\frac{\partial c}{\partial t} = \underbrace{\frac{\partial}{\partial x} \left( D_{1}(x) \frac{\partial c}{\partial x}\right)}_{1}+\underbrace{\frac{\partial}{\partial x} \left( D_{2}(x) \frac{\partial c}{\partial x}\right)}_{2},
\end{equation}
with corresponding boundary conditions
\begin{equation}
 (D_1(x)+D_2(x))\frac{\partial c}{\partial x} \Bigr|_{x=a} = D\frac{\partial c}{\partial x} \Bigr|_{x=a}=0 \quad \text{ and }\quad 
 (D_1(x)+D_2(x))\frac{\partial c}{\partial x} \Bigr|_{x=b}= D\frac{\partial c}{\partial x}\Bigr|_{x=b}=0.
\end{equation}
In addition we specify the initial condition $c(x,0)=c_0(x)$. It is 
straightforward to show that, because $D_{1}(x)=0$ in $[I_{2},b]$, the operator 
indicated by $1$ in equation
\eqref{eq:main_split} does not influence the concentration of $c$ in that region. In a similar way, because $D_{2}(x)=0$ in $[a,I_1]$, the operator indicated by $2$ in equation
\eqref{eq:main_split} does not influence the concentration of $c$ in that region. Now let  $\phi^{1}_{\tau}, \phi^{2}_{\tau}
$ be the flow maps associated with the propagation of the operators  $1$ and $2$ in equation \eqref{eq:main_split} until time $\tau$. Specifically this means that the solution of the following equations 
\begin{subequations}
\begin{align} \label{eq:parts_1}
\frac{\partial c^{(1)}}{ \partial t} &= \frac{\partial }{ \partial x} 
\left(D_{1}(x) \frac{\partial c^{(1)}}{\partial x} \right), \quad 
D_{1}(a)\frac{\partial c^{(1)}}{\partial x} 
\Bigr|_{x=a}=D_{1}(I_{2})\frac{\partial c^{(1)}}{\partial x} \Bigr|_{x=I_{2}}=0,
\\
\frac{\partial c^{(2)}}{ \partial t} &= \frac{\partial }{ \partial x} 
\left(D_{2}(x) \frac{\partial c^{(2)}}{\partial x} \right), \quad 
D_{2}(I_{1})\frac{\partial c^{(2)}}{\partial x} 
\Bigr|_{x=I_{1}}=D_{2}(b)\frac{\partial c^{(2)}}{\partial x} \Bigr|_{x=b}=0, \label{eq:parts_2}
\end{align}
\end{subequations}
subject to initial conditions $c^{(i)}(x,0)=c^{(i)}_{0}(x)$ can be written as 
$c^{(i)}(x,\tau) = \phi^{(i)}_{\tau}(c^{(i)}_{0})(x), \ \text{for} \ i=1,2$,  respectively\footnote{Note that due to the choice of blending functions the boundary condition at $I_2$ in \eqref{eq:parts_1}  and at $I_1$ in \eqref{eq:parts_2} are automatically satisfied.}.

The idea behind splitting methods is that one can now obtain an 
approximation for the solution of equation \eqref{eq:main_split} at time $\tau$ by 
using an appropriate composition of the flow maps $\phi^{(1)}_{\tau}$ and 
$\phi^{(2)}_{\tau}$. In particular, the simplest splitting method is given by
\begin{equation} \label{eq:split_idea}
c(x,\tau) \approx (\phi^{(1)}_{\tau} \circ \phi^{(2)}_{\tau})(c_{0})(x),
\end{equation} 
where we note that the ordering of the composition is unimportant.

At a first glance this seems like an unnecessarily complicated approach for 
obtaining an approximation for the solution of equation \eqref{eq:main}. 
However, choosing the flow maps $\phi^{1}_{\tau}$ and $\phi^{2}_{\tau}$ to 
represent propagation operators for two different model types allows us to 
seamlessly blend the distinct numerical update rules of the different modelling 
regimes described in Section \ref{section:modelling_at_different_scales}. For 
example, when coupling the PDE to the compartment based model, $\phi^{1}_{\tau}$ might represent an update operator for the numerical solution of the PDE up to time $\tau$, whilst $\phi^{2}_{\tau}$ might represent steps of the position-jump Markov processes described in Section \ref{sect:Modelling_meso} up until time $\tau$.

Due to the properties of the diffusion functions $D_{i}(x)$, the two models only 
co-exist in the blending region $[I_{1},I_{2}]$. Therefore, in applying the operator splitting update illustrated in equation \eqref{eq:split_idea}, 
we only need to worry about how the concentration of the numerical solution of the PDE in the blending region translates to particle 
numbers for the  compartment-based approach and vice versa. We must ensure that any PDE solution update in the blending region implemented by operator $\phi^{1}_{\tau}$ is also reflected in the compartment-based solution. Equivalently, any update to the compartment-based solution in the blending region implemented via $\phi^{2}_{\tau}$ must be reflected in the PDE solution. In a similar way, when coupling the compartment-based model to Brownian dynamics, one need only worry about how the particle numbers for the compartment-based approach in the blending region impact on the particle positions of the off-lattice Brownian dynamics and vice versa. Outside the two blending regimes the two representations are effectively decoupled in terms up their update operators.

\subsection{Conversion rules}\label{subsection:conversion_rules}
In this section we illustrate how to couple two distinct representations of reaction-diffusion processes in the blending region. First we tackle a PDE-compartment-based hybrid pairing, followed by a coupling between compartment-based and Brownian-based particle dynamics.

\paragraph{Conversion between PDE and compartment-based model:} 
We assume that the numerical solution of the PDE is calculated on the discrete 
mesh\footnote{Note that we describe the coupling between the two regimes 
in the blending region using the terminology of the finite volume PDE 
discretisation that we employ in our numerical examples (see Section 
\ref{section:results}). However, we also note that finite volume voxels can be 
substituted for finite difference or finite element mesh points in a 
straightforward manner.} (see figure \ref{fig:pde_discretisation} in Appendix \ref{section:appendixA} for an illustration) of size 
$\Delta x$ in $[a,I_{2}]$ and that compartment-based dynamics are simulated with 
compartments of size $h$ in $[I_{1},b]$. It is natural to assume that $h  
\geq \Delta  x$, as a fine discretisation of the PDE mesh is required in order 
to minimise the error between the numerical solution and the exact solution it 
approximates. Note, however, that this is not a limitation of our 
algorithm and that $h \leq \Delta x$ would also be possible. There are 
$n_{1}=\frac{I_{2}-I_{1}}{\Delta x}$ PDE solution voxels in the overlap region 
$[I_{1},I_{2}]$ and $n_{2}=\frac{I_{2}-I_{1}}{h}$ compartments in the same 
region, where $n_1,n_2\in \mathbb{N}$. For ease of computation we assume that $n_{1}=\gamma n_{2}$, with $\gamma 
\in \mathbb{N}$ so that there are an integer number of PDE solution voxels 
per compartment. There are also $n_p=(I_1-a)/\Delta x$ PDE solution voxels in 
the purely PDE region, $[a,I_1]$, and $n_c=(b-I_2)/h$ compartments in the 
purely compartment-based regime $[I_2,b]$. The numerical solution of the PDE in 
voxel $i$ is labelled $q_i$ for  $i=1, \dots, n_p+n_1$ and the number of 
particles in compartment $i$ is labelled $C_i$ for $i=1, \dots, n_2+n_c$.

In each time interval of length $\Delta t_p$ we assume, without loss of generality, 
that the PDE solution is updated first and the compartment-based solution 
second.
After the propagation of the discrete PDE solution operator in the time interval 
$[t,t+\Delta t_p]$, assume that the concentrations in PDE voxels of the blending 
region  have changed. Consequently it is necessary to modify the 
corresponding compartment-based description in the blending region 
$[I_{1},I_{2}]$ before propagating the compartment-based model in the region 
$[I_{1},b]$. More precisely, for compartment $i$ in the blending region, set
\begin{equation}
\label{eq:conv_rule1}
C_{i}=\sum_{j=1}^{\gamma}q_{n_p+\gamma (i-1)+j} \Delta x.
\end{equation}
Because we are required to synchronise the representations of the solutions in the two regimes according to equation \eqref{eq:conv_rule1}, the number of particles 
contained in the $i$-th compartment in the blending region is no longer an integer.  
Nevertheless, when it comes to performing the stochastic simulation algorithm we 
work with these non-integer values to calculate the time until the next event. 
This could potentially be an issue when the copy numbers in a compartment are 
low, but arguably
this would imply that we were using the PDE description to represent concentrations in a region of the domain for which this is not appropriate. 
A similar synchronisation is implemented once the compartment-based model has 
been propagated  and the number of particles in the blending region has changed. 
In particular, if $\delta C_{i}$ corresponds to the integer change in particle 
numbers in the compartment $i$ in the blending region, then one adds uniformly 
$\delta C_{i}/\gamma \Delta x$ to the PDE solution in each of the PDE voxels, 
\emph{i.e}
\begin{equation} \label{eq:conv_rule2}
q_{n_p+\gamma i+j}=q_{n_p+\gamma i+j}+\frac{\delta C_{i}}{\gamma \Delta x}, \quad j=1,\cdots, \gamma.
\end{equation} 

Reactions in the blending region are always implemented according to the 
compartment-based paradigm. If reactions occur then particle numbers in 
compartments are updated and the corresponding change is also implemented in the 
appropriate PDE voxels, as in equation \eqref{eq:conv_rule2}.

\begin{figure}[h!]
\begin{center}
\subfigure[][]{
\includegraphics[width=0.47\textwidth]{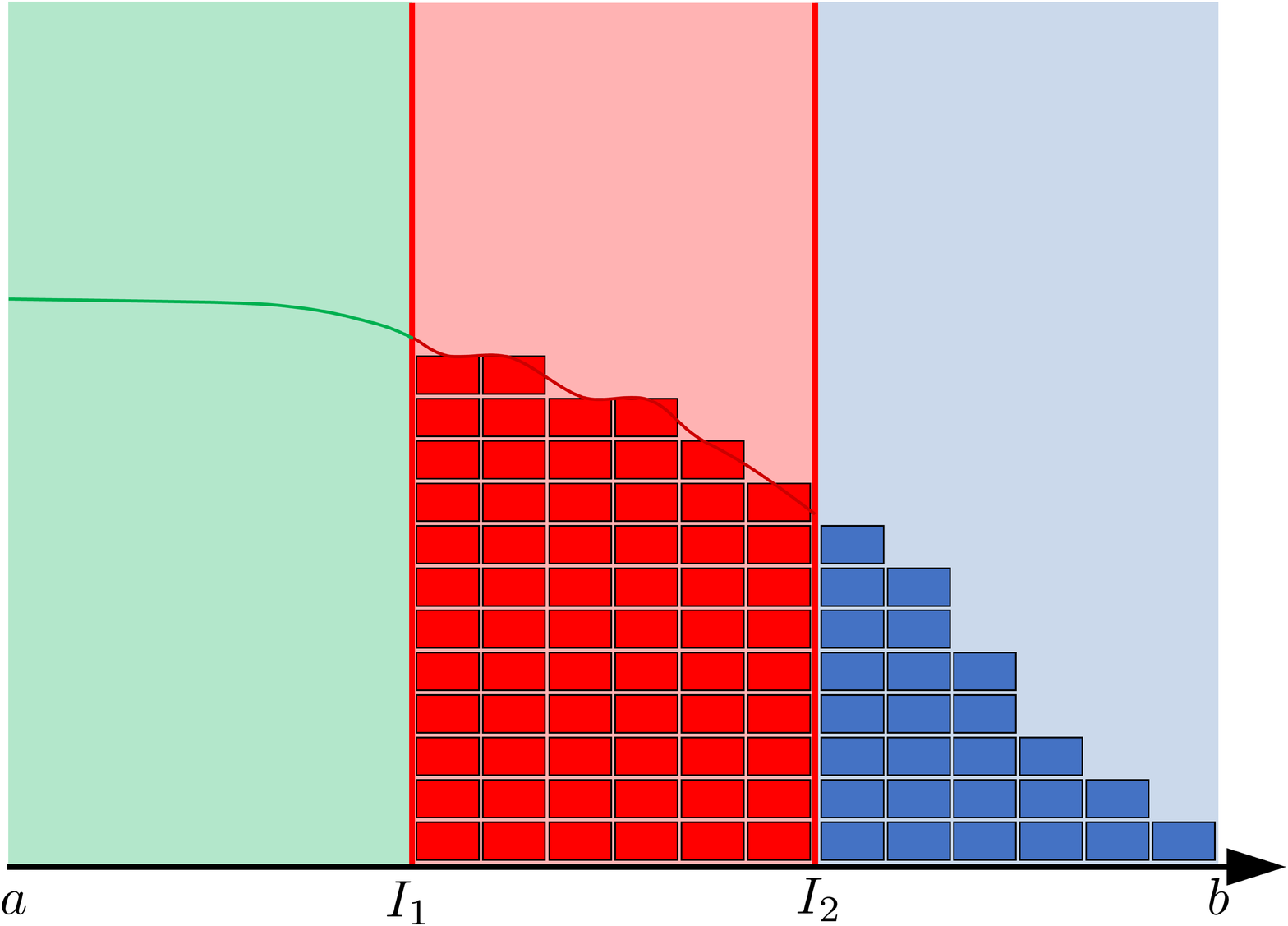}
\label{figure:blending_macro_meso_schematic}}
\subfigure[][]{
\includegraphics[width=0.47\textwidth]{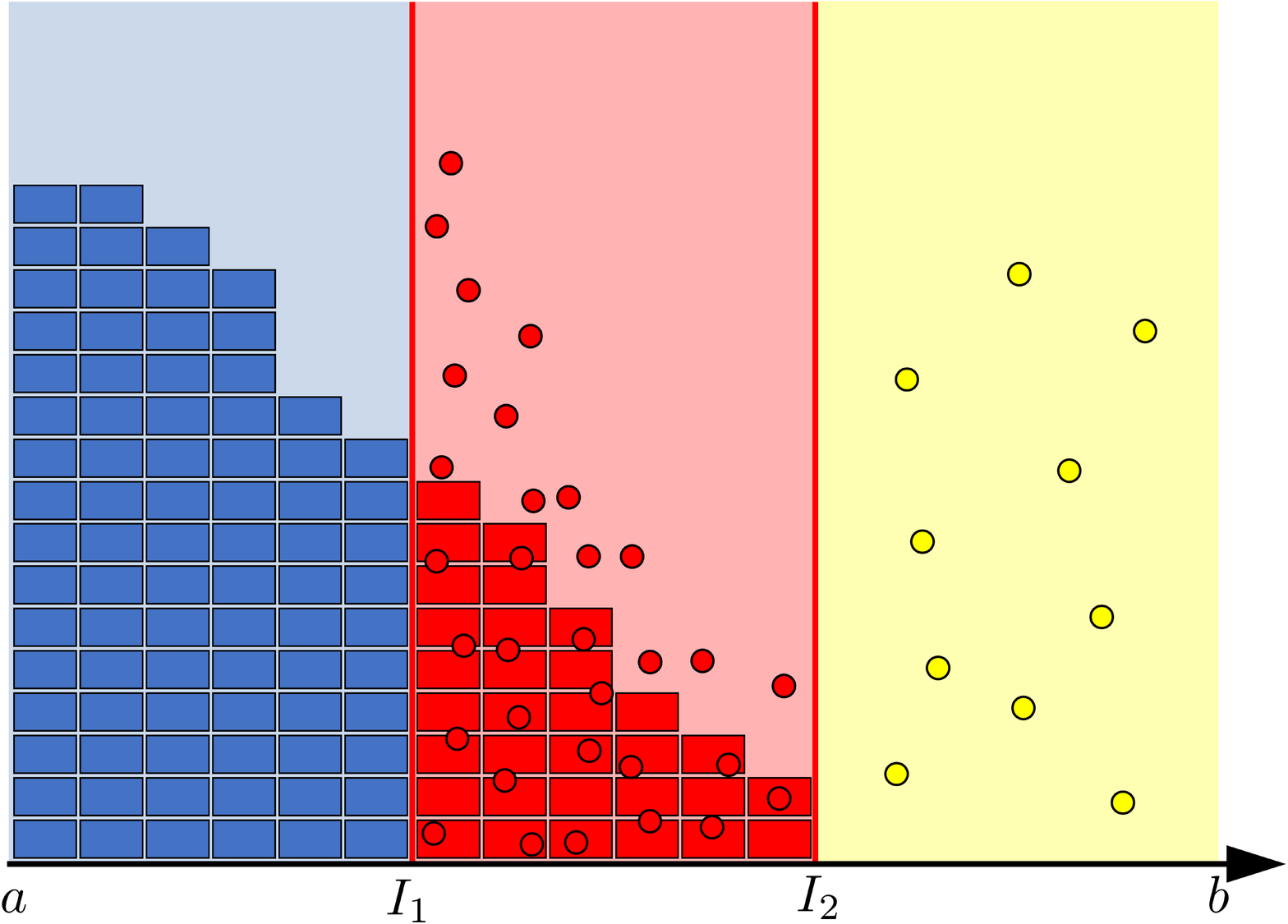}
\label{figure:blending_meso_micro_schematic}}
\end{center}
\caption{Schematic representations of \subref{figure:blending_macro_meso_schematic} the PDE-compartment hybrid and \subref{figure:blending_meso_micro_schematic} the compartment-Brownian hybrid. In panel \subref{figure:blending_macro_meso_schematic} the green curve in the green region $[a,I_1]$ represents the PDE solution in the purely PDE region of the domain. The red curve and the red boxes represent equivalent PDE- and compartment-based representations of the mass in the red blending region. The blue boxes in the blue region of the domain represent the number of particles in each compartment in the purely compartment region of the domain. In panel \subref{figure:blending_meso_micro_schematic} the blue boxes in the blue region of the domain represent the number of particles in each compartment in the purely compartment region of the domain. The red boxes and the red circles represent equivalent compartment- and Brownian-based representations of the mass in the red blending region. The yellow circles in the yellow region of the domain represent individual particles in the purely Brownian region of the domain. Note that we have given each Brownian particle a different height to aid clarity of visualisation, but  in reality all particles lie on the $x$-axis in these one-dimensional simulations.}
\label{figure:blending_schematics}
\end{figure}

\paragraph{Conversion between compartment based and individual particle models}

Without loss of generality assume that the compartment-based model is 
employed in $[a,I_{2}]$ and the Brownian-based model is employed in $[I_1,b]$ 
with the two models being simultaneously employed in the blending region 
$[I_1,I_2]$. Compartment-based dynamics are simulated with compartments of size 
$h$ in $[a,I_{2}]$. There are $n_c=(I_1-a)/h$ compartments in the purely 
compartment-based region, $[a,I_1]$, and $n_2=(I_2-I_1)/h$ compartments in the 
overlap region $[I_1,I_2]$. The number of particles in compartment $i$ is, as 
before, labelled $C_i$ for $i=1, \dots, n_c+n_2$. Brownian particles are 
simulated off-lattice with positions updated according to the dicretised SDE 
\eqref{equation:edge_SDE} in $[I_1,b]$.

In each time interval of length $\Delta t_b$ assume, without loss of 
generality, that the compartment-based solution is updated first, followed by 
the Brownian-based dynamics.
During the propagation of the compartment-based solution  it is likely that the 
numbers of particles in the compartments of the blending region have changed. 
Consequently we need to alter the positions of Brownian particles in the blending 
region. If a particle jumps from compartment $i$ to a neighbouring compartment 
$j$ in the hybrid region, then we select a Brownian particle uniformly at random 
from amongst the particles which currently reside in compartment $i$ and move it 
a distance $\pm h$ with the sign of the displacement corresponding to the 
direction of the compartment-based particle's jump i.e.

\begin{equation}
\label{equation:Brownian_synchronisation}
y_k=y_k\pm h,
\end{equation}
where $k$ indexes the randomly selected Brownian particle from compartment $i$. 

If a particle in compartment $n_c+1$ (the first compartment in the blending region) jumps leftwards out of the blending region (according to the compartment-based jump rates) and into the purely compartment-based region then a Brownian particle in the compartment $n_c+1$ is selected uniformly at random and removed from the simulation (as well as particle numbers in the affected compartments being updated). Conversely, if a compartment-based particle jumps to the right, out of the last compartment in the purely compartment-based regime into the first compartment in the blending region, then a Brownian particle is added with its position chosen uniformly at random in this compartment, $[I_1,I_1+h]$ (as well as particle numbers in the affected compartments being updated). Note that the jump rates in the compartment-based model, which implement diffusion corresponding to equation \eqref{eq:model_pde}, are such that, with our chosen blending diffusion coefficients, the rate of jumping to the right out of the final compartment is zero, so that no compartment-based particles can erroneously jump into the purely-Brownian regime. Similarly, the diffusion coefficient of the Brownian particles at the pure-compartment/blending region interface is zero. Technically, with our finite time-step implementation of diffusion it might be possible for Brownian particles to erroneously jump over the interface into the purely compartment-based regime\footnote{Whilst there do exist integrators for diffusion processes which can guarantee that this situation does not happen \cite{beskos2005esd}, implementing such an approach is beyond the scope of the article.}. On the rare occasions that a Brownian particle is chosen to jump over the interface (as an artefact of the numerical implementation) we simply reflect it back into the blending region. Since the diffusion coefficient is low in the boxes close to the interface this very rarely happens, and when it does the error caused by reflecting the particle is minimal.

Once the particle-based method has been propagated, it is usually necessary 
to update the number of particles in the compartments of the blending region, 
$C_i$ for $i=n_c+1,\dots,n_c+n_2$. Rather than tracking  every 
Brownian-particle movement to see whether it has crossed over a compartment 
boundary, instead we simply sum the number of Brownian-based particles in each 
compartment at the end of the Brownian update to find the numbers of particles 
in each compartment of the blending region:
\begin{equation}
\label{equation:compartment_based_synchronisation}
C_{n_c+i}=\sum_{k=1}^N \mathbb{I}_{y_k\in[I_1+(i-1)h,I_1+ih]},\quad \text{for} \quad i=1,\dots,n_2,
\end{equation}
where $\mathbb{I}_{y\in[I_1+(i-1)h,I_1+ih]}$ is the indicator function which takes the value 1 if the Brownian particle lies in the $(n_c+i)$th compartment and 0 otherwise.

Reactions in the blending region (similarly to the PDE-compartment hybrid method)  are always implemented using the compartment-based paradigm. If a reaction occurs in the hybrid region then the appropriate Brownian particles are added (with positions chosen uniformly at random across the corresponding compartment) or removed (with the particle(s) selected uniformly at random from amongst those in the compartment).

\subsection{Coupling algorithms}
\label{subsec:algo}
Having established the conversion rules in the previous section we are now in the position to present two hybrid algorithms. In particular,  Algorithm \ref{algo1} is the algorithm that couples diffusion in the PDE and compartment-based models, while Algorithm \ref{algo2} is the algorithm that couples diffusion in the compartment-based models with Brownian-based dynamics. We have presented both of these algorithms with maximum generality in order to emphasise that the specific simulation methodologies are not important. In the next section we implement these algorithms with a finite volume PDE solver, the spatial Gillespie algorithm for compartment-based dynamics and the $\lambda-\rho$ Brownian reaction-diffusion paradigm for the Brownian dynamics, in order to provide concrete examples of their implementation. Algorithms for the implementation of these three methods are given in Appendices A, B and C respectively.

\begin{algorithm}[h]
  \KwIn{PDE mesh size -- $\Delta  x$; compartment size -- $h$; time-step for the 
solution of the PDE -- $\Delta t_p$; left and right ends of the blending region -- 
$I_{1}, I_{2}$ ; initial concentration for the PDE -- $c_{init}$ ; initial 
particle numbers -- $\bs{C}_{init}$ ; final time -- $T$.}
\SetAlgoLined
Set $t = 0$.
\\
\While{$t<T$}{
Simulate diffusion due to the PDE  in  $[a,I_{2}]$ (and reactions due to the PDE in $[a,I_1]$) between $t$ and $t+\Delta t_p$ for diffusion coefficient given by $D_{1}(x)$ using an appropriate numerical solver.

Update the compartment-based particle numbers in $[I_{1},I_{2}]$ according to equation \eqref{eq:conv_rule1}.

Simulate diffusion and reactions due to the compartment-based approach in $[I_{1},b]$ between $t$ and $t+\Delta t_p$ for diffusion coefficient given by $D_{2}(x)$ using an appropriate stochastic simulation algorithm,  taking as an initial condition the updated particle numbers from line 4.

Update the PDE solution in $[I_{1},I_{2}]$  according to equation \eqref{eq:conv_rule2}.

Set $t=t+\Delta t_p$.

}
\caption{Coupling a PDE solution with a compartment-based approach}\label{alg:pde_comp}
\label{algo1}
\end{algorithm}



\begin{algorithm}[h]
\KwIn{Compartment size -- $h$; time-step for the solution update of the Brownian 
dynamics -- $\Delta t_b$; left and right ends of the blending region -- $I_{1}, 
I_{2}$; initial particle numbers -- $\bs{C}_{init}$ ; initial Brownian particle 
positions -- $\bs{y}$; final time -- $T$.}
\SetAlgoLined
Set $t = 0$.
\\
\While{$t<T$}{
Simulate diffusion and reactions due to the compartment-based approach in $[a,I_2]$ between $t$ and $t+\Delta t_b$ for diffusion coefficient given by $D_{1}(x)$ using an appropriate stochastic simulation algorithm.

Update the Brownian particle positions in $[I_{1},I_{2}]$  according to equation \eqref{equation:Brownian_synchronisation} (if appropriate).

Simulate diffusion due to the Brownian particle dynamics in  $[I_{1},b]$ (and reactions due to the Brownian particle dynamics in $[I_2,b]$) between $t$ and $t+\Delta t_b$ for diffusion coefficient given by $D_{2}(x)$ using an appropriate numerical solver, taking as an initial condition the updated concentration from line 4.

Update the compartment-based particle numbers in $[I_{1},I_{2}]$ according to equation \eqref{equation:compartment_based_synchronisation}.

Set $t=t+\Delta t_b$.

}
\caption{Coupling a compartment-based approach with Brownian dynamics.}\label{alg:comp_brown}
\label{algo2}
\end{algorithm}

\section{Results}
\label{section:results}
In this section we demonstrate that our proposed algorithms correctly 
simulate four test problems of increasing complexity. The first two 
problem are simulations  of pure diffusion with different initial conditions, 
demonstrating that the fluxes over the interface of the hybrid model are 
consistent with the expected behaviour of the finer-scale representation in 
each 
hybrid model. The third problem, one of morphogen gradient formation,  
evidences the successful implementation of reactions in our hybrid algorithms. 
Finally, in the fourth test problem we implement a second-order reaction system 
in three dimensions, demonstrating the applicability of the method to more 
complicated scenarios.

For each of the first three test problems, the one-dimensional domain we employ 
is $\Omega=[a,b]=[0,1]$, with $I_1=1/3$ and $I_2=2/3$. The remainder of the 
parameter values for examples 1 and 2 are specified in table 
\ref{table:example_1_parameters}, for example 3 in table 
\ref{table:example_3_parameters} and for example 4 in table 
\ref{table:example_4_parameters}.
The blending functions for these three problems (and by simple extension for the fourth problem) are defined as the simple linear functions
\begin{align}
f_1(x)=2-3x,\\
f_2(x)=3x-1,
\end{align}
which scale the contribution of each method to the diffusion coefficient linearly between 0 and 1 across the blending region.  These, in conjunction with equations \eqref{equation:diffusion_coefficient_1} and \eqref{equation:diffusion_coefficient_2}, define the diffusion coefficients for both regimes across the whole domain.
 For each of the first three example and for both couplings we 
will quantify the qualitative comparisons (provided by density comparison 
snapshots) with error plots displaying the evolution of the difference between 
the averaged profiles of our hybrid methods the mean-field PDE (see equations \eqref{PDE_compartment_relative_mass_error_PDE}-\eqref{compartment_Brownian_relative_mass_error_Brownian}). In the fourth example (for which the PDE is not an exact description of the mean behaviour of the individual-based methods) we will compare the averaged profiles of our hybrid methods with the averaged profiles of the finer scale `ground truth' (e.g. mesoscale or miscroscale) simulations (see equations \eqref{PDE_compartment_relative_mass_error_PDE_3D}-\eqref{Brownian_compartment_relative_mass_error_Brownian_3D}).

\begin{table}[h!!!!!!!!!!]
\begin{center}
\begin{tabular}{|c|c|c|}
\hline
Parameter &  Value & Description \\
\hline
$N$ & 1000 & Number of particles \\
$\Omega$ & $[0,1]$ & Spatial domain \\
$D$ &1 & Diffusion coefficient\\
$K$ & $20$ & Number of compartments\\
$h$ & $1/30$ & Compartment width \\
$\Delta x$ & $1/300$ & PDE voxel width\\
$ \Delta t_p$ & $10^{-4} $  & PDE time-step\\
$ \Delta t_b$ & $10^{-4} $  & Brownian time-step\\
M & 500 & Number of repeats\\
\hline
\end{tabular}
\captionof{table}{Table of Parameter values used for the pure diffusion 
simulation of test problems 1 and 2.}
\label{table:example_1_parameters}
\end{center}
\end{table}

\subsection{Test Problem 1: Uniform distribution}\label{Test Problem 1}
The first test of our hybrid algorithms is to determine whether, when 
simulating diffusion, they are 
capable of maintaining the uniform steady state distribution across the domain 
without introducing any bias. We initialise particles uniformly across the 
domain and implement zero-flux boundary conditions.

\begin{figure}[h!]
\begin{center}
\subfigure[][]{
\includegraphics[width=0.31\textwidth]{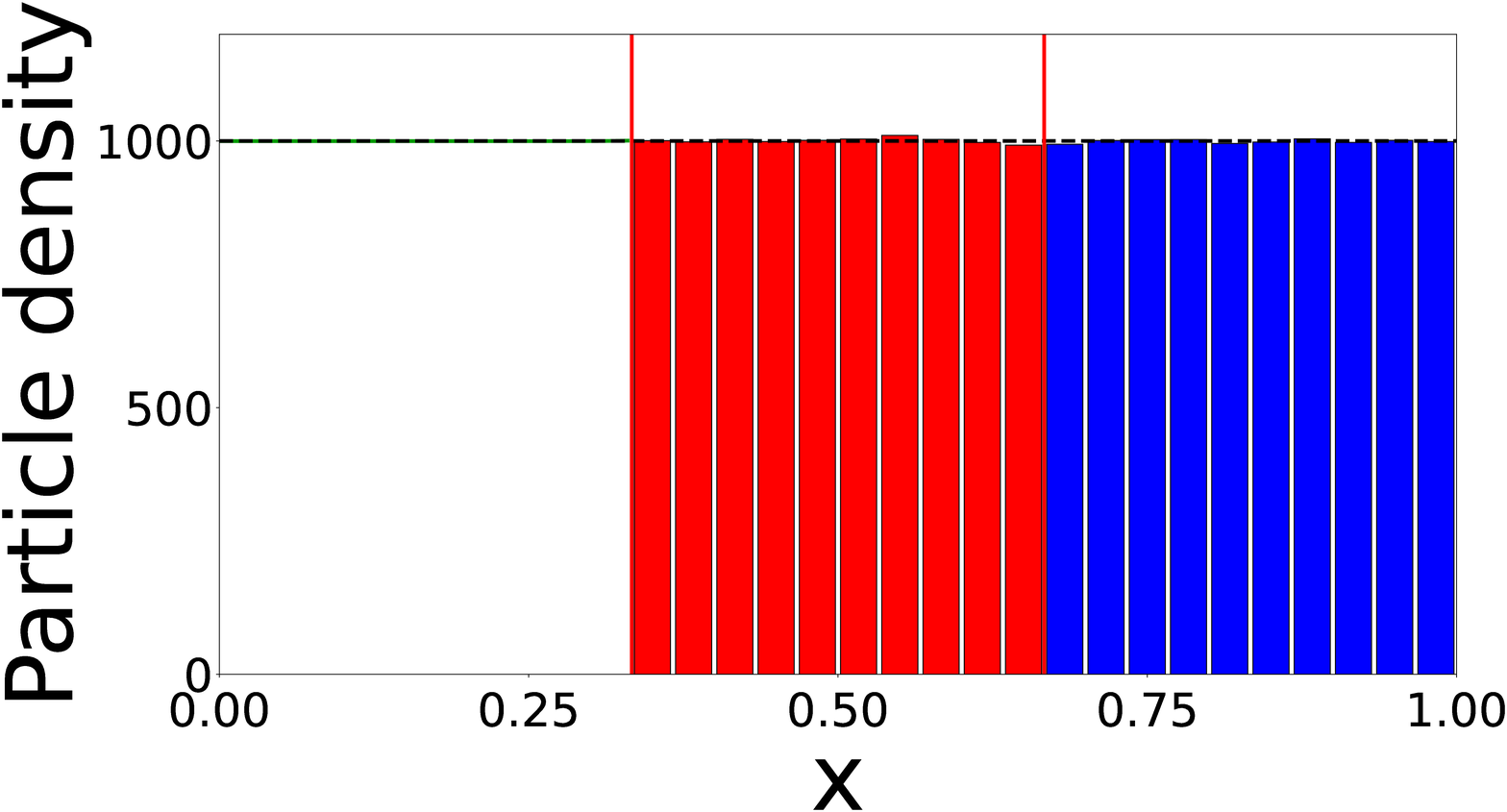}
\label{figure:uniform_PDE_0.1}
}
\subfigure[][]{
\includegraphics[width=0.31\textwidth]{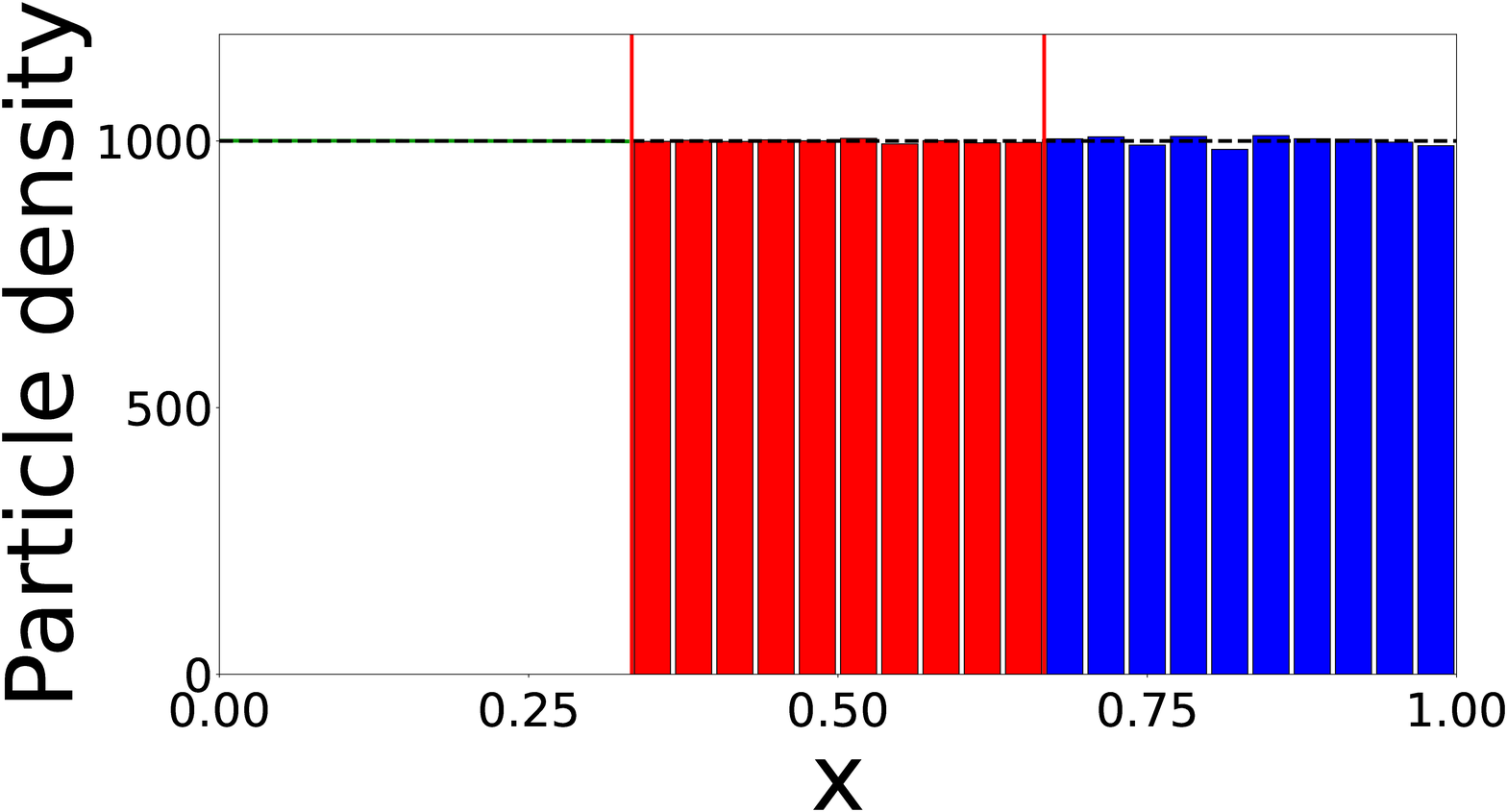}
\label{figure:uniform_PDE_1}
}
\subfigure[][]{
\includegraphics[width=0.31\textwidth]{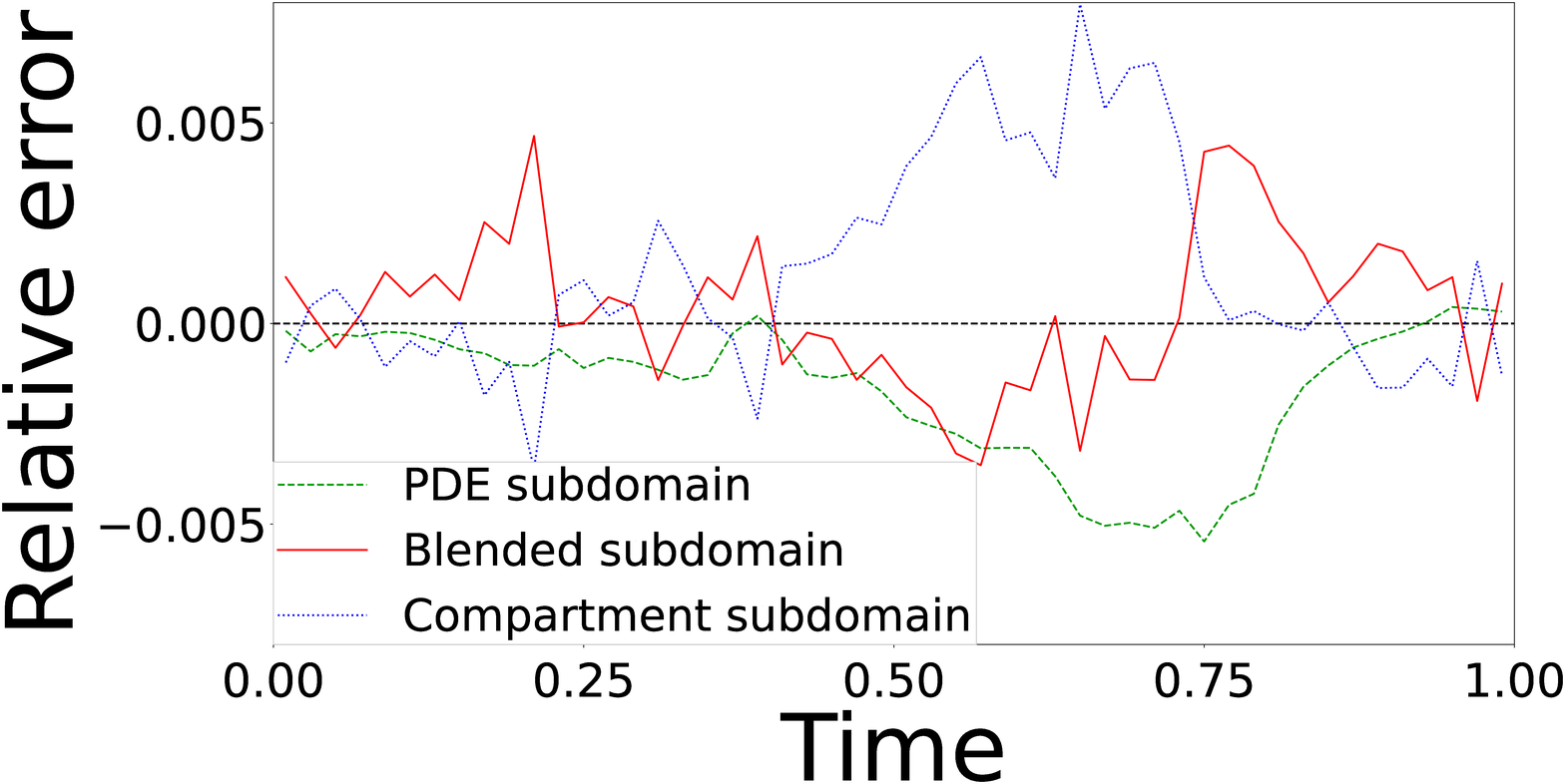}
\label{figure:uniform_PDE_rel_error}
}
\subfigure[][]{
\includegraphics[width=0.31\textwidth]{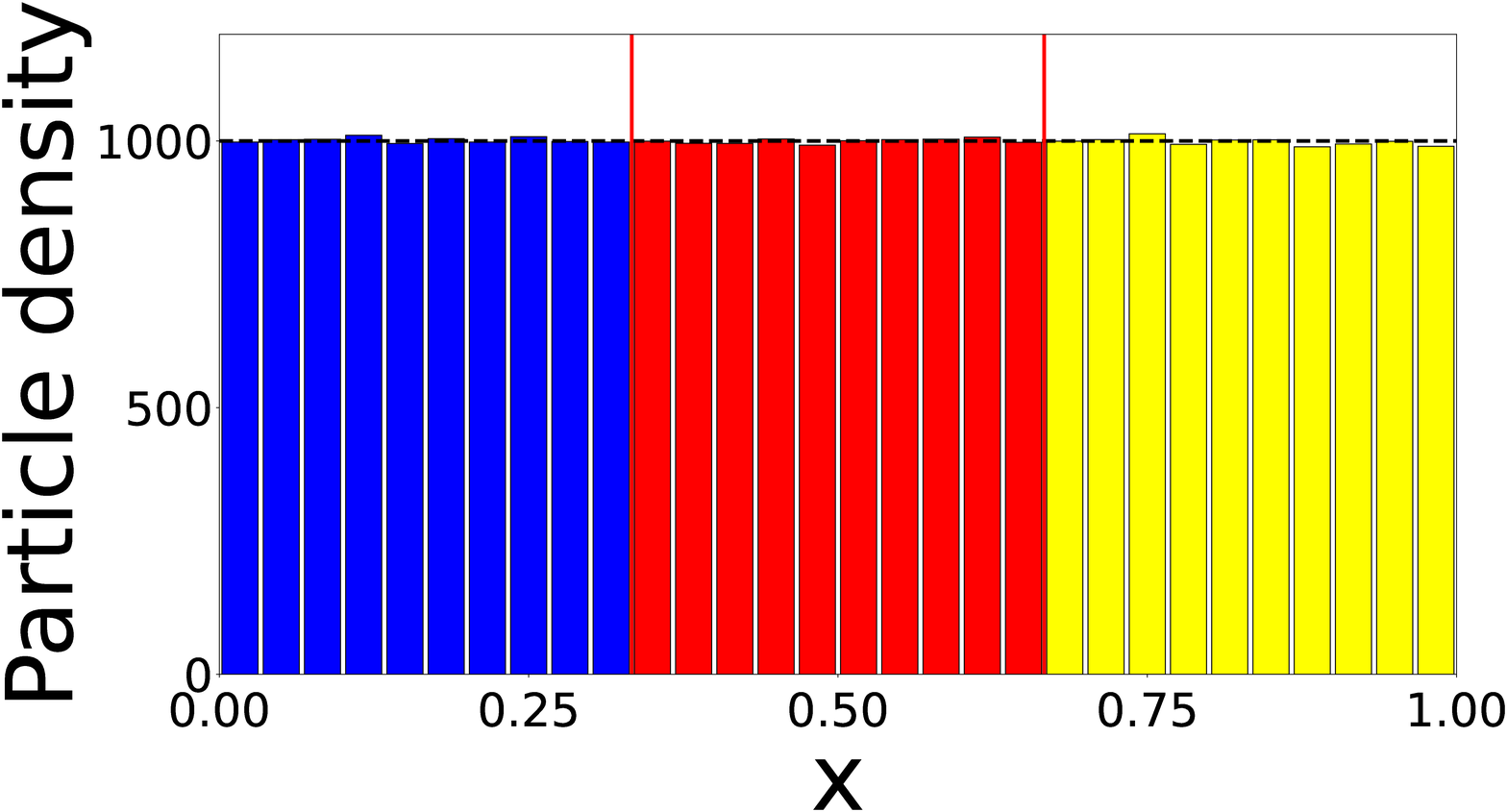}
\label{figure:uniform_Brown_0.1}
}
\subfigure[][]{
\includegraphics[width=0.31\textwidth]{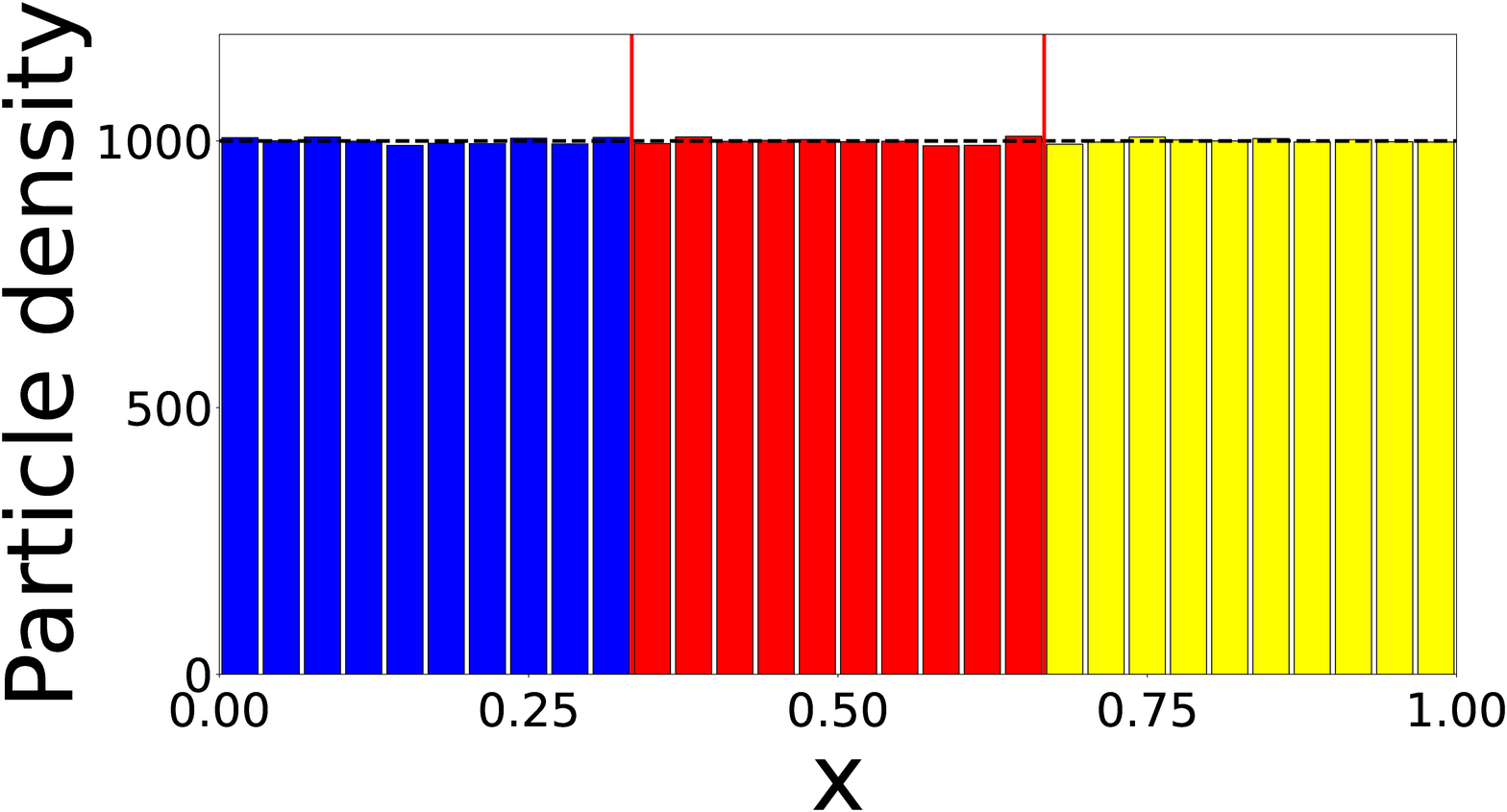}
\label{figure:uniform_Brown_1}
}
\subfigure[][]{
\includegraphics[width=0.31\textwidth]{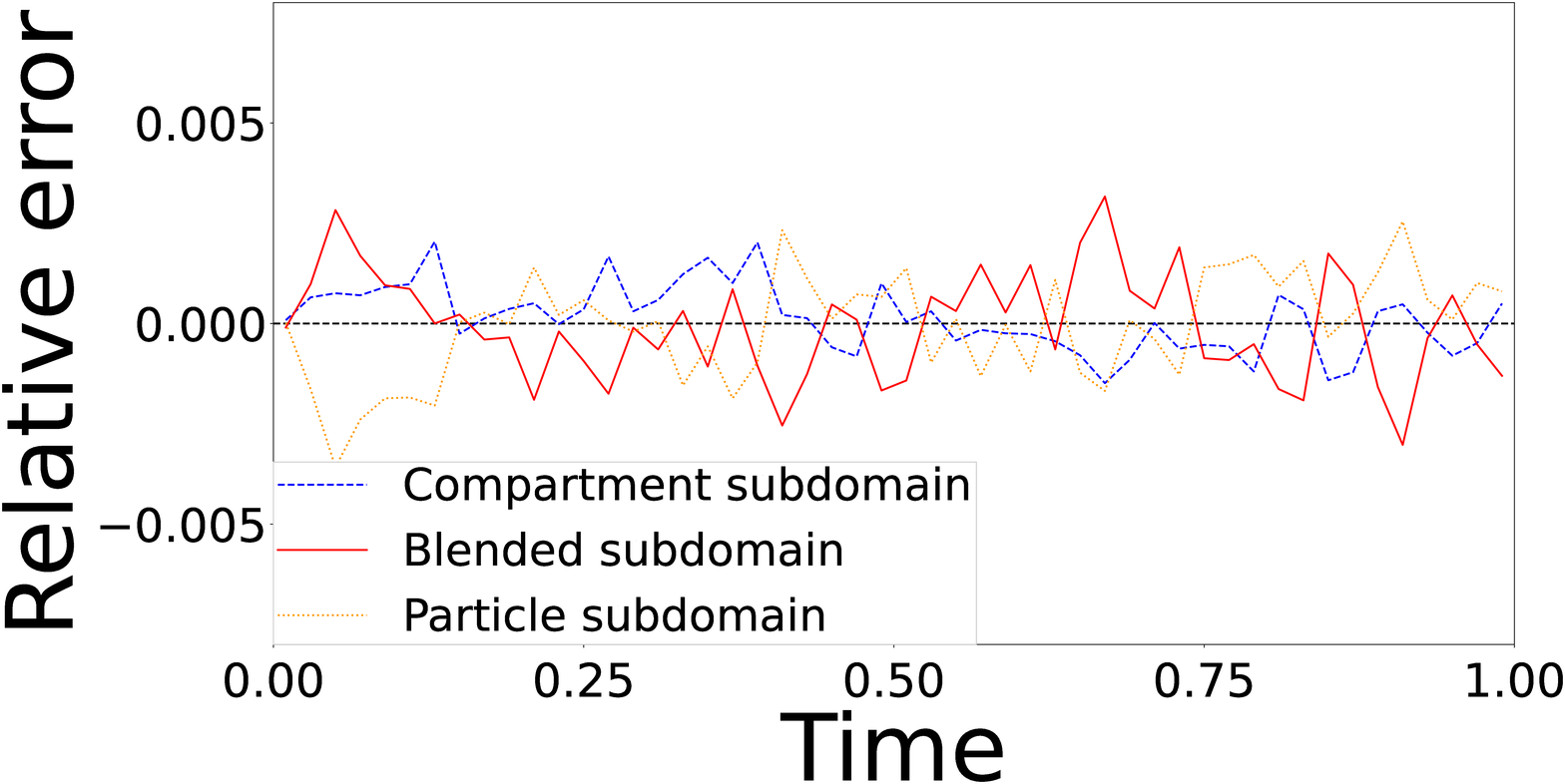}
\label{figure:uniform_Brown_rel_error}
}
\end{center}
\caption{Density and error plots for test problem 1 - pure diffusion with a 
uniform initial condition. Panels 
\subref{figure:uniform_PDE_0.1}-\subref{figure:uniform_PDE_rel_error} are for 
the PDE-compartment hybrid method and 
\subref{figure:uniform_Brown_0.1}-\subref{figure:uniform_Brown_rel_error} for 
the compartment-Brownian hybrid method. Panels \subref{figure:uniform_PDE_0.1} 
and \subref{figure:uniform_Brown_0.1} are snapshots at time  0.1, and 
\subref{figure:uniform_PDE_1} and \subref{figure:uniform_Brown_1} at time 1. In 
panels \subref{figure:uniform_PDE_0.1} and \subref{figure:uniform_PDE_1} the 
green line is the PDE part of the hybrid method, the red bars represent the 
number of particles in each compartment in the blending region and the blue 
bars represent the number of particles in each compartment in the purely 
compartment-based region. In panels \subref{figure:uniform_Brown_0.1} and 
\subref{figure:uniform_Brown_1} the blue bars represent the number of particles 
in each compartment in the purely compartment-based region, the red bars represent the 
number of particles in each compartment in the blending region and the yellow 
bars the number of particles  (appropriately binned for visualisation purposes) 
in the purely Brownian region. In all four density comparison panels the black 
dashed line represents the analytical solution of the diffusion equation with 
given initial condition. Vertical red lines mark the position of the 
interfaces.
Panel \subref{figure:uniform_PDE_rel_error} shows the relative error (described 
in the main text) between the density given by PDE-compartment hybrid method 
and 
the density given by the analytical solution of the diffusion equation with the 
same initial condition. Similarly panel \subref{figure:uniform_Brown_rel_error} 
shows the relative error (described in the main text) between the density given 
by compartment-Brownian hybrid method and the density given by the analytical 
solution of the diffusion equation with the same initial condition. Results 
shown are for $N=1000$ particles and are averaged over 500 repeats. All other 
parameters are given within table \ref{table:example_1_parameters}.}
\label{figure:uniform_IC}
\end{figure}

In figure \ref{figure:uniform_IC} (as well as for figures 
\ref{figure:step_IC}-\ref{figure:morphogen}) the top three figures are for the 
PDE-compartment coupling and the bottom three figures for the 
compartment-Brownian coupling. The left-most panels display the density profile 
of the hybrid methods at time $t=0.1$ and the middle panels the density profile 
at $t=1$. In both left and middle panels the mean-behaviour of the stochastic 
model simulated across the whole of the domain is displayed as a black, dashed 
line for comparison. The right-most panels display the evolution through time 
of 
the relative mass error of each region of the domain: $[a,I_1]$, $[I_1,I_2]$ 
and 
$[I_2,b]$. For the PDE-compartment coupling the relative mass error (RME) is the 
difference between the average (over 500 repeats - unless otherwise stated) 
number of particles in the given region in the hybrid method and the 
corresponding number in the same region in the analytical solution of the PDE, $u(x,t)$, 
divided by the number of particles in the relevant region of the analytical 
solution of the PDE (to normalise):
\begin{align}
RME_P(t)=&\frac{\int_{\Omega_P} \bar{c}(x,t) dx-\int_{\Omega_P}u(x,t) dx}{\int_{\Omega_P}u(x,t) dx}, 
\label{PDE_compartment_relative_mass_error_PDE}\\
RME_H(t)=&\frac{\sum_i \bar{C}_i(t) \mathbb{I}_{c_i\in\Omega_H} -\int_{\Omega_H}u(x,t) dx}{\int_{\Omega_H}u(x,t) dx},\label{PDE_compartment_relative_mass_error_blending}\\
RME_C(t)=&\frac{\sum_i \bar{C}_i(t) \mathbb{I}_{c_i\in\Omega_C} -\int_{\Omega_C}u(x,t) dx}{\int_{\Omega_C}u(x,t) dx},\label{PDE_compartment_relative_mass_error_compartments}
\end{align}
where $\Omega_P=[a,I_1]$ is the purely PDE region of the domain, $\Omega_H=[I_1,I_2]$ is the blending region and $\Omega_C=[I_2,b]$ is the purely compartment region of the domain. The averaged solution of the PDE component of the hybrid method at position $x$ at time $t$ is denoted $\bar{c}(x,t)$ and the averaged compartment particle numbers in voxel $i$ of the hybrid method are denoted $\bar{C}_i$. The positions $c_i$ are the centres of the compartments.

For the compartment-Brownian coupling the relative 
mass error is the difference between the average (over 500 repeats - unless 
otherwise stated) number of particles in each region given by the hybrid method and the 
number of particles in the analytical solution of the mean-field PDE model in 
the corresponding region, divided by the number of particles in the relevant 
region of the analytical solution of the PDE (to normalise). In the pure compartment and blending regions these are given by equations \eqref{PDE_compartment_relative_mass_error_compartments} and \eqref{PDE_compartment_relative_mass_error_blending} respectively, with the altered definition of $\Omega_C=[a,I_1]$ for equation \eqref{PDE_compartment_relative_mass_error_compartments}.
For the purely Brownian region the RME is given by
\begin{equation}
RME_B(t)=\frac{\bar{B} -\int_{\Omega_B}u(x,t) dx}{\int_{\Omega_B}u(x,t) dx},\label{compartment_Brownian_relative_mass_error_Brownian}
\end{equation}
where $\Omega_B=[I_2,b]$ and $\bar{B}$ represents the average number of Brownian particles in the purely Brownian regime.

Figure \ref{figure:uniform_IC} demonstrates that both of our hybrid blending 
methods pass this most-basic test of maintaining a uniform distribution across 
the domain. The interfaces between the different modelling regimes are 
effectively undetectable. Qualitatively, the density plots all show good 
agreement between the hybrid methods and the analytical solution to the 
mean-field diffusion equation. This is confirmed by the relative error plots 
(panels \ref{figure:uniform_PDE_rel_error} and 
\ref{figure:uniform_Brown_rel_error}) which demonstrate low errors which 
fluctuate around zero with no discernible long-term bias.

\subsection{Test Problem 2: Particle redistribution}
\begin{figure}[h!]
\begin{center}
\subfigure[][]{
\includegraphics[width=0.31\textwidth]{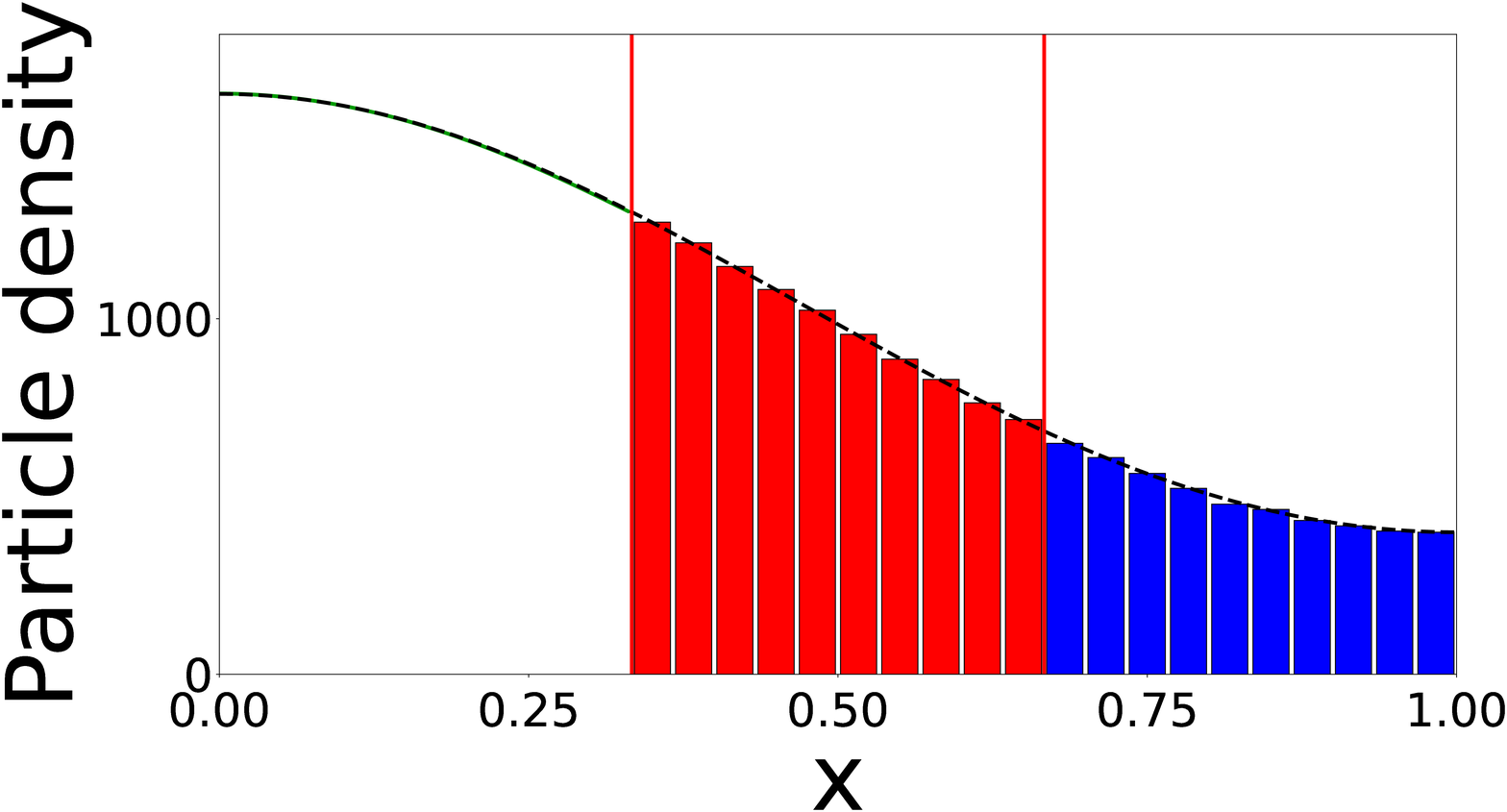}
\label{figure:step_PDE_0.1}
}
\subfigure[][]{
\includegraphics[width=0.31\textwidth]{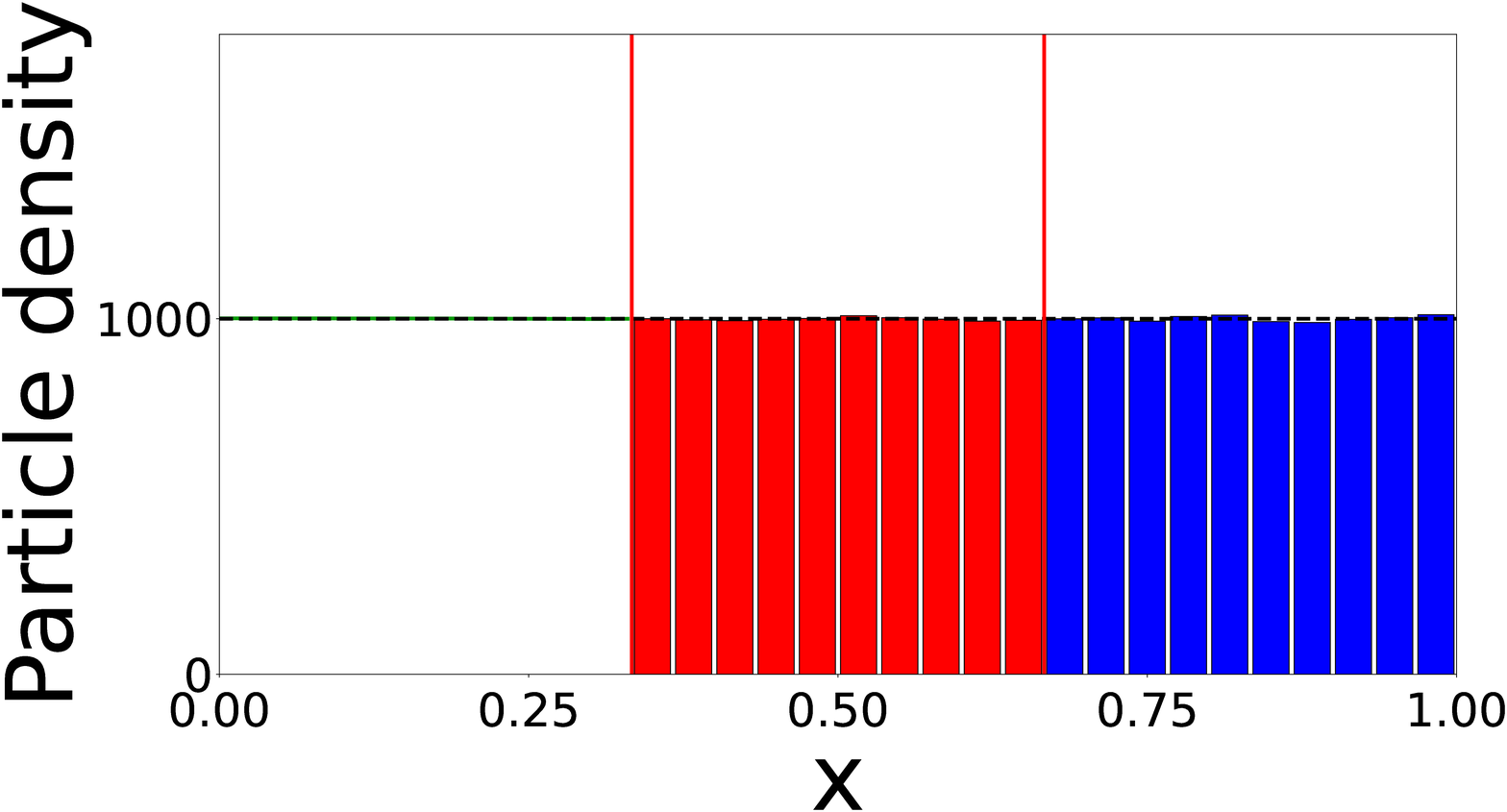}
\label{figure:step_PDE_1}
}
\subfigure[][]{
\includegraphics[width=0.31\textwidth]{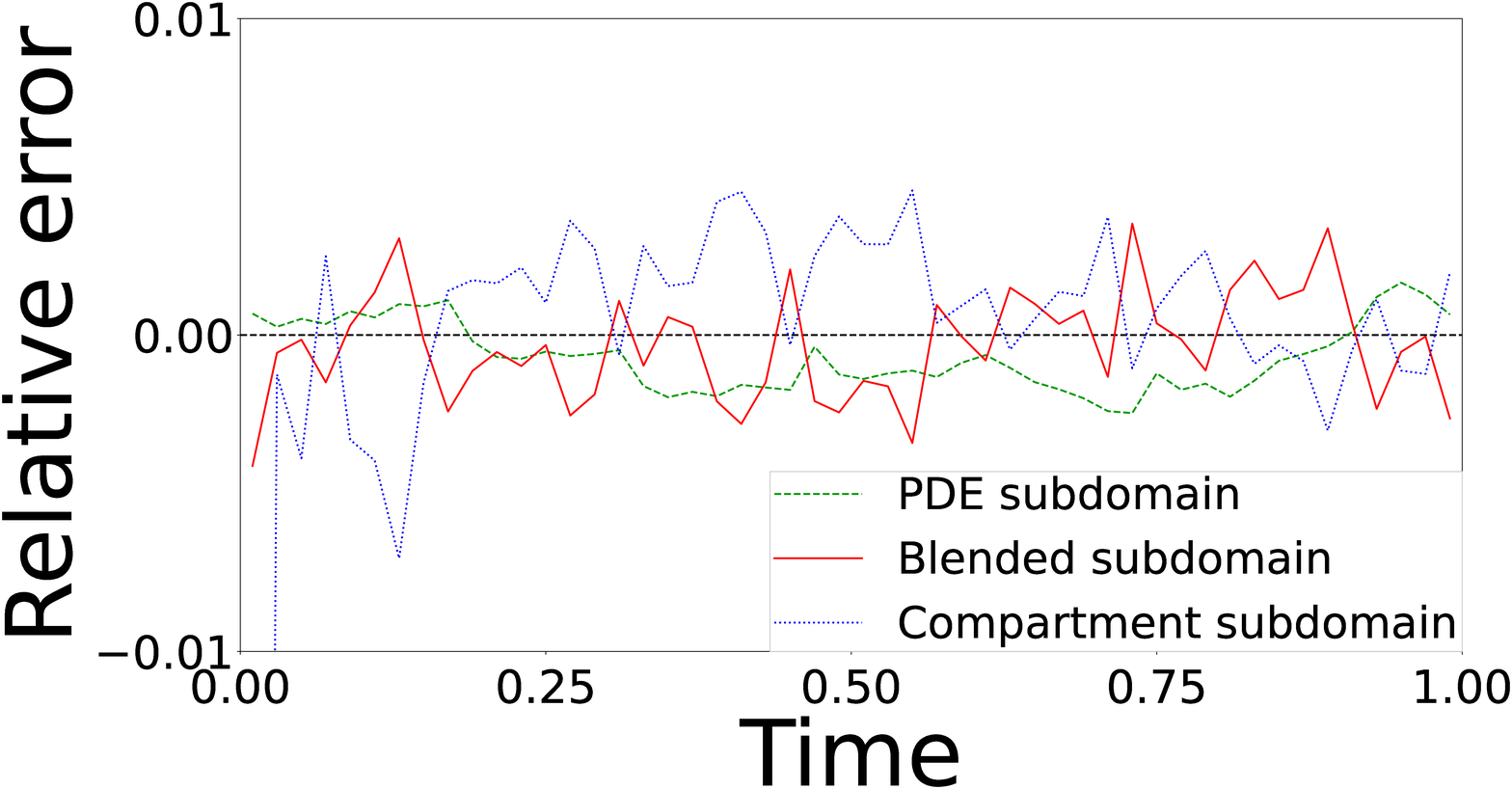}
\label{figure:step_PDE_rel_error}
}
\subfigure[][]{
\includegraphics[width=0.31\textwidth]{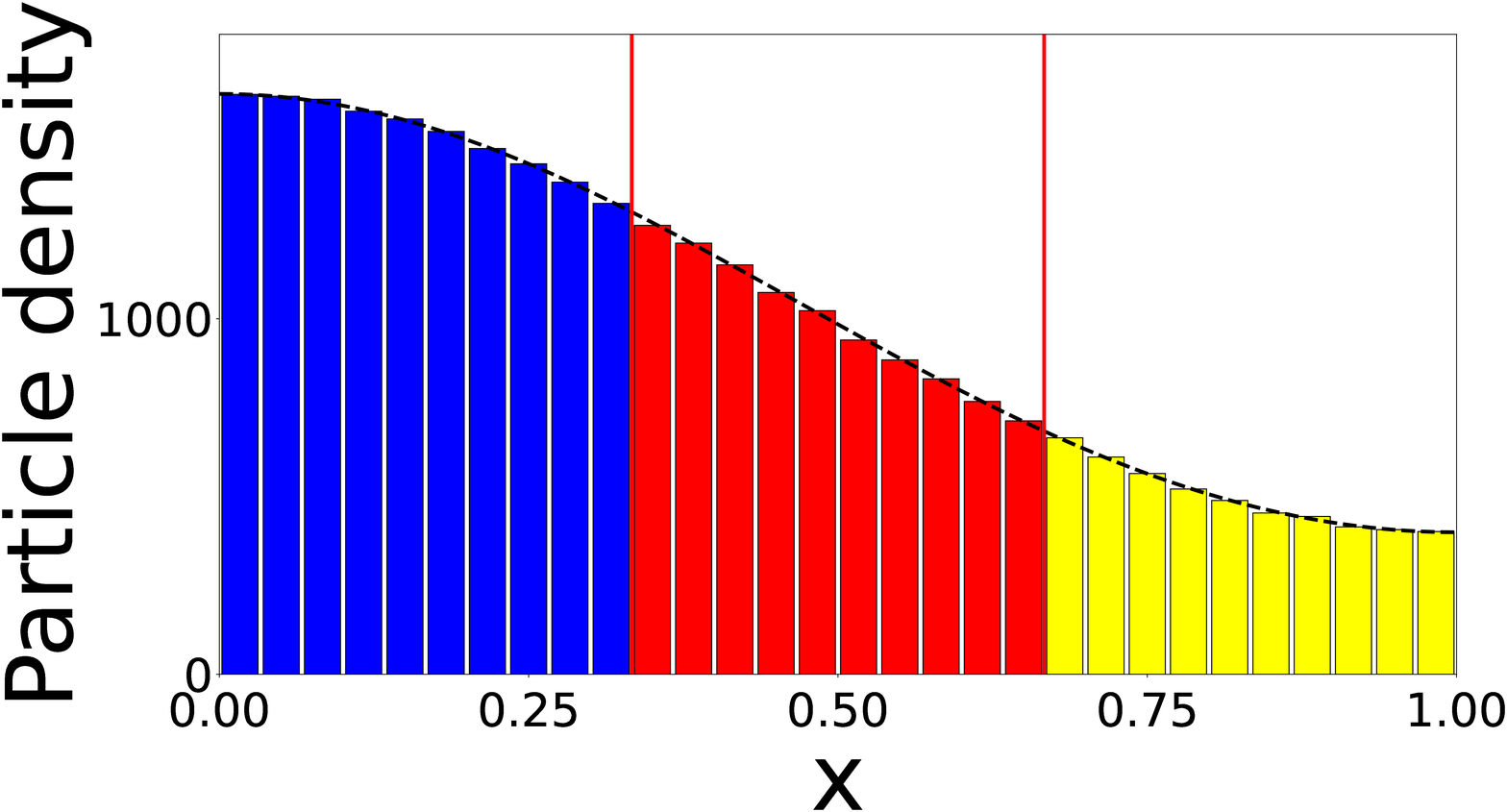}
\label{figure:step_Brown_0.1}
}
\subfigure[][]{
\includegraphics[width=0.31\textwidth]{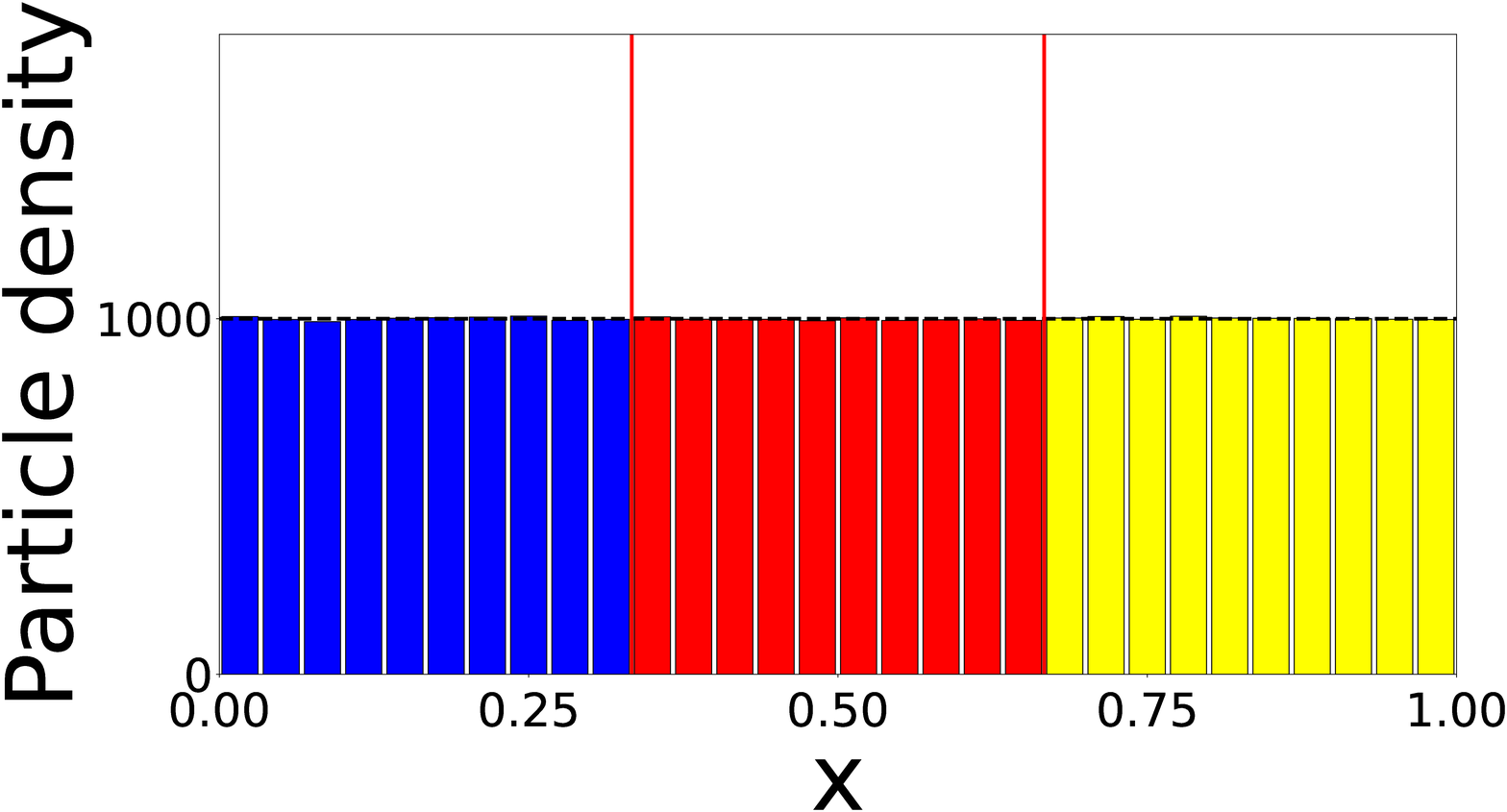}
\label{figure:step_Brown_1}
}
\subfigure[][]{
\includegraphics[width=0.31\textwidth]{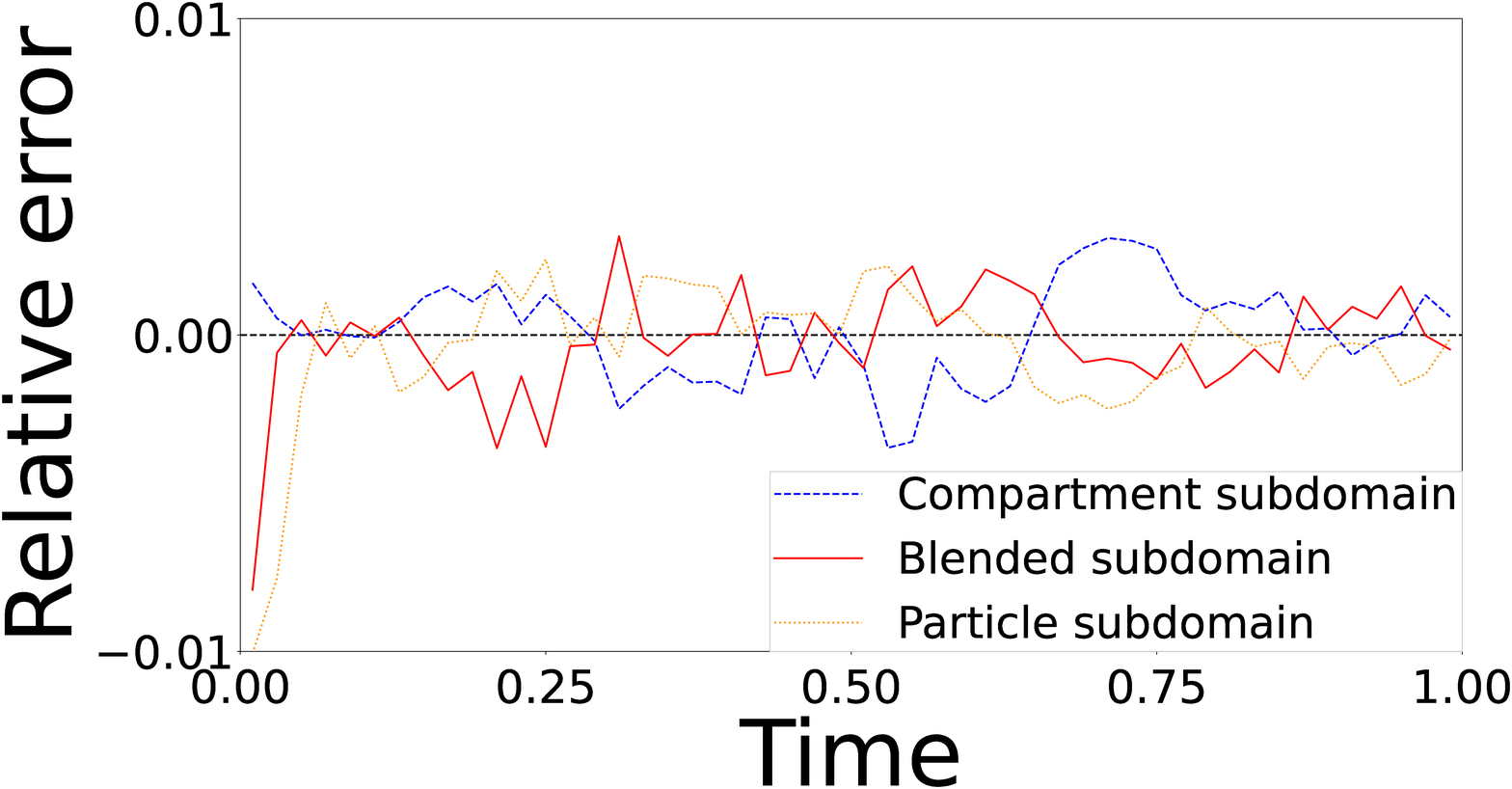}
\label{figure:step_Brown_rel_error}
}
\end{center}
\caption{Density and error plots for test problem 2 - pure diffusion with a 
step 
function initial condition in $[a,I_1]$. Descriptions, including definitions of 
relative errors are as in figure \ref{figure:uniform_IC}.}
\label{figure:step_IC}
\end{figure}

The second test problem is designed to determine whether the hybrid methods can 
cope with high levels of flux across their interfaces. As with the previous 
example, we model pure diffusion with no reactions, but this time with a 
different initial condition.
All the particles are distributed uniformly within $[a,I_1]$ and the system is 
allowed to equilibrate. The results of these simulations for both the 
PDE-compartment hybrid and the compartment-Brownian hybrid are given in figure 
\ref{figure:step_IC}.

In figure \ref{figure:step_IC} we have initialised the particles uniformly in 
the left-hand-most third of the domain, corresponding to the purely PDE region 
in the PDE-compartment hybrid and the purely compartment-based region in the 
compartment-Brownian hybrid\footnote{We see similarly agreeable 
results when particles are initialised in the third of the domain $[I_2,b]$ 
corresponding to the purely compartment or purely Brownian regions 
respectively.}. As in test problem 1, both of our hybrid methods 
correctly match the evolution of the density of the mean-field diffusion 
equation, as evidenced quantitatively  by the relative error plots 
\ref{figure:step_IC}\subref{figure:step_PDE_rel_error} and 
\ref{figure:step_IC}\subref{figure:step_Brown_rel_error}. 

\subsection{Test Problem 3: A morphogen gradient formation model}
\begin{figure}[h!]
\begin{center}
\subfigure[][]{
\includegraphics[width=0.31\textwidth]{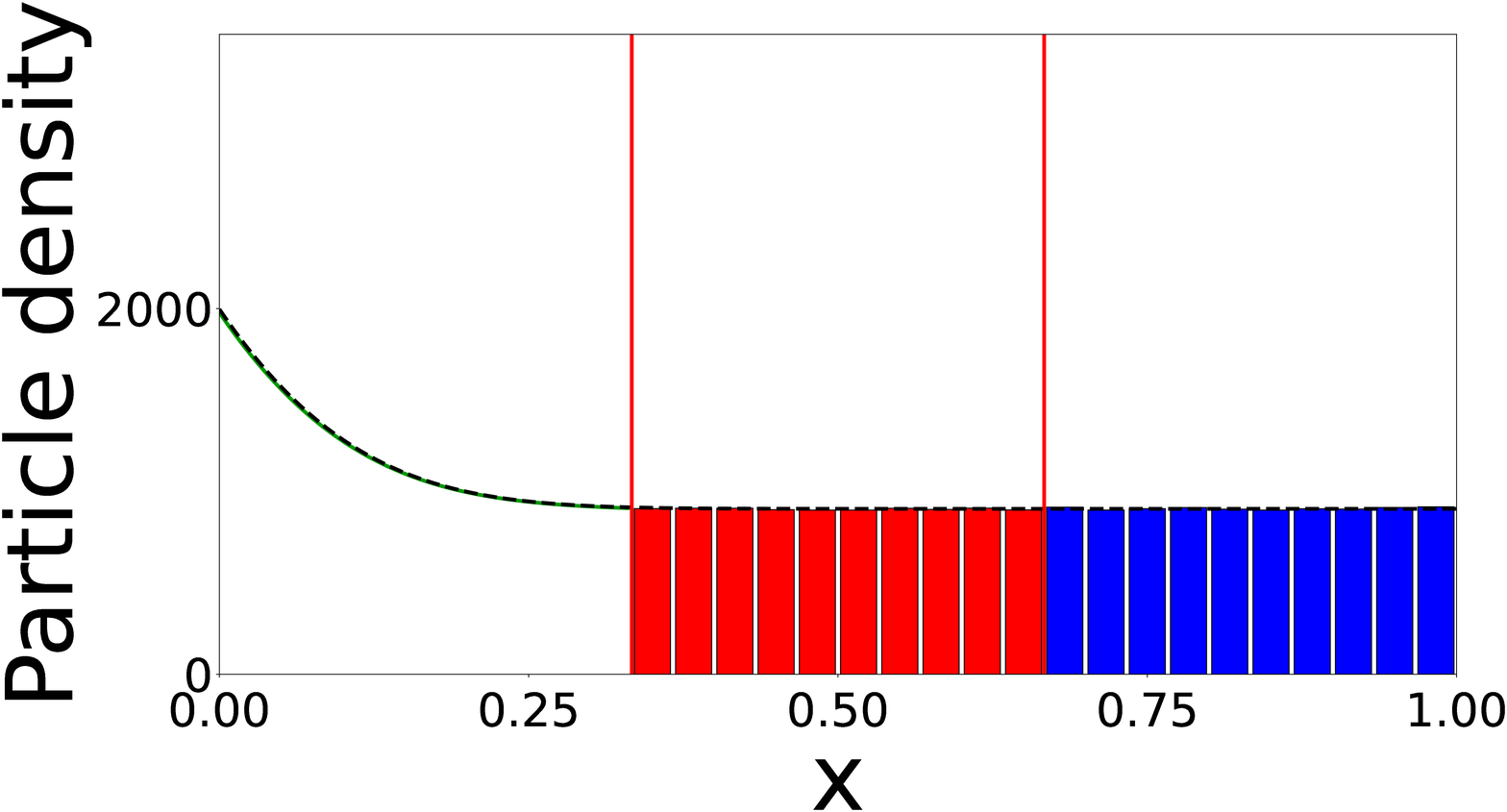}
}
\subfigure[][]{
\includegraphics[width=0.31\textwidth]{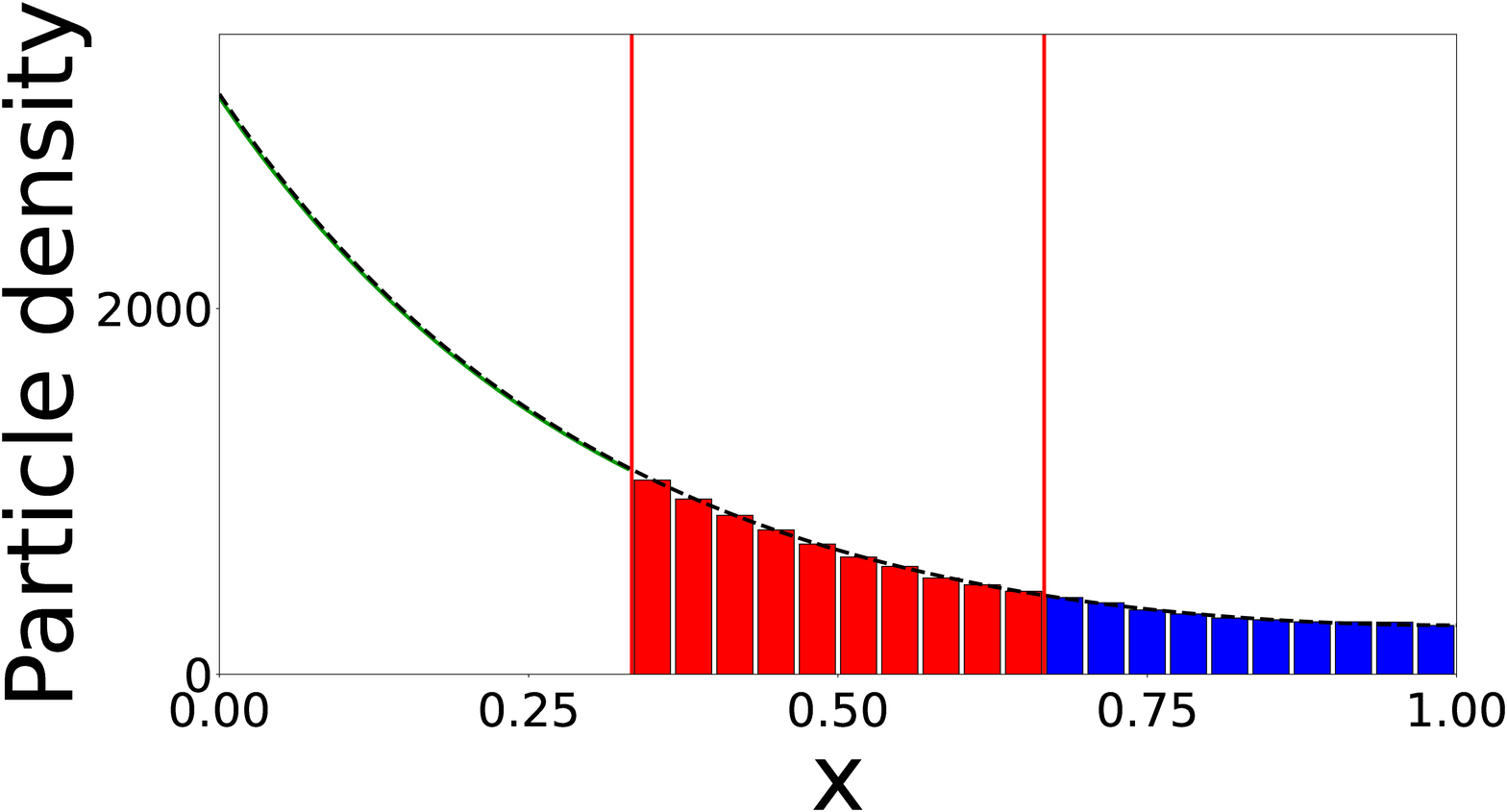}
}
\subfigure[][]{
\includegraphics[width=0.31\textwidth]{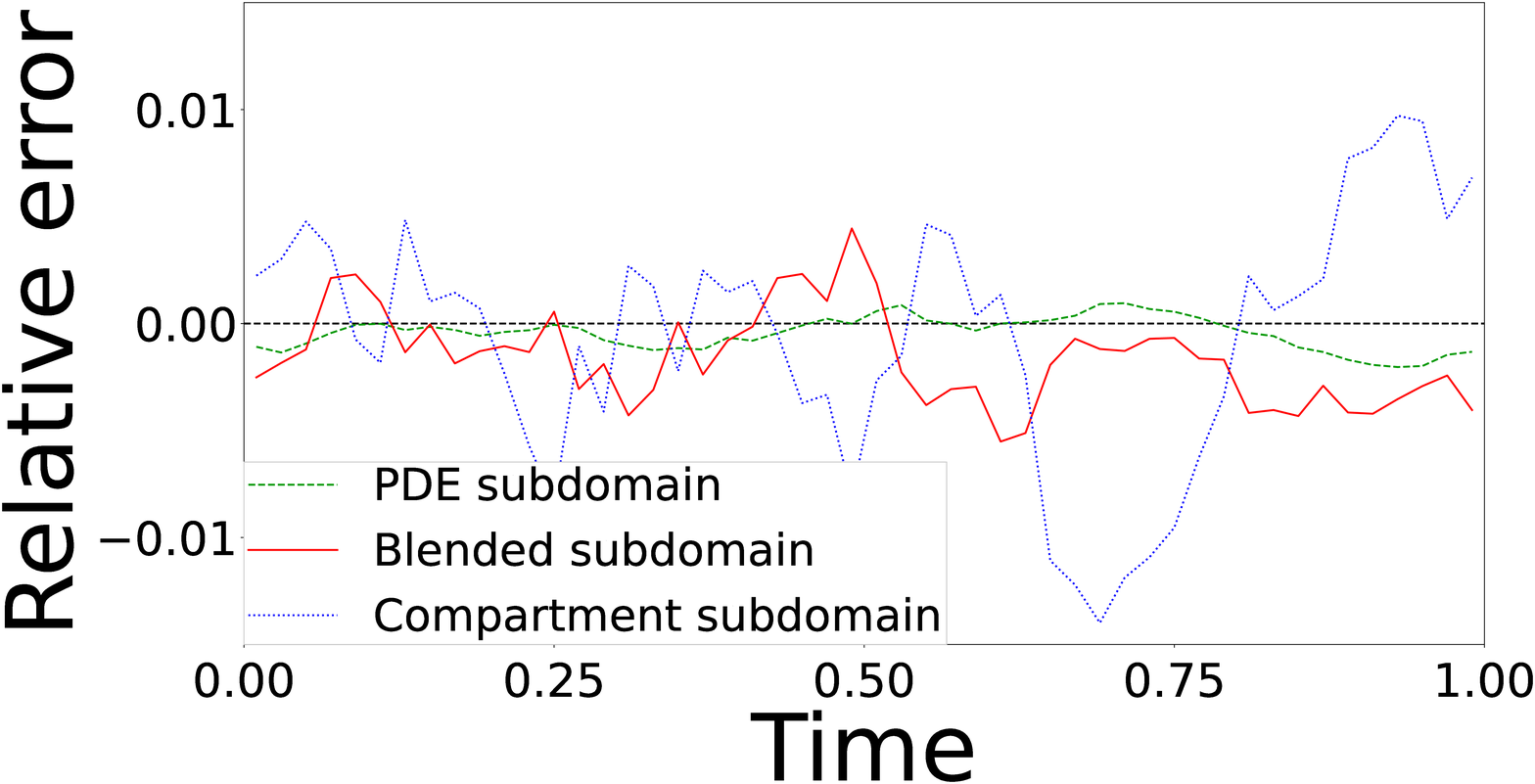}
\label{figure:PDE_comp_rel_error}
}
\subfigure[][]{
\includegraphics[width=0.31\textwidth]{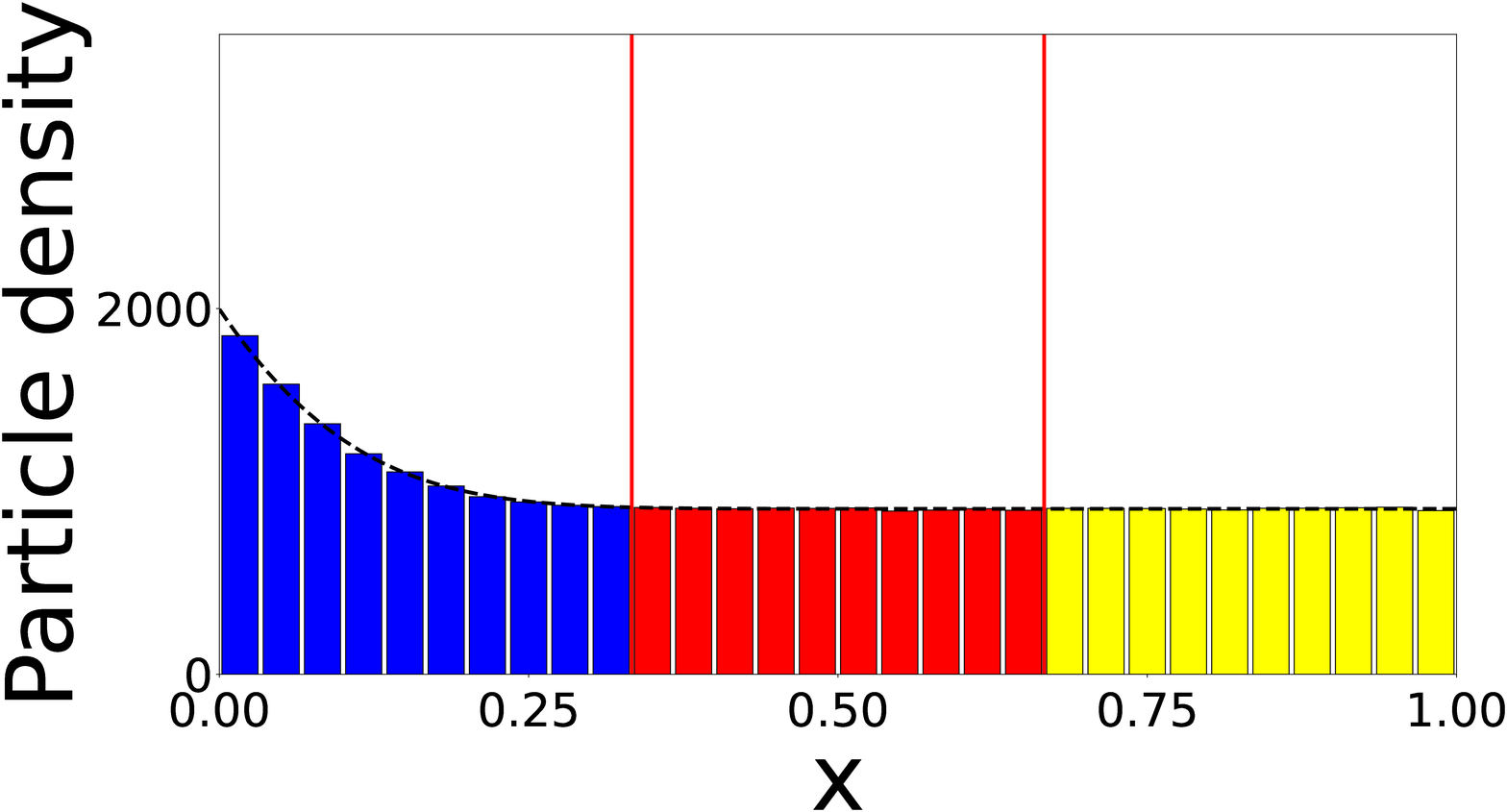}
}
\subfigure[][]{
\includegraphics[width=0.31\textwidth]{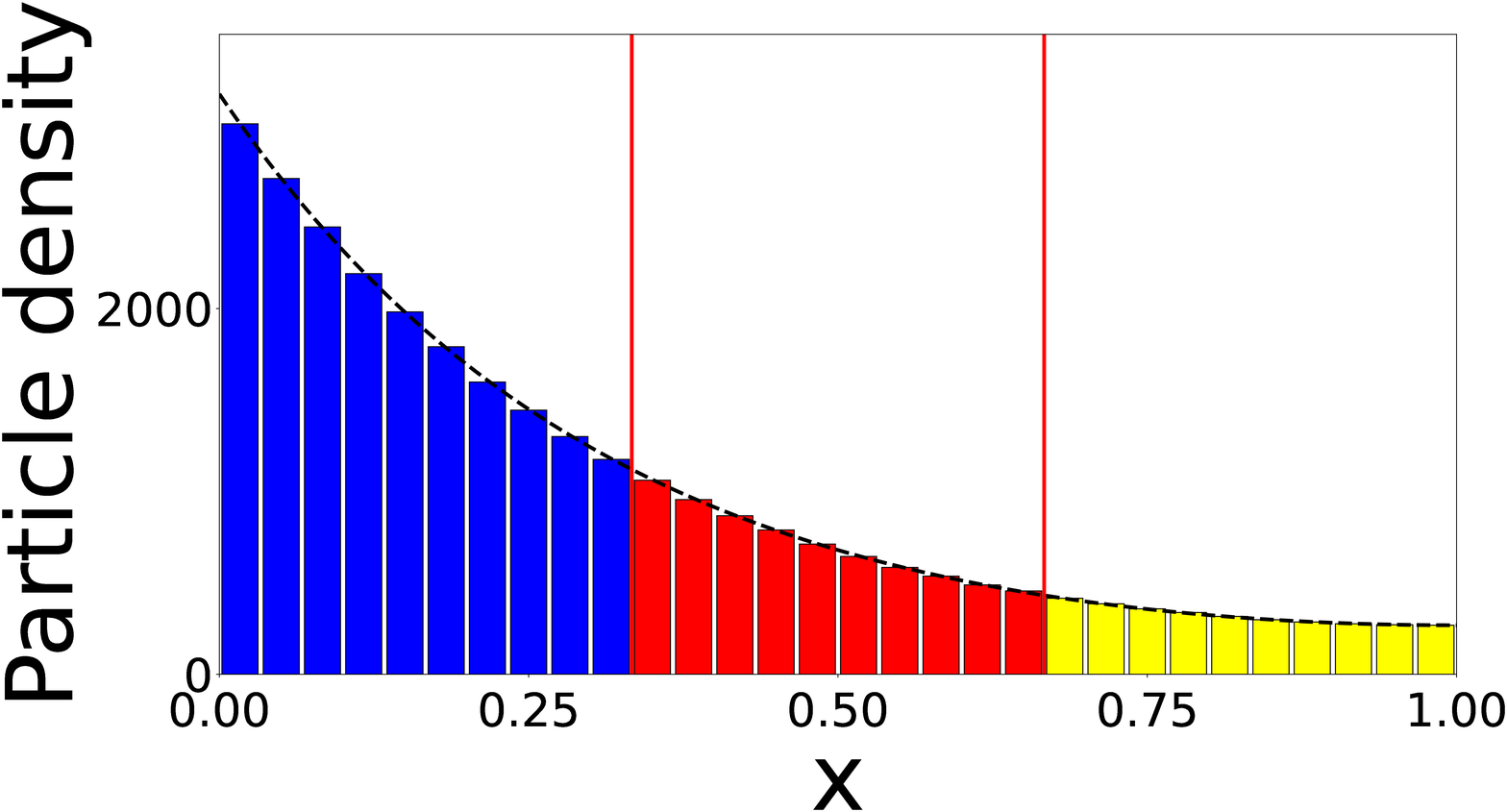}
}
\subfigure[][]{
\includegraphics[width=0.31\textwidth]{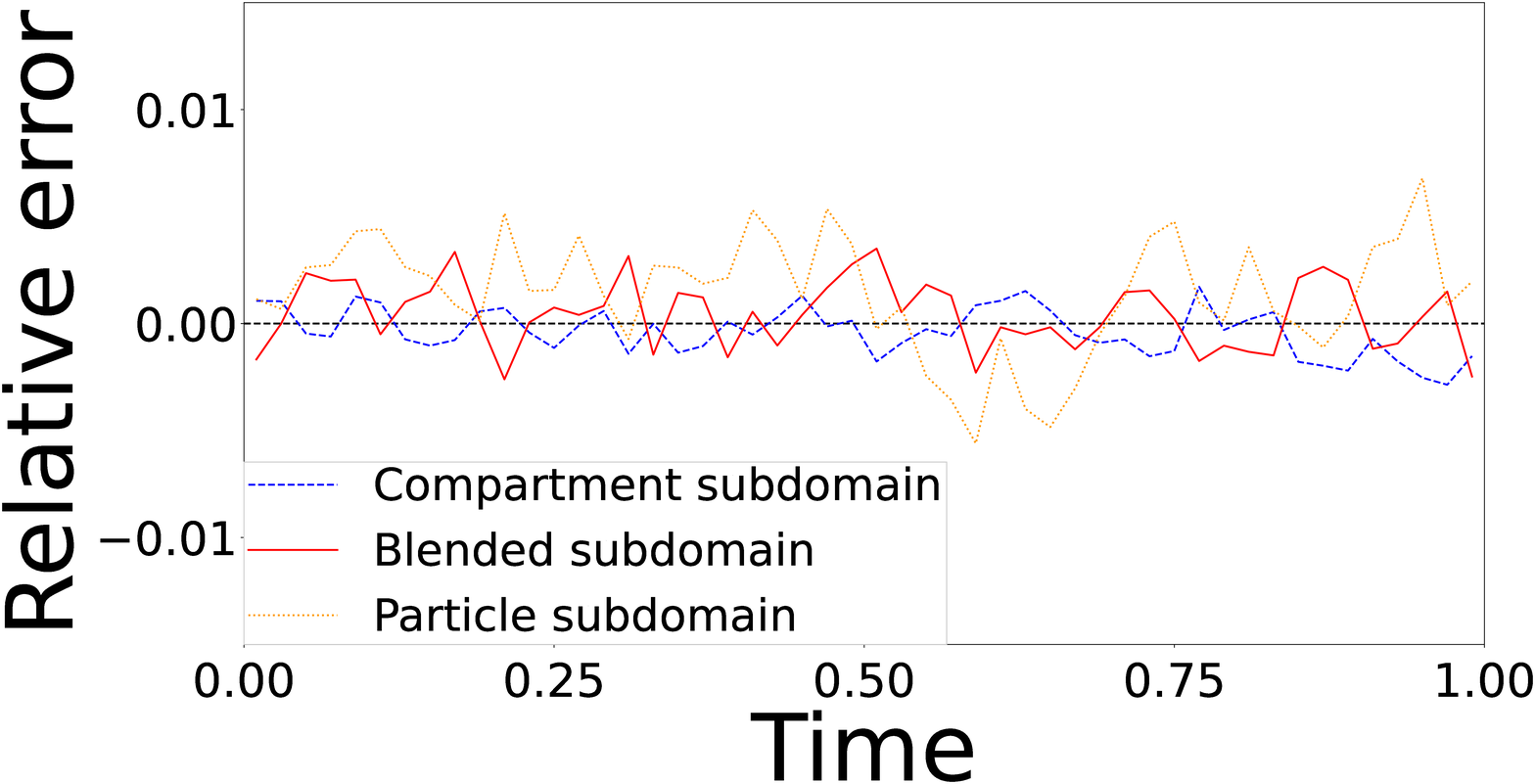}
\label{figure:comp_part_rel_error}
}
\end{center}
\caption{Density and error plots for test problem 3 - morphogen gradient 
formation with a uniform initial condition. Descriptions, including definitions 
of relative errors are as in figure \ref{figure:uniform_IC} except that panels (a) and (d) are density profiles evaluated at $t=0.01$ rather than at $t=0.1$. All figure 
parameters are given in table \ref{table:example_3_parameters}.}
\label{figure:morphogen}
\end{figure}

The formation of a morphogen gradient from a uniform initial condition 
constitutes the third test of our hybrid simulation algorithms. Particles are 
allowed to diffuse freely throughout the domain and  degrade at a rate $\mu$. 
To 
counteract the  degradation and ensure a non-trivial steady state, particles 
are 
introduced at the left-hand boundary, $x=a=0$, with flux $DJ$, and a 
zero-flux boundary condition is implemented at $x=b=1$. Since the reactions we 
have introduced are first order, the continuum mean-field model corresponding 
to 
the described set up is governed by the following PDE: 
\begin{align}
\frac{\partial c}{\partial t}=D\frac{\partial^2  c }{{\partial x}^2} - \mu 
c,\quad\text{ for } x\in(0,1) \text{ and } 
t\in(0,T),\label{eqn:morphogen_gradient_mean_field_model}
\end{align}
with boundary conditions
\begin{equation}
\frac{\partial c}{\partial x}(0,t) = -J,\quad \frac{\partial c}{\partial 
x}(1,t) = 0, \quad t\in(0,T), 
\label{equation:morphogen_gradient_boundary_condtions}
\end{equation}
and initial condition
\begin{equation}
\quad c(x,0) = c_0, \text{ for } x\in [0,1], \text{ where } 
c_0=\frac{DJ}{\mu}.
 \end{equation}
The initial condition is chosen so that we begin with the same number of 
particles as there will be at steady state, but distributed uniformly across 
the domain\footnote{Note that we have chosen this initial condition to ensure the PDE-compartment algorithm functions appropriately. Whilst the compartment-to-Brownian algorithm can deal naturally with low particle numbers, as noted earlier, there is the potential for low particle numbers to break the PDE-compartment algorithm. Potentially, when particle numbers are low in the blending region, fractional particle numbers in a compartment could cause a particle to be chosen to jump out of one compartment even though there is not sufficient mass for this to occur. The solution to this problem, as will be proposed in the discussion, is to introduce adaptive blending regimes, which ensure the PDE representation is only employed in regions of the domain where particle concentrations are sufficiently high to justify its use.}. The parameters we employ for the simulations shown in figure 
\ref{figure:morphogen} are given in table \ref{table:example_3_parameters}.
Specifically, influx parameter, $J$, and degradation parameter, $\mu$, are chosen to ensure an average of 1000 particles populating the domain throughout the simulation.
\begin{table}
\begin{center}
\begin{tabular}{|c|c|c|}
\hline
Parameter &  Value & Description  \\
\hline
$N(0)$ & 1000 & Initial number of particles \\
$\Omega$ & $[0,1]$ & Spatial domain \\
$D$&1 & Diffusion coefficient\\
$J$ & $10,000$ & Rate of influx at the left boundary\\
$\mu$ & 10 & Rate of particle decay \\
$K$ & $20$ & Number of compartments\\
$h$ & $1/30$  & Compartment width \\
$\Delta x$ & $1/300$ & PDE voxel width\\
$ \Delta t_p$ & $10^{-4} $  & PDE time-step\\
$ \Delta t_b$ & $10^{-4} $  & Brownian time-step\\
M & 1000& Number of repeats\\
\hline
\end{tabular}
\end{center}
\caption{Table of parameters for the morphogen gradient simulation (Test problem 3).}
\label{table:example_3_parameters}
\end{table}

Figure \ref{figure:morphogen} illustrates the solutions of our two hybrid 
methods and those of the corresponding mean-field model (given by equation 
\eqref{eqn:morphogen_gradient_mean_field_model}). As with the previous two test 
problems, qualitative density profiles are in close agreement and quantitative 
error plots show low error and no sustained bias about zero.

\subsection{Test Problem 4: bimolecular production-degradation}
\label{section:bimolecular_production}

The final scenario we will use to demonstrate the accuracy of our hybrid 
methods 
is a system of diffusing particles interacting through the following pair of 
chemical reactions: 

\begin{equation}
2A \xrightarrow{\kappa_1} \emptyset,\quad \emptyset \xrightarrow{\kappa_2} A, 
\label{eqn:TP4_System}
\end{equation}
which occur within the cuboidal domain $\Omega\subseteq\mathbb{R}^3$ of volume 
$V$, where $\Omega=[0,10]\times[0,1]\times[0,1]$. 

The blending hybrid method is extended to this three-dimensional example in the natural way.
As in the one-dimensional 
test problems, the domain is divided in to three equally sized subdomains, this time with 
planar interfaces, $I_1$ at $x=10/3$ and $I_2$ at $x=20/3$. The 
compartment-based region for each hybrid method is divided into a lattice of cuboidal 
compartments of size $h_x\times h_y\times h_z$. The blending region is itself a cuboidal region in which both the coarse and fine models co-exist as equivalent representations of the mass in that region. For this translationally invariant example the blending functions are simply a function of $x$. This means that only diffusion parallel to the $x$-axis is impacted in the blending region. Of course, for differently shaped domains and interfaces, the blending functions may be functions of all three coordinates chosen to scale-diffusion as required providing $f_1(x,y,z)+f_2(x,y,z)=D$ for all $(x,y,z)\in \mathcal{B}$, the blending region, and both $f_1(\bs{I_1})=f_2(\bs{I_2})=D$ and $f_1(\bs{I_2})=f_2(\bs{I_1})=0$ are satisfied, where $\bs{I_1}\in\mathbb{R}^3$ and $\bs{I_2}\in \mathbb{R}^3$ are surfaces specifying the interfaces which form the boundaries of the blending region.

\begin{figure}[h!]
\begin{center}
\subfigure[][]{
\includegraphics[width=0.31\textwidth]{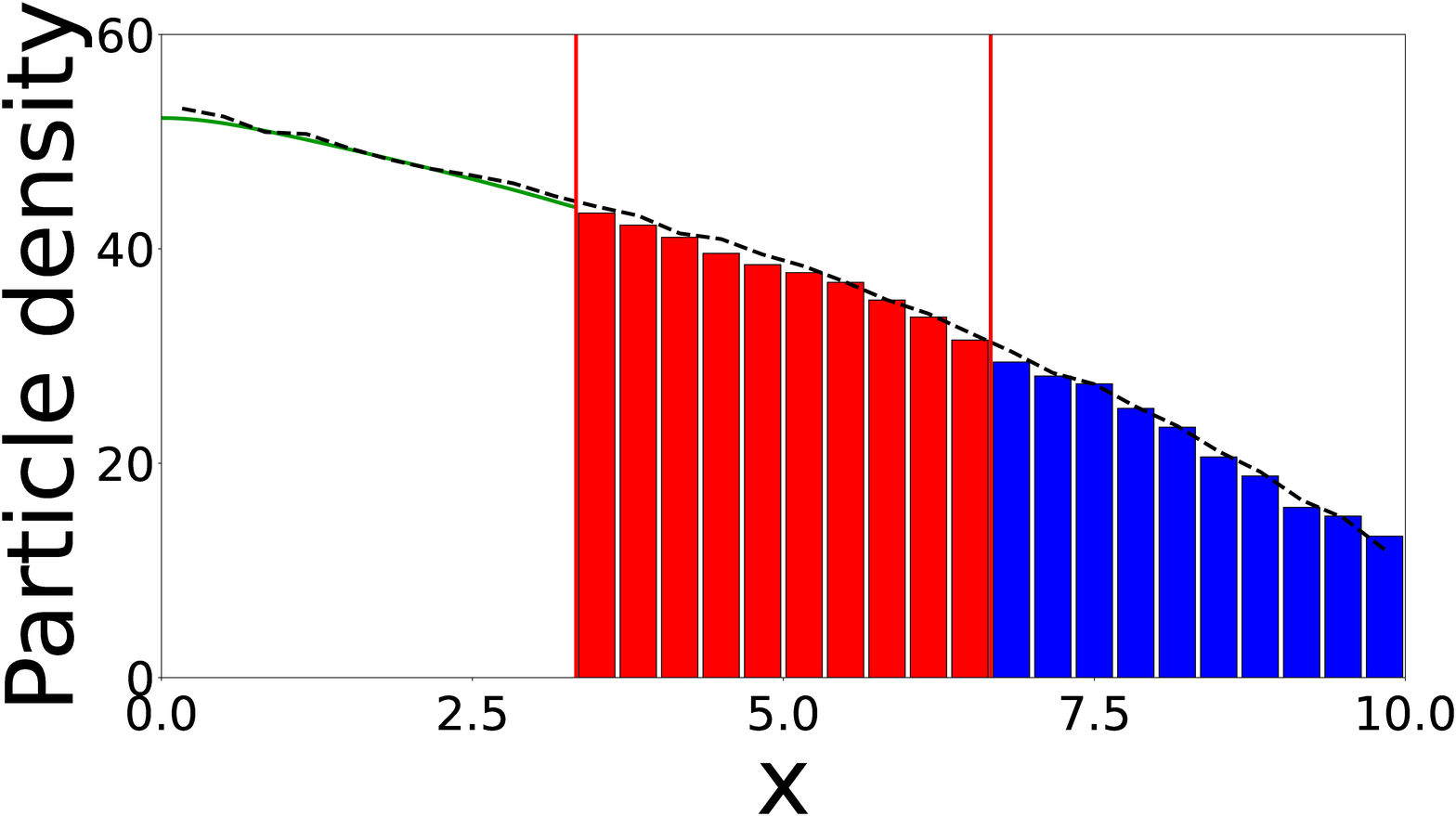}
\label{figure:second_PDE_0.1}
}
\subfigure[][]{
\includegraphics[width=0.31\textwidth]{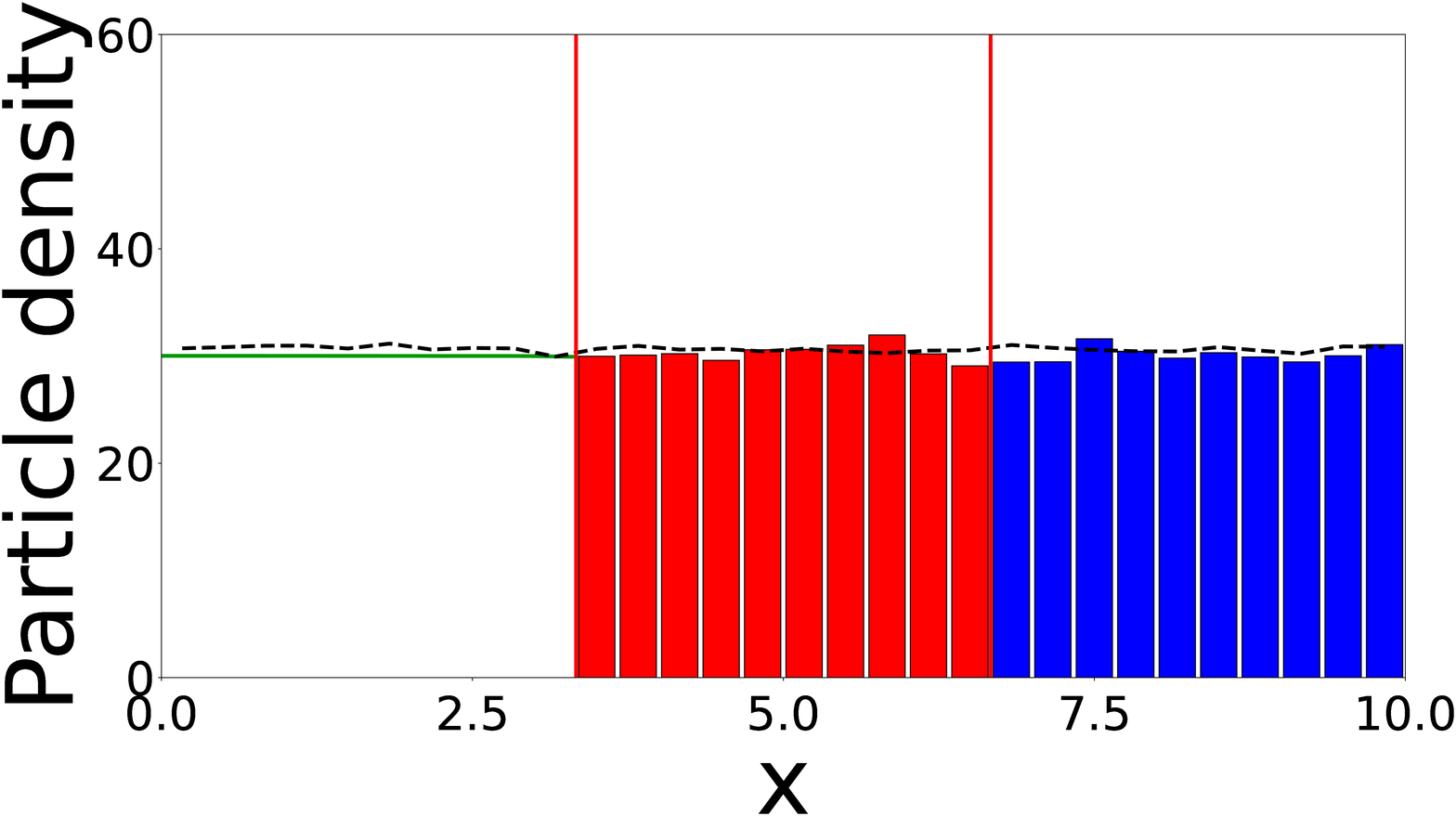}
\label{figure:second_PDE_1}
}
\subfigure[][]{
\includegraphics[width=0.31\textwidth]{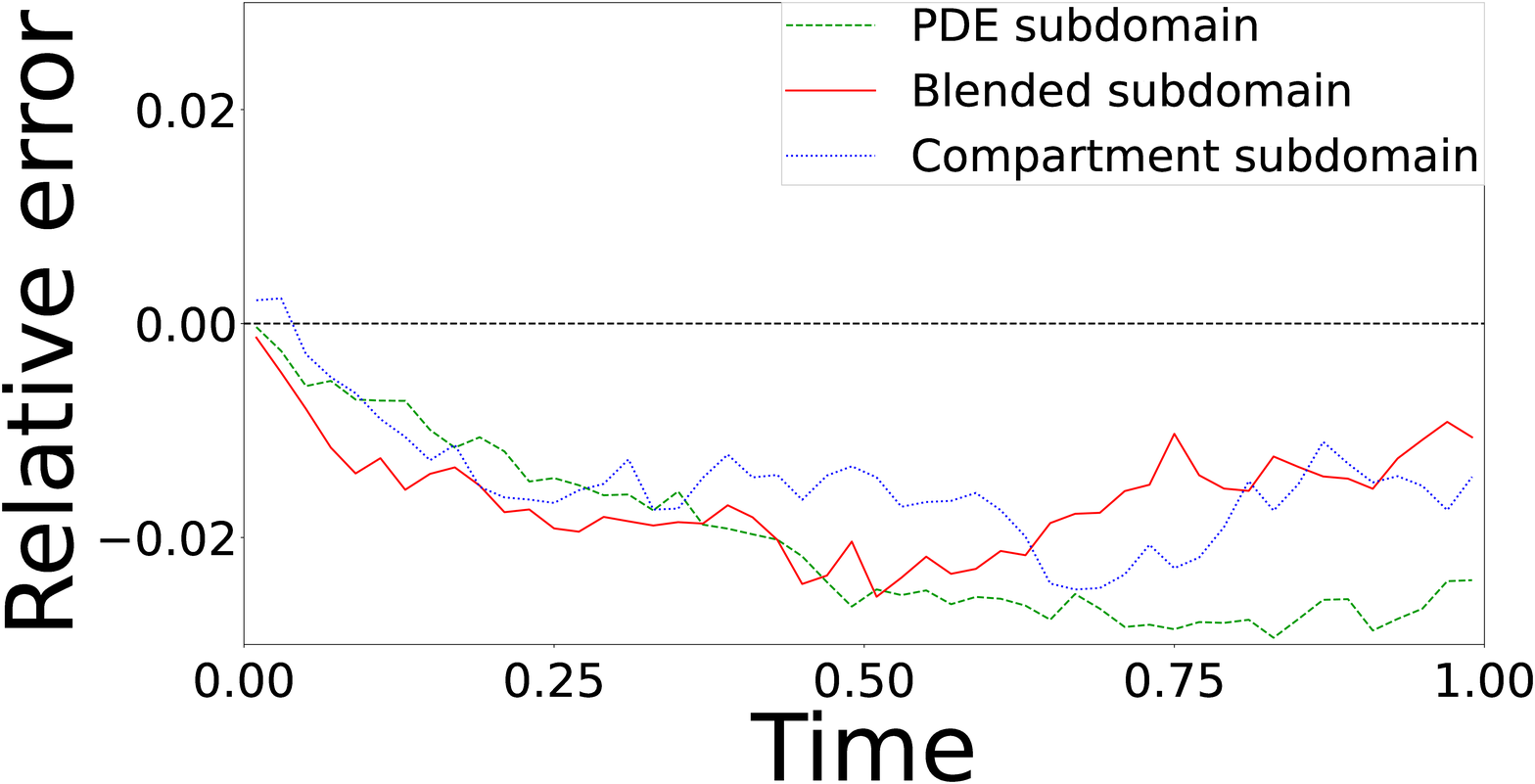}
\label{figure:second_PDE_rel_error}
}
\subfigure[][]{
\includegraphics[width=0.31\textwidth]{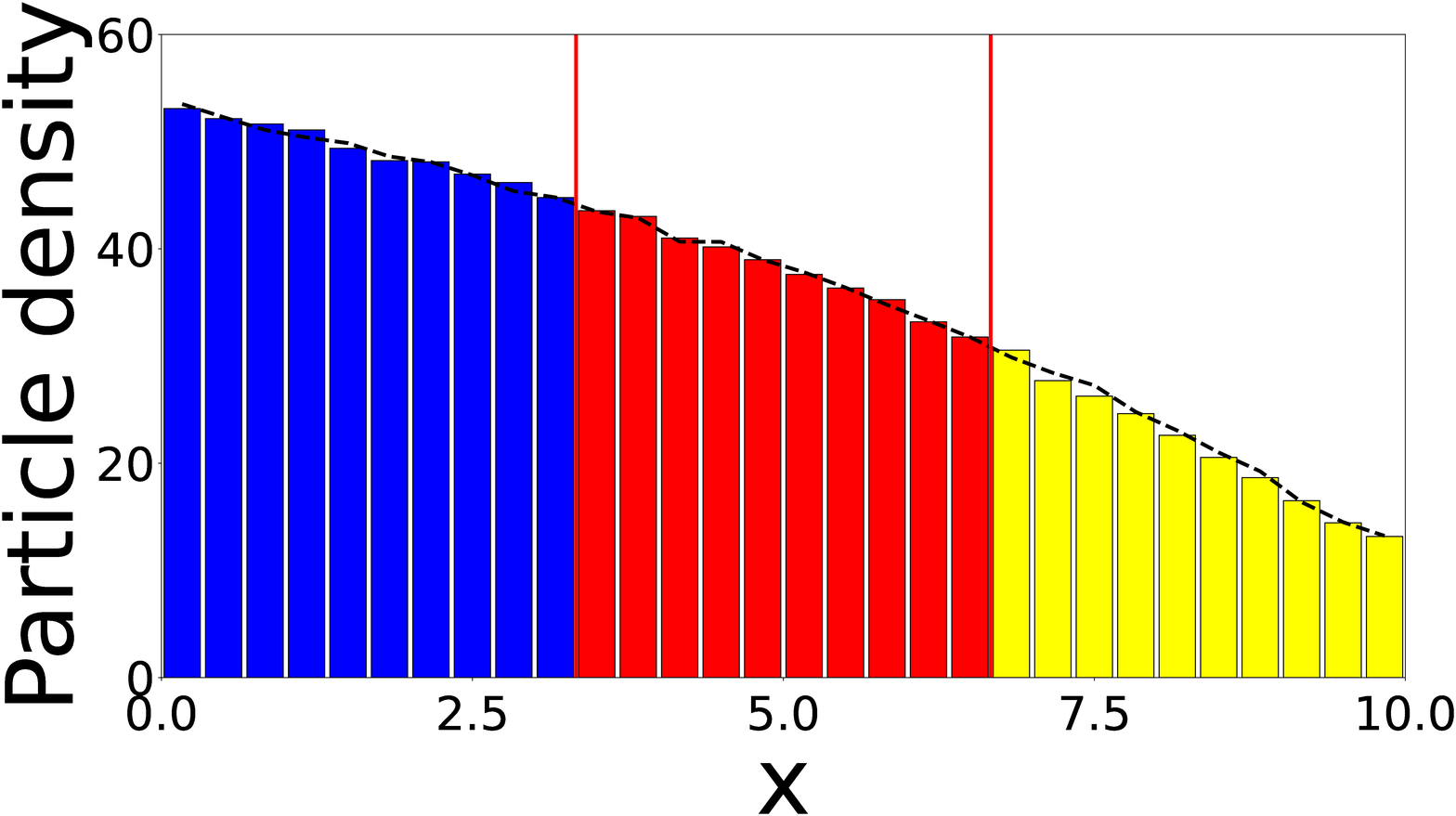}
\label{figure:second_Brown_0.1}
}
\subfigure[][]{
\includegraphics[width=0.31\textwidth]{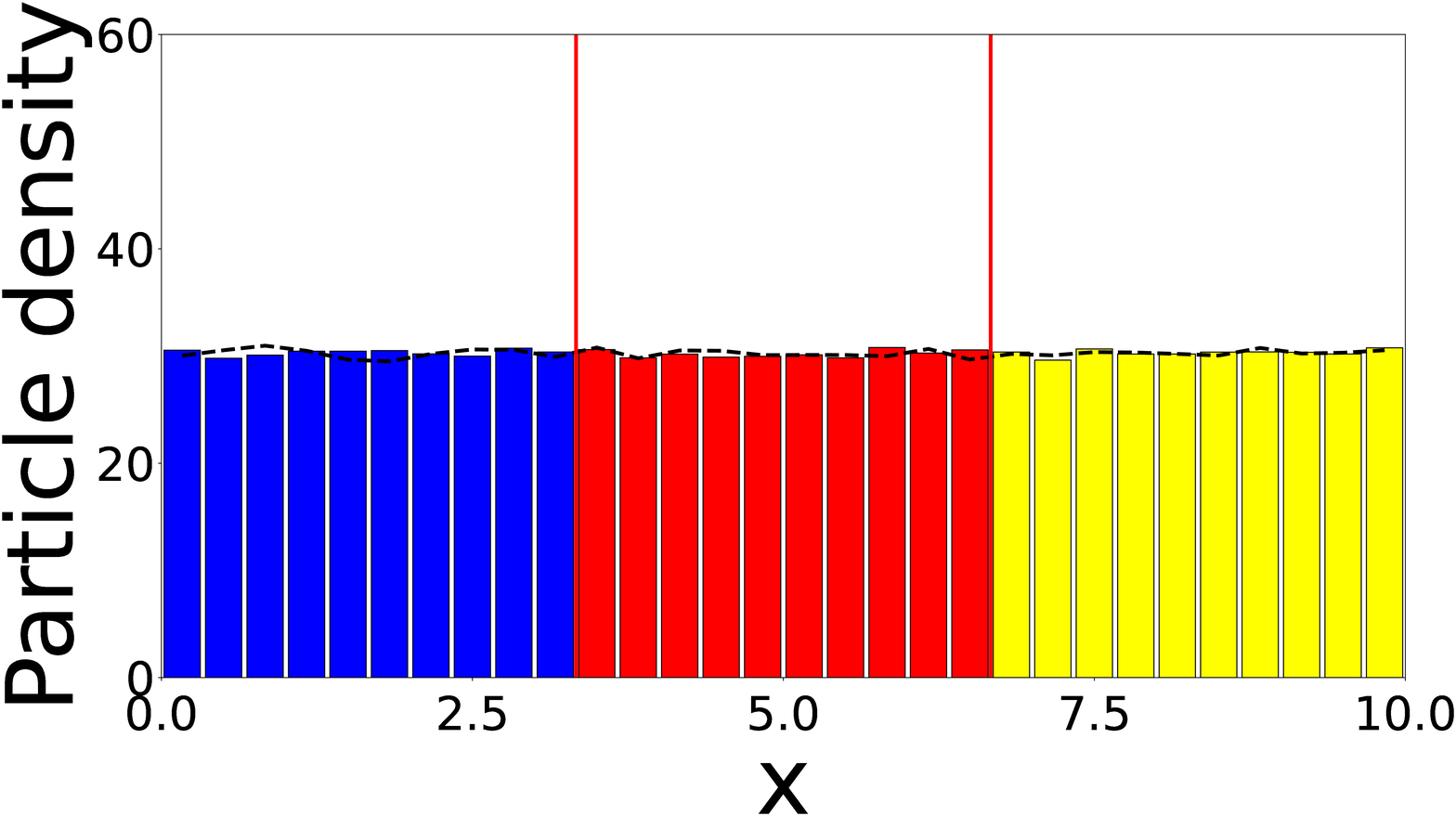}
\label{figure:second_Brown_1}
}
\subfigure[][]{
\includegraphics[width=0.31\textwidth]{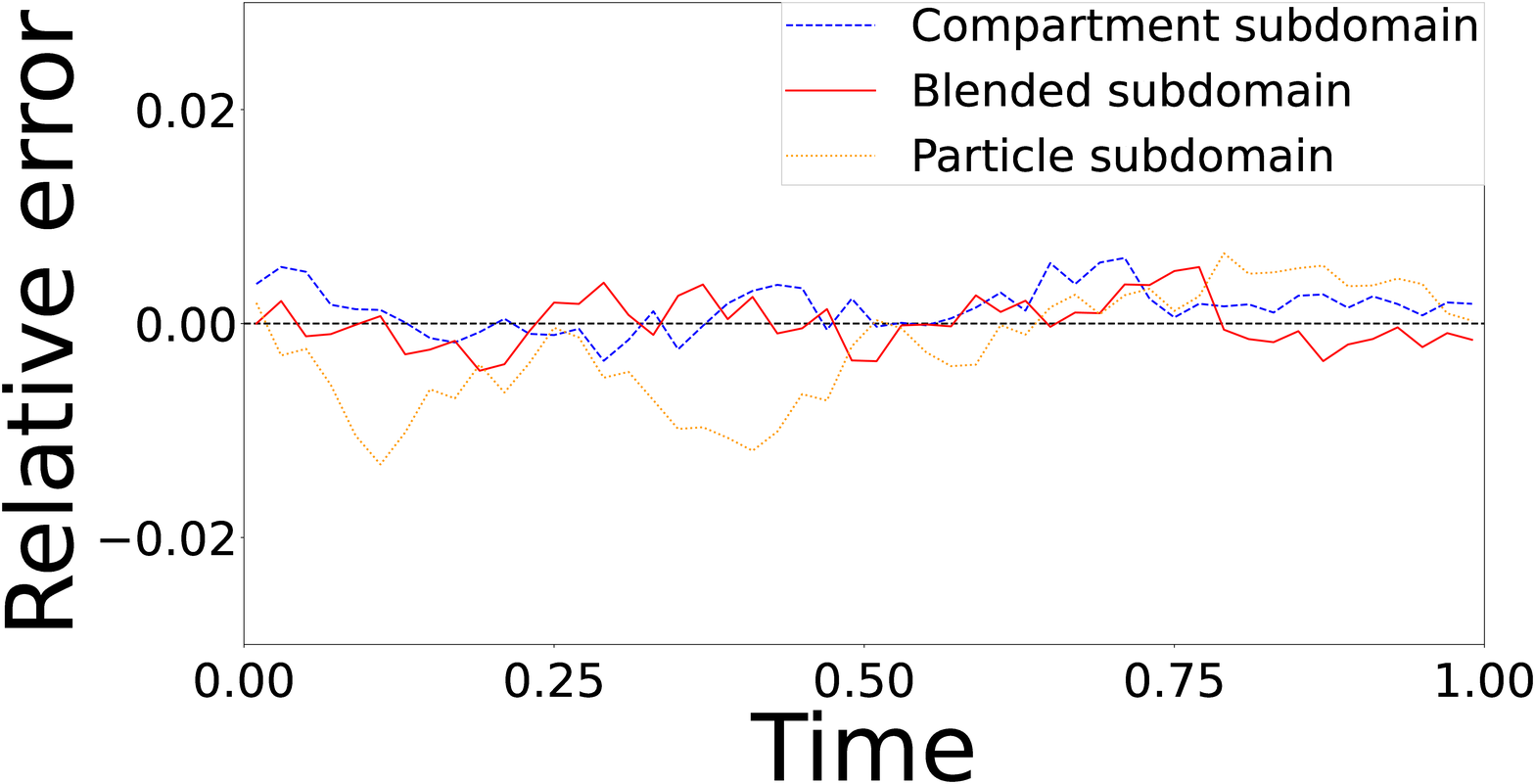}
\label{figure:second_Brown_rel_error}
}
\end{center}
\caption{Density and error plots for test problem 4 with an initial condition 
which exhibits a constant gradient. Descriptions, excluding definitions of 
relative errors are as in figure \ref{figure:uniform_IC}. All parameters are 
given in table \ref{table:example_4_parameters}.}
\label{figure:second_order}
\end{figure}

The mean-field PDE that corresponds to the reaction system 
\eqref{eqn:TP4_System} under the Poisson moment closure assumption is given 
by
\begin{equation}
\frac{\partial c}{\partial t} = D {\nabla}^2  c -\kappa_1 c^2 + \kappa_2, 
\label{equation:second_order_PDE_example}
\end{equation}
with corresponding zero-flux boundary conditions on each of the domain's boundaries:
\begin{equation}
\left.\frac{\partial c}{\partial n}\right|_{\partial \Omega} = 
0.\label{equation:second_order_PDE_BCs}
\end{equation}
For the simulations whose results are displayed in figure 
\ref{figure:second_order}, we initialise the particles according to a linear 
gradient so that the initial density decreases in the positive $x$-direction. 
Explicitly particle density profiles are initialised according to the following 
density profile: 
\begin{equation}
c(x,y,z)=\frac{183-18x}{2}, \text{ for } (x,y,z)\in 
[0,10]\times[0,1]\times[0,1],
\end{equation}
giving $N=465$ particles initially. The PDE part of the hybrid simulation 
can be initialised exactly according to this profile. For the regions of the domain 
modelled by stochastic components of the hybrid method (e.g. in  compartment-based 
regions or Brownian-based regions) the density profile is normalised and used 
as a probability density function (pdf) to assign positions to the appropriate number of particles corresponding to that region of the domain. In the blending 
regions, particles are initialised according to the finer-scale simulation 
method and the coarse scale density is matched appropriately. For example in 
the compartment-Brownian hybrid method we initialise, on average, one third of the 
particles with $y$ and $z$ coordinates chosen uniformly at random in $[0,1]$ 
and 
$x$-coordinates chosen from the pdf 
\begin{equation}
 {\displaystyle P(x)={
 \begin{cases} 0& \text{ for } \quad 0\leq x<10/3,\\
 \frac{183-18x}{310}& \text{ for } \quad 10/3\leq x<20/3,\\
 0 & \text{ for } \quad  20/3\leq x<10.
 \end{cases}}}
\end{equation}
Once the positions of the Brownian particles have been specified, the particles 
can then be binned into compartments to determine the compartment-based initial 
condition in that region. 

\begin{figure}[h!]
\begin{center}
\subfigure[][]{
\includegraphics[width=0.45\textwidth]{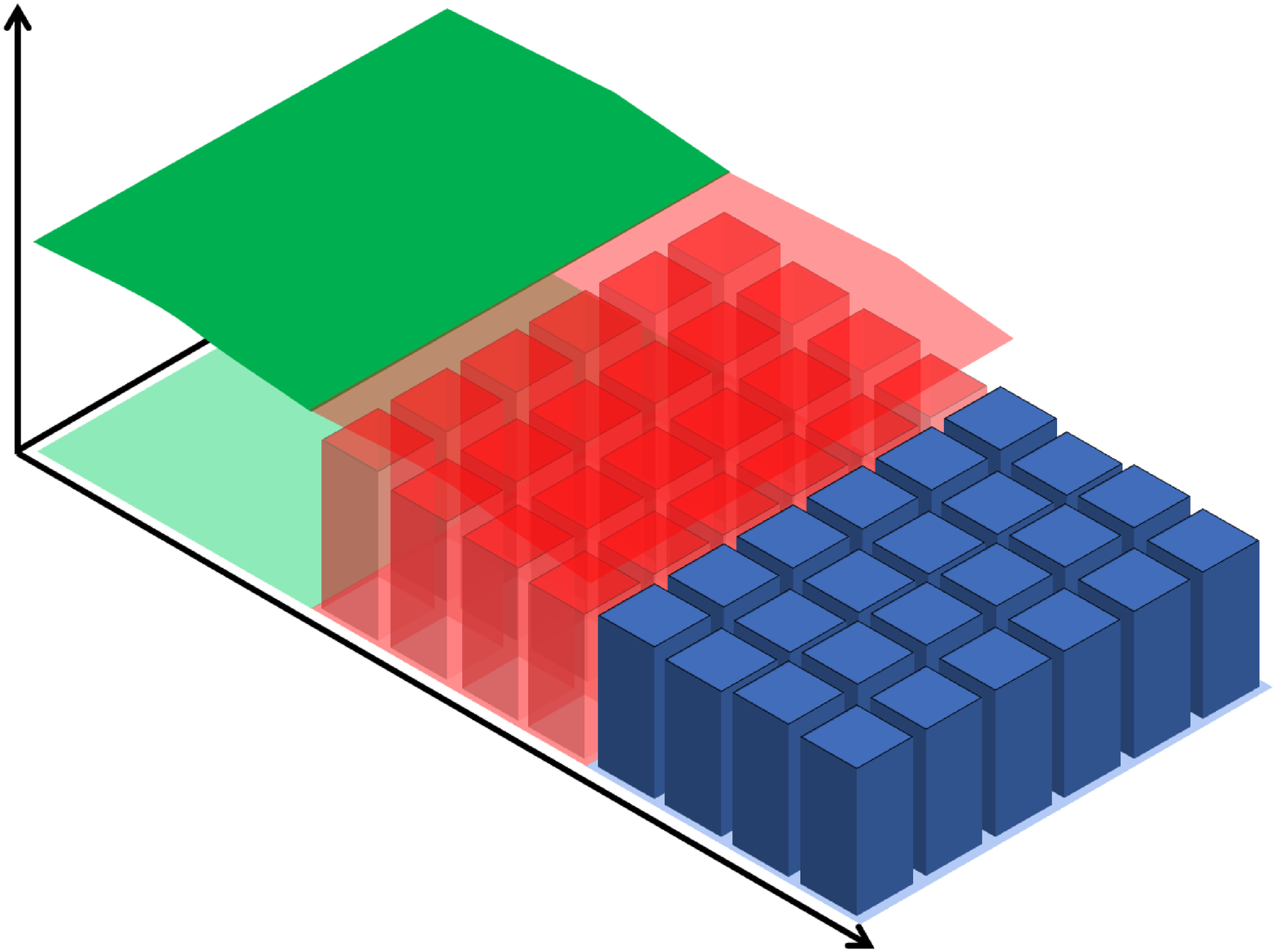}
\label{figure:2d_macro_meso_schematic}
}
\subfigure[][]{
\includegraphics[width=0.45\textwidth]{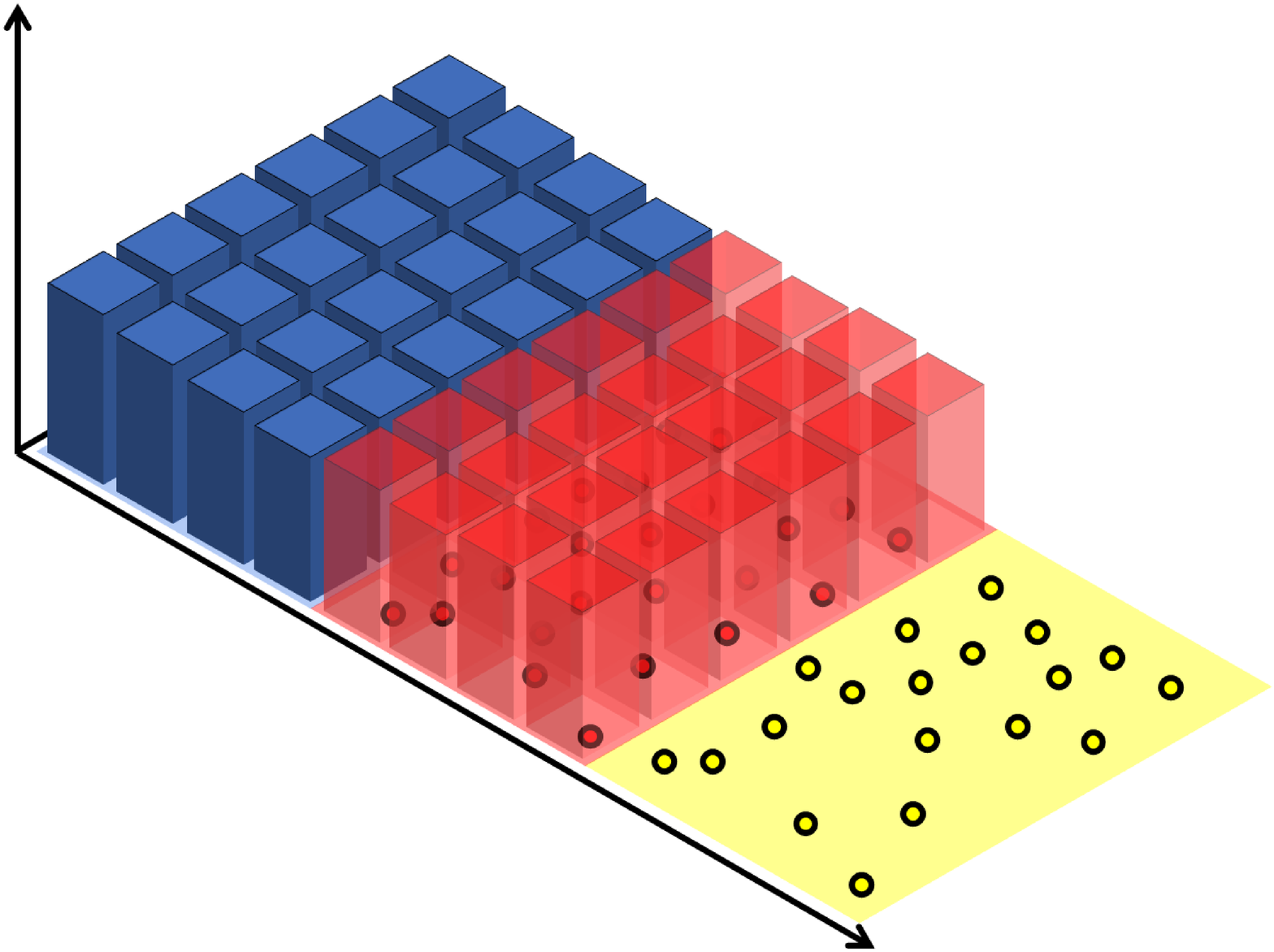}
\label{figure:2d_meso_micro_schematic}
}
\end{center}
\caption{Schematic representations of the two-dimensional \subref{figure:2d_macro_meso_schematic}  PDE-compartment hybrid and \subref{figure:2d_meso_micro_schematic} compartment-Brownian hybrid. In panel \subref{figure:2d_macro_meso_schematic} the green surface in the green region represents the PDE solution in the purely PDE region of the domain. The red surface and the red columns represent equivalent PDE- and compartment-based representations of the mass in the red blending region. The blue columns in the blue region of the domain represent the number of particles in each compartment in the purely compartment region of the domain. In panel \subref{figure:2d_meso_micro_schematic} the blue boxes in the blue region of the domain represent the number of particles in each compartment in the purely compartment region of the domain. The red boxes and the red circles represent equivalent compartment- and Brownian-based representations of the mass in the red blending region. The yellow circles in the yellow region of the domain represent individual particles in the purely Brownian region of the domain.}
\label{figure:2d_schematics}
\end{figure}

The hybrid method in three dimensions proceeds in an entirely analogous way to the one-dimensional algorithms described above with full three-dimensional simulation of the PDE, compartment-based method and Brownian-based method. As before, in the blending region the two different descriptions are kept in sync with each other at every time step. Figure \ref{figure:2d_schematics} provides schematic representations of the two coupling methods in two dimensions (in order to illustrate how the method generalises from one dimension).

\begin{table}[h!!!!!!!!!!]
\begin{center}
\begin{tabular}{|c|c|c|}
\hline
Parameter &  Value & Description   \\
\hline
$N(0)$ & 465 &  Initial number of particles \\
$\Omega$ & $[0,10]\times[0,1]\times[0,1]$ & Spatial domain \\
$D$&1 & Diffusion coefficient\\
$\kappa_1$ & $0.1$ & Rate of degradation reaction (see system \eqref{eqn:TP4_System}) \\
$\kappa_2$ & $89.7$ & Rate of production reaction (see system \eqref{eqn:TP4_System})\\
$\rho$ & $0.06$ & Particle interaction radius \\
$ \Delta t_b$ & $10^{-4} $  & Brownian time-step\\
$P_{\lambda}$ & $2.5\times 10^{-5}$ & Probability of reaction when inside the interaction radius\\
$K$ & $20$& Number of compartments\\
$h_x$ & $1/3$  & Compartment width\\
$h_y$ & 1  & Compartment depth\\
$h_z$ & 1  & Compartment height\\
$\Delta x$ & $1/300$ & PDE voxel width\\
$ \Delta t_p$ & $10^{-4} $  & PDE time-step\\
M & 500 & Number of repeats\\
\hline
\end{tabular}
\end{center}
\captionof{table}{Table of parameters for the bimolecular reaction simulation \eqref{eqn:TP4_System} (test problem 4).}
\label{table:example_4_parameters}
\end{table}

The layout for figure \ref{figure:second_order} is the same as for figures 
\ref{figure:uniform_IC}-\ref{figure:morphogen}. The only difference is the 
calculation of the relative mass error. For this final example, which includes 
second-order reactions, the solution of the mean-field PDE model will no longer 
match the mean behaviour of either the compartment-based model or the 
Brownian-based model. Consequently, in order to calculate the relative mass 
error, we use the average behaviour of the finest-scale model in each hybrid 
representation (e.g. the compartment-based representation in the 
PDE-compartment 
model and Brownian-based representation in the compartment-Brownian model) 
simulated across the whole domain as the ground truth. For the 
PDE-compartment 
coupling the relative mass error is, then, the difference between the average 
(over 500 repeats) number of particles in the given region in the hybrid method 
and the corresponding average (over 500 repeats) number in the same region in 
the purely compartment-based simulation, divided by the number of particles 
in the relevant region of the purely compartment-based simulation (to 
normalise):
\begin{align}
RME_P(t)=&\frac{\int_{\Omega_P} \bar{c}(x,y,z,t) dx-\sum_{i,j,k} \bar{F}_{i,j,k}(t) \mathbb{I}_{c_{i,j,k}\in\Omega_P}}{\sum_{i,j,k} \bar{F}_{i,j,k}(t) \mathbb{I}_{c_{i,j,k}\in\Omega_P}}, 
\label{PDE_compartment_relative_mass_error_PDE_3D}\\
RME_H(t)=&\frac{\sum_{i,j,k} \bar{C}_{i,j,k}(t) \mathbb{I}_{c_{i,j,k}\in\Omega_H} -\sum_{i,j,k} \bar{F}_{i,j,k}(t) \mathbb{I}_{c_{i,j,k}\in\Omega_H}}{\sum_{i,j,k} \bar{F}_{i,j,k}(t) \mathbb{I}_{c_{i,j,k}\in\Omega_H}},\label{PDE_compartment_relative_mass_error_blending_3D}\\
RME_C(t)=&\frac{\sum_{i,j,k} \bar{C}_{i,j,k}(t) \mathbb{I}_{c_{i,j,k}\in\Omega_C} -\sum_{i,j,k} \bar{F}_{i,j,k}(t) \mathbb{I}_{c_{i,j,k}\in\Omega_C}}{\sum_{i,j,k} \bar{F}_{i,j,k}(t) \mathbb{I}_{c_{i,j,k}\in\Omega_C}},\label{PDE_compartment_relative_mass_error_compartments_3D}
\end{align}
where, as before, $\Omega_P$ is the purely PDE region of the domain, $\Omega_H$ is the blending region and $\Omega_C$ is the purely compartment region of the domain. The averaged solution of the PDE component of the hybrid method at position $(x,y,z)$ at time $t$ is denoted $\bar{c}(x,y,z,t)$, the averaged compartment particle numbers in compartment $(i,j,k)$ of the hybrid method are denoted $\bar{C}_i,j,k$ and the averaged compartment particle numbers in compartment $(i,j,k)$ of the fully compartment-based `ground truth' simulation are denoted $\bar{F}_{i,j,k}$. The positions $c_{i,j,k}$ are the centres of the compartments indexed $(i,j,k)$.

For the compartment-Brownian coupling the relative mass error is 
the difference between the average (over 500 repeats) number of particles in each 
region given by the hybrid method and the average number of particles in the 
same region in the purely Brownian-based simulation, divided by the number 
of particles in the relevant region of the purely Brownian-based simulation 
(to normalise):
\begin{align}
RME_C(t)=&\frac{\sum_{i,j,k} \bar{C}_{i,j,k}(t) \mathbb{I}_{c_{i,j,k}\in\Omega_H}-\bar{E}_C(t)}{\bar{E}_C(t)}, 
\label{compartment_Brownian_relative_mass_error_compartment_3D}\\
RME_H(t)=&\frac{\sum_{i,j,k} \bar{C}_{i,j,k}(t) \mathbb{I}_{c_{i,j,k}\in\Omega_H} -\bar{E}_H(t)}{\bar{E}_H(t)},\label{Brownian_compartment_relative_mass_error_blending_3D}\\
RME_B(t)=&\frac{\bar{B}(t) -\bar{E}_B(t)}{\bar{E}_B(t)},\label{Brownian_compartment_relative_mass_error_Brownian_3D}
\end{align}
where, as before, $\Omega_B$ is the purely Brownian region of the domain and $\bar{B}(t)$ represents the mean number of Brownian particles in the purely Brownian region of the hybrid method and $\bar{E}_C(t)$, $\bar{E}_H(t)$ and $\bar{E}_B(t)$ represent the mean number of Brownian particles in $\Omega_C$, $\Omega_H$ and $\Omega_B$ respectively at time $t$ in the fully Brownian `ground truth' simulations.

There are some special points to note about the models which incorporate 
second-order reactions. Firstly, as noted above, the solution of mean-field 
PDE, 
which we will employ in the PDE region of the PDE-compartment hybrid method, 
will not correspond to the mean behaviour of the compartment-based method. This 
is a result of the moment-closure approximation which must be used in order to 
derive a closed PDE for the mean behaviour. As a consequence, we might expect 
some disparity between the solution of the hybrid method and the solution of 
the 
fully-compartment-based simulation that we take to be the ground truth in the 
PDE-compartment-based hybrid.  Fortunately, for our compartment-Brownian hybrid 
method, \cite{erban2009smr} provide a method for matching reaction rates in 
compartment-based simulations to those in Brownian-based simulations, which we 
make use of.

We must also be careful to choose our parameters carefully in the 
compartment-Brownian method.
If compartment-sizes are too small in the compartment-based method then 
particles can 
become too sparsely distributed and second-order reactions lost. 
\cite{erban2009smr} provide a way to alter the reaction rate (depending on the 
compartment size) to maintain the same overall reaction rate as a well mixed 
system. This correction, however, only holds down to a certain compartment 
size, beyond which second-order reactions are irrevocably lost. It is worth noting that \cite{isaacson2013crd} postulated the convergent reaction-diffusion master equation representation (in which particles can interact with others in neighbouring boxes), which is consistent with the spatially continuous Doi model for reaction-diffusion even as box sizes become small. \cite{hellander2015rrm} numerically approximate mesoscopic reaction rates that are consistent with the popular Smoluchowski Brownian dynamics model up to a given lower limit on mesh size.

In the Brownian-based method we need to ensure that the time-step is chosen to 
be sufficiently small that particles do not jump `too far' between position 
updates. If particles jump large distances in each time-step then it is 
possible 
that particles which should have been given the opportunity to react with each 
other may not come into close enough proximity and some second-order reactions 
may be lost. Choosing the reaction radius of particles to be large may help to 
mitigate this somewhat, but brings its own problems. The size of the 
interaction 
radius is calculated by considering particles in free-space \cite{erban2009smr}. In reality, in our 
simulations, particles are often close to boundaries. The proportion of the 
particle's interaction radius that overlaps the exterior of the domain is not 
able to interact with particles inside the domain, so the rate of second-order 
reactions is again effectively reduced. 
Since, for a given reaction rate, the size of the interaction radius increases 
with the time-step, reducing the time-step is often sufficient to solve both of 
these problems.  We note that, with the exception of the PDE not matching the mean behaviour of 
the compartment-based method, these issues are all inherent to the individual 
modelling paradigms we have chosen to couple, and are  not specific to the 
hybrid methods we have developed. With sensible  simulation parameter choices 
these issues can be overcome.

The results of our simulations are plotted in figure \ref{figure:second_order}. In 
Figures \ref{figure:second_order}\subref{figure:second_Brown_0.1} and 
\ref{figure:second_order}\subref{figure:second_Brown_1}, which compare densities for the  the 
compartment-Brownian hybrid paradigm we have good qualitative agreement with 
the 
ground truth (the ubiquitously Brownian-based model). These qualitative results 
are further corroborated in figure \ref{figure:second_order}\subref{figure:second_Brown_rel_error} in which the low and unbiased relative mass error over time are demonstrated. 

The density plots in figures \ref{figure:second_order}\subref{figure:second_PDE_0.1} and \ref{figure:second_order}\subref{figure:second_PDE_1} for the 
PDE-compartment hybrid coupling also appear to demonstrate good qualitative 
agreement. However, when considering the relative mass error in the different 
regions, in figure \ref{figure:second_order}\subref{figure:second_PDE_rel_error},  we observe that, although low, the 
relative mass errors appear to be biased. This, as discussed above, should not 
be a surprise since the  mean-field PDE does not capture the mean behaviour of 
the compartment-based model, which we assume to be the ground truth for the 
relative mass error calculations. The overall mass expected in the fully 
compartment-based model at equilibrium would exceed that predicted by the 
mean-field PDE. In agreement with this expectation we find that the total mass 
in 
all three regions of the domain is less than it would be in the fully compartment-based simulations 
with the problem being particularly acute in the PDE region. A simple 
comparison 
of the expected densities at time $t=1$ shows that the maximum magnitude of the PDE relative 
error with respect to the compartment based model is roughly $3 \times 
10^{-2}$, demonstrating that the size of the relative error we find between our 
hybrid method and the solution of the fully compartment-based simulations is of 
an appropriate order or magnitude, as it is similar to the difference in the concentration when comparing the equilibrium profile of the full PDE to the fully compartment based method, adjusted for the specific voxel size.


\section{Discussion}\label{section:discussion}
When modelling multiscale phenomena it is often the case that concentrations 
vary spatially to such a degree that in one region of the domain a coarse, 
computationally inexpensive model can be tolerated, whereas, in another region 
of the domain a more accurate, but more expensive representation is required.
 
In this paper we have proposed a general hybrid blending mechanism which 
facilitates the spatial coupling of two reaction-diffusion modelling paradigms 
at different levels of detail in order to accommodate the modelling of such 
multiscale phenomena. Our method employs a blending region and a corresponding 
blending function. The blending function scales up or down (respectively) the 
relative contribution to diffusion of a coarse or fine (respectively) 
representation of the reaction-diffusion process across the blending region 
such that diffusion is handled to a different degree by each modelling representation. 

Specifically, we have developed an algorithm which couples a PDE 
representation of a reaction-diffusion process to a compartment-based 
representation and, separately, an algorithm which couples a compartment-based 
representation at the coarse scale to a Brownian-based representation at the 
fine scale. Other algorithms exist to achieve such couplings 
\cite{yates2015pcm,flegg2012trm,flegg2015cmc,franz2012mrd}. Some of these algorithms are conceptually complex - relying variously on artificially introduced `psuedo-compartments', `ghost cells' and `overlap regions' - technically challenging to implement, and strongly parameter dependent - working only in specific parameter regimes. We believe our blending method 
provides a conceptually simple and easily implementable coupling methodology - requiring only an intuitively defined blending function to couple the two regimes together. 
This methodology might be readily employed to couple other modelling regimes 
(for example PDE and Brownian modelling regimes) to form novel hybrid methods 
under a unified framework or implemented simply by non-experts for physical and 
biological applications.

We have demonstrated, through four representative examples, that both of our coupling algorithms are able to handle a wide range of reaction-diffusion processes from simple diffusion through to reaction-diffusion processes incorporating first- and second-order reactions. The hybrid methods are capable of representing these processes accurately  (low error) and without bias (in the situation for which there is no discrepancy between mean-field behaviour of the coupled models) or with the expected bias (when such a discrepancy exists). Due to the computational savings afforded by coupling a cheap coarse model with an expensive fine-scale model, we can scale up particle numbers in our simulations in order to demonstrate that the hybrid algorithms perform arbitrarily well in comparison to the full finest-scale model. For this reason we do not provide explicit time comparisons of our methods, but rather focus on their accuracy.

There are several directions in which we intend to extend this work, but which are not appropriate for inclusion in this initial proof-of-principle paper. Firstly, and perhaps most straight-forwardly we would like to extend these hybrid methods to deal with more complex domain geometries. Although we have demonstrated that our blending hybrid methods can cope with three dimensional reaction-diffusion processes, in real biological scenarios boundaries are likely to be curved and there is the potential for the requirement that interfaces between coarse and fine regimes are non-planar. 

Secondly, the dynamic nature of many biological processes mean that 
concentrations change significantly over time. If we are to ensure that the 
coarse modelling regime represents regions of high concentration and the fine 
modelling regime regions of low concentration, then it is necessary for 
interfaces that border the blending region and the blending region itself to be dynamic. The main challenge associated with dynamic interfaces 
is the conversion of one particle type into another. Fortunately, this challenge 
has been overcome previously by a number of different hybrid methods, whose 
dynamic interface methodologies we might readily adapt to our hybrid paradigm in a follow up 
work \cite{robinson2014atr,harrison2016hac,spill2015ham}. Related concerns are 
the need for the creation or removal of multiple interfaces in scenarios in 
which particle concentrations oscillate in space and time. Similarly, 
reaction-diffusion simulations in which more than just a single species are 
interacting may require different interfaces for each of the different species. 
This raises potentially difficult questions about how to carry-out reactions 
between species represented by distinct modelling paradigms in the same region 
of space. 

A final direction in which we would like to extend this work is by considering entirely new hybridisation methods. For example, rather than having the two distinct modelling paradigms representing the same particles (as we have in the blending region) requiring both regimes to be updated when one changes, it might be practicable to have the two modelling paradigms co-existing across the whole of the domain, but representing different proportions of the particles depending on the concentration. Such a method would remove the requirement for interfaces between the regions of the domain, effectively doing away with many of the concerns related to dynamically and spatially changing concentrations raised earlier in this section.

Since biological and physical experiments can be carried out at increasingly high levels of detail, we are gaining more intricate and specific information about a wide variety of multiscale processes. In order to test experimentally generated hypotheses about such processes we need to have modelling frameworks which are capable of replicating experimental behaviour to a high degree of accuracy.
The blending hybrid methods presented in this paper provide a straightforward 
way to couple modelling paradigms with different levels of detail, which will 
facilitate more accurate and more efficient multiscale modelling. Consequently, 
we expect that both our own future work and the work of others, building on just 
such hybrid paradigms, will enable biochemical simulations which go beyond what 
is tractable with current approaches.


\appendix
\section*{Appendices}
\section{Numerical simulation of the PDE}
\label{section:appendixA}

We now  provide more details on the specifics of the macroscopic 
model that we employ throughout the paper, including an algorithm for its 
implementation. There exist a number of well-developed, efficient numerical methods for the solution of  such reaction-diffusion PDEs \cite{smith1985nsp,morton2005nsp,eymard2000fvm,brenner2004fem}. Typically to 
implement these algorithms we discretise the PDE on a spatial mesh. This 
results in a system of ordinary differential equations (ODEs). These ODEs can 
then be integrated forwards in time using standard numerical techniques.

\begin{figure}[h!!!!!!!!!!!!!]
\begin{center}
\includegraphics[width=\columnwidth]{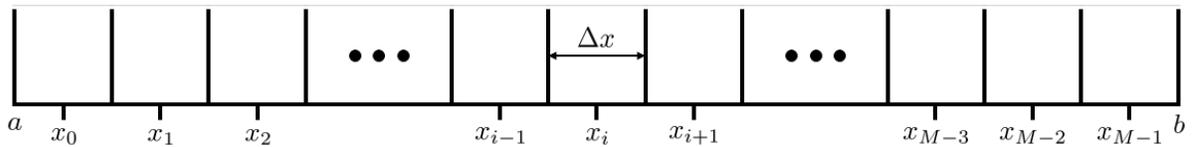}
\caption{Schematic illustrating a spatial discretisation of the one-dimensional 
domain $[a,b]$, which is used to simulate equation  \eqref{eq:model_pde} 
numerically. $\Delta x$ represents the size of the voxels and $x_i$ for 
$i=0,\dots,M-1$ represent their  centres.}
\label{fig:pde_discretisation}
\end{center}
\end{figure}

For the PDE \eqref{eq:model_pde}  we start by  dividing  $[a,b]$  into $M$ voxels each 
of size $\Delta x=(b-a)/M$ and we define $x_{j}=\Delta x(j-1/2)$, so that 
$x_{j}$ is the centre of the voxel $j$ (see figure \ref{fig:pde_discretisation}). Typically the grid spacing of the PDE 
solution method is very fine (much finer than the discretisation of space 
in the compartment-based method) in order to mimic the true continuous-space 
PDE solution as closely as possible. We discretise the PDE using the finite 
volume method over the grid in figure \ref{fig:pde_discretisation}. For time 
integration we use the simple $\theta$-method \cite{morton2005nsp}.


Below we provide a detailed implementation algorithm for the finite 
volume PDE simulation method which is designed to replace line 3 in Algorithm 
\ref{alg:pde_comp}. We start by introducing
 \begin{equation}
 q_{j}(t)=\frac{1}{\Delta x} \int_{x_{j-1/2}}^{x_{j+1/2}} c(t,x) \, \ud x,
 \end{equation}
which corresponds to the average concentration per voxel, and $D_{j}=D(x_{j})$, where we note that $j$ is not necessarily integer valued.
By integration of PDE \eqref{eq:model_pde} over the finite volume voxels we  
then  obtain the  semi-discrete approximation,
\begin{equation}
 \frac{d 
\boldsymbol{q}}{dt}=A \boldsymbol{q}+\boldsymbol{b}+R(\boldsymbol{q}),
\end{equation}
where 
\[
A=\frac{1}{\Delta x^{2}}  \left( \begin{array}{ccccc}
 -D_{1/2} &D_{1/2} & & &  \\
 D_{1/2} &-(D_{1/2}+D_{3/2}) &  \quad \quad D_{3/2}& & \\
 \\
 \\
 &\ddots&  \ddots& \ddots &  \\
 \\
 \\
 & &  D_{M-5/2}& -(D_{M-5/2}+D_{M-3/2}) &  D_{M-3/2}\\ 
 & &  &  D_{M-3/2} &   -D_{M-3/2}
 \end{array} \right) ,
 \]
$\boldsymbol{b}=\Delta x^{-1}(-J_{a},0,\cdots,0,J_{b})$ and $\boldsymbol{q}(t)=(q_{0}(t),q_{1}(t), \cdots,q_{M-1}^{T}(t))$. 
We now solve the semi-discrete approximation using the $\theta$-method\footnote{We employed a fully implicit method (i.e. $\theta = 1$) for the test problems with linear reaction terms. For the test problem with non-linear reaction terms we integrated the dynamics explicitly (i.e. $\theta =0$).}. The 
complete method is described in Algorithm \ref{alg:pde}.

\begin{algorithm}[H]
\KwIn{PDE mesh size -- $\Delta x$; time-step for the solution of the PDE -- 
$\Delta t_p$; left and right ends of the domain -- $a, b$ ;  
initial concentration for the PDE -- $c_{\text{init}}$;  final integration 
time -- $T$ ; value of $\theta$.}
 \SetAlgoLined
 Set $t = 0$, calculate $\boldsymbol{\hat{c}}$ such that 
 \[
 \hat{c}_{j}= \frac{1}{\Delta x} \int_{x_{j-1/2}}^{x_{j+1/2}} c_{\text{init}}(x) \, \ud x,
 \] and set $\boldsymbol{q}_{0}=\boldsymbol{\hat{c}}, n=0 $.
\\
 \While{$t<T$}{
 \[
 \frac{\boldsymbol{q}_{n+1}-\boldsymbol{q}_{n}}{\Delta t_{p}}=(1-\theta)A \boldsymbol{q}_{n} +\theta A  \boldsymbol{q}_{n+1}+\boldsymbol{b}+(1-\theta)R(\boldsymbol{q}_{n})+\theta R(\boldsymbol{q}_{n+1})\]
 \\
 Set $t=t+\Delta t_p, n=n+1$.
 }
 \caption{An algorithm for the numerical solution of equation \eqref{eq:model_pde} using a first-order finite volume method}\label{alg:pde}
 \end{algorithm}

\section{Simulation of the compartment-based method}
\label{section:appendixB}

We now provide more details on the specifics of the mesoscopic 
model that we employ throughout the paper, including an algorithm for its 
implementation. More precisely, we have used an event-driven approach in order to simulate our compartment-based dynamics.
The most commonly used event-driven algorithm for simulating Markov processes 
is 
the Gillespie direct method \cite{gillespie1977ess}. Each event is 
characterised by a propensity function which specifies the rate parameter of 
the 
exponentially distributed waiting time until the next `firing' of that event. 
It 
can be shown that the time until the next reaction of any type (i.e. the 
minimum 
waiting time) is also exponentially distributed with a rate which is the sum of 
the rates of the individual reactions. Gillespie's algorithm first generates an 
exponentially distributed minimum waiting time and subsequently, with 
probabilities proportional to their propensity functions, chooses a reaction to 
fire. Alternatively, time-driven algorithms can be employed, in which a 
sufficiently small time-step is chosen such that the probability of more than 
one reaction/movement event firing in that time interval is negligible. 
Time-driven algorithms tend to be inefficient due to the small time-step 
required during which, typically no change to the state is implemented. 
Consequently, exact event-driven algorithms tend to be favoured for the 
simulation of compartment-based dynamics.

Here we discuss the implementation of a compartment-based reaction-diffusion 
model in one dimension with spatially varying diffusion 
coefficient. Although we present our 
algorithm in one dimension it is straightforward to extend it to higher 
dimensions with planar interfaces. We give one such three-dimensional example 
in Section \ref{section:bimolecular_production}. 

\begin{figure}
\begin{center}
\includegraphics[width=\columnwidth]{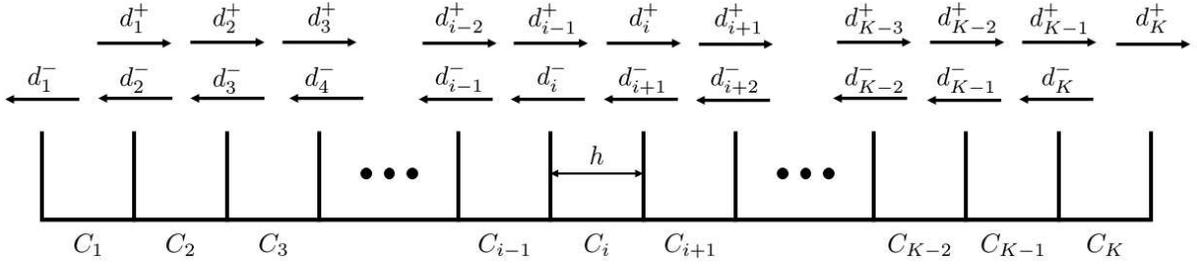}
\caption{Schematic illustrating the jump rates, $d_i^{\pm}$ between compartments 
when simulating the mesoscopic reaction-diffusion paradigm on the domain $[a,b]$ 
with compartment-size $h$.}
\label{fig:discrete}
\end{center}
\end{figure}

We first discretise the region $[a,b]$ into $K$ compartments, each 
of size $h=(b-a)/K$. In order to replicate the density dependent diffusion specified in 
the macroscopic model described by equation \eqref{eq:model_pde} we require 
that 
the rates at which particles jump to the left and the right are not equal in 
regions in which the diffusion coefficient is non-constant. Specifically, we 
must evaluate the jump rates based on the diffusion coefficient at compartment boundaries \cite{othmer1997abc}.
This is visualised in figure \ref{fig:discrete} where we denote 
the jumping rate  of a particle in compartment $i$ into compartment $i+1$ with $d^{+}_{i}$, while 
we denote the jumping rate of a compartment $i$ particle into 
compartment $i-1$ with $d^{-}_{i}$. 
Without loss of generality, assuming that the left-hand boundary of the 
compartment-based regime is at $a$, the left jump rate for compartment $i$ is 
given by 
\begin{equation}
d_i^-=\frac{D(a+(i-1)h)}{h^2}, \text{ for } i=2, \dots, 
K.\label{equation:left_jump_rate}\\
\end{equation}
and the right jump rate is given by 
\begin{equation}
d_i^+=\frac{D(a+ih)}{h^2}, \text{ for } i=1, \dots, 
K-1.\label{equation:right_jump_rate}
\end{equation}
Jump rates $d_1^-$ and $d_K^+$ at the boundaries can be chosen in order to 
replicate the chosen boundary conditions \cite{taylor2015dab}. In the case of 
zero-flux boundary conditions (equivalent to setting $J_a=J_b=0$ in equations \eqref{equation:PDE_boundary_conditions}), these jump rates are simply chosen to be 
$d_1^-=d_K^+=0$. 
Reaction propensity functions are specified to bring about the desired reaction 
rate\footnote{It should be noted that the rate of second- and higher-order 
reactions depends, non-trivially, on the compartment size, $h$, and that the 
desired rate of such higher-order reactions may not be implementable for some 
particularly small compartment sizes \cite{erban2009smr}.}.  Once all the 
event rates have been specified then one can simulate the system using 
Gillespie's direct method \cite{gillespie1977ess}.

We next provide a detailed implementation for the spatial Gillespie 
algorithm over a time interval of size $\Delta t$. This algorithm is designed to 
replace line 5 of Algorithm \ref{alg:pde_comp} and line 3 of algorithm 
\ref{alg:comp_brown}. Without loss of generality assume the compartment-based 
region occupies $[a,I_2]$, as in the compartment-Brownian hybrid method. 
However, we note the caveat that for the PDE-compartment hybrid method the 
compartment-based region would occupy $[I_1,b]$. As already noted, left and 
right jumping rates from compartment $i$ are different and given in equations 
\eqref{equation:left_jump_rate} and \eqref{equation:right_jump_rate}, 
respectively.

\begin{algorithm}[H]
  \KwIn{Compartment size -- $h$; time-step for the solution of the PDE or Brownian update time-step -- 
$\Delta t$; left and right ends of the blending region -- $I_{1}, I_{2}$; the 
total number of compartments -- $K$; number of reactions -- $M$; initial 
particle numbers -- $\bs{C}_{init}$; compartment-based diffusion coefficient --
$D_1(x)$.}
\SetAlgoLined
Set $t = 0$.
\\
\While{$t<\Delta t$}{
Using equations \eqref{equation:left_jump_rate} and 
\eqref{equation:right_jump_rate}, respectively, calculate the propensity 
functions corresponding to the left jumps, $\alpha_i=d_i^-$, and right jump, 
$\alpha_{i+K}=d_i^+$, respectively, for $i=1,\dots, K$. Calculate also the 
propensity functions for each reaction, $m$, in each compartment, 
$\alpha_{(m+1)K+i}$ for $m=1,\dots,M$ and $i=1,\dots,K$.

Calculate the sum of the propensity functions
\begin{equation}
 \alpha_0=\sum_{i=1}^K\left( \alpha_i +\alpha_{i+K}+\sum_{m=1}^M \alpha_{(m+1)K+i}\right).
\end{equation}

Draw the time, $\tau$, until the next reaction from an exponential distribution with parameter $\alpha_0$:
\begin{equation}
 \tau=-\frac{1}{\alpha_0}\ln(u),
\end{equation}
where $u$ is drawn from a uniform distribution with support $(0,1)$.

Update the time: $t=t+\tau$.

\If{$t>\Delta t$}{
  Break
  }

Choose the $j^{\text{th}}$ reaction to fire. Each reaction, $j$, (for $j=1,\dots, (2+M)K$) is chosen with probability $\alpha_j/\alpha_0$ (proportional to its propensity function).

Implement the particle movement or reaction specified by reaction $j$ by updating the corresponding particle numbers.

}
\caption{Simulating the spatial Gillespie algorithm for a time-interval $\Delta t$.}\label{alg:spatial_gillespie}
\end{algorithm}

Note that the Gillespie algorithm steps forwards in discrete time-steps. 
However, the time-steps themselves are drawn form a continuous distribution so 
that the solution time-points of the Gillespie algorithm do not match up with 
those of the fixed time-step algorithms for PDE and Brownian-based simulation. 
Consequently, our technique to couple the two simulation methodologies is 
to simulate the compartment-based dynamics until such a time as $\Delta t$ is 
exceeded for the first time. Since a PDE or Brownian update step is due at time 
$\Delta t$ we do not implement the final Gillespie reaction whose time-step took 
us over the $\Delta t$ time limit. Instead we implement a PDE or Brownian update 
step accordingly and correspondingly update the propensity functions ready to 
begin Algorithm \ref{alg:spatial_gillespie} again.

\section{Simulation of Brownian dynamics}
\label{section:appendixC}
In this appendix we provide a provide a detailed implementation algorithm for Brownian-based dynamics, which is designed to replace line 5 in Algorithm \ref{alg:comp_brown}.

\vskip 2pt
\setlength{\algoheightrule}{0pt} 
\SetNlSty{textbf}{}{}
\hrule height 0.8pt 
\begin{algorithm}[H]
  \KwIn{Brownian update time-step -- $\Delta t_b$; left and right ends of the 
blending region -- $I_{1}, I_{2}$; number of reactions -- $M$; number of zeroth- 
first- and second-order reactions -- $Z$, $F$ and $S$ (respectively); 
probability of zeroth-, first- and second-order reactions in time-step $\Delta 
t$ -- $P_z$, $P_f$ and $P_s$ (respectively); initial position of particle $i$ 
for $i=1,\dots, N$ -- $\bf{y}_i$; Brownian-based diffusion coefficient -- 
$D_2(x)$.}
\SetAlgoLined
\For{$i=1:N$}{
Use equation \eqref{equation:edge_SDE} to update particle positions, implementing reflective boundary conditions for any particles that cross the interface at $I_1$ or the boundary at $b$.
}
\If{there are zeroth-order reactions}{
\For{z=1,\dots, Z}{
\If {$u_z<P_z$}{
Position a new particle of the appropriate type uniformly across the purely Brownian domain.}
}
}
\If{there are first-order reactions}{
\For{each appropriate particle}{
\For{f=1,\dots, F}{
\If {$u_f<P_f$}{
implement reaction $f$ by removing reacting particle and/or placing product particle(s) as appropriate.}
}
}
}
\caption{Simulating a Brownian update step of length $\Delta t_b$.}\label{alg:brownian_update}
\end{algorithm}

\setlength{\algoheightrule}{0pt} 
\setlength{\algotitleheightrule}{0pt}
\SetNlSty{textbf}{}{}
\begin{algorithm}
  \LinesNumbered
\setcounter{AlgoLine}{19}
\SetAlgoVlined

\If{there are second-order reactions}{
Calculate the distances between each pair of particles capable of reacting with each other.

\For{each appropriate reaction pair}{
\For{s = 1, \dots, S}{
\If{particle pair are within $\rho_s$ of each other}{
\If{$u_s$<$P_{s}$}{
implement reaction $s$ by removing reacting particles and/or placing product particles as appropriate.
}
}
}
}
}
In the above $u_z$, $u_f$ and $u_s$ are drawn from a uniform distribution with support $(0,1)$.
\hrule height 0.8pt 
\end{algorithm}

\newpage
For zeroth- and first-order reactions, respectively, reaction probabilities, 
$P_z$ and $P_f$, respectively, are calculated simply by multiplying the rate of 
reaction with the time-step, $\Delta t_b$.
The calculation of $P_s$ for second-order reaction, $s$, is somewhat more complicated and depends on the choice of reaction radius, $\rho_s$. For more details on this and the placement of new particles after reaction see \cite{erban2009smr}. Note that Brownian reactions are only implemented in the region $[I_2,b]$, since outside this region reactions are implemented using the compartment-based regime.

In theory, the fact that the diffusion coefficient of the Brownian-based 
particles falls to zero at $I_1$ should mean that particles cannot cross the 
interface there, rendering the implementation of reflecting boundary conditions 
at $I_1$ in step 2 redundant. However, in practice, the finite time-step we use 
to update the Brownian particles means that, with low probability, particles can 
jump across the interface and must consequently be reflected back.

\section*{Acknowledgements}
Part of this work was conceived during the authors' stay at the Newton Institute for the program Stochastic Dynamical Systems in Biology: Numerical Methods and Applications. This work was supported by EPSRC grant no EP/K032208/1. This work was partially supported by a grant from the Simons Foundation. CAS  is supported by a scholarship from the EPSRC Centre for Doctoral Training in Statistical Applied Mathematics at Bath (SAMBa), under the project EP/L015684/1. AG was partially supported by a summer research placement from the Institute for Mathematical Innovation at the University of Bath.


\begin{thebibliography}{10}

\bibitem{alexander2002ars}
F.~Alexander, A.~Garcia, and D.~Tartakovsky.
\newblock {Algorithm refinement for stochastic partial differential equations:
  I. Linear diffusion}.
\newblock {\em J. Comput. Phys.}, 182(1):47--66, 2002.

\bibitem{alexander2005ars}
F.~Alexander, A.~Garcia, and D.~Tartakovsky.
\newblock {Algorithm refinement for stochastic partial differential equations:
  II. Correlated systems}.
\newblock {\em J. Comput. Phys.}, 207(2):769--787, 2005.

\bibitem{andrews2004ssc}
S.~Andrews and D.~Bray.
\newblock Stochastic simulation of chemical reactions with spatial resolution
  and single molecule detail.
\newblock {\em Phys. Biol.}, 1(3-4):137--151, 2004.

\bibitem{baker2009fmm}
R.~Baker, C.~Yates, and R.~Erban.
\newblock {From microscopic to macroscopic descriptions of cell migration on
  growing domains}.
\newblock {\em Bull. Math. Biol.}, 72(3):719--762, 2010.

\bibitem{benson1993apf}
D.~Benson, P.~Maini, and J.~Sherratt.
\newblock Analysis of pattern formation in reaction diffusion models with
  spatially inhomogenous diffusion coefficients.
\newblock {\em Math. Comput. Model.}, 17(12):29--34, 1993.

\bibitem{beskos2005esd}
A.~Beskos and G.~Roberts.
\newblock Exact simulation of diffusions.
\newblock {\em The Annals of Applied Probability}, 15(4):2422--2444, 2005.

\bibitem{brenner2004fem}
S.~Brenner and C.~Carstensen.
\newblock {\em Finite element methods}, chapter~1.
\newblock Encyclopedia of Computational Mechanics. John Wiley \& Sons, Ltd,
  2004.

\bibitem{chiam2006hss}
K.-H. Chiam, C.~Tan, V.~Bhargava, and G.~Rajagopal.
\newblock Hybrid simulations of stochastic reaction-diffusion processes for
  modeling intracellular signaling pathways.
\newblock {\em Phys. Rev. E}, 74(5):051910, 2006.

\bibitem{dobramysl2015pbd}
U.~Dobramysl, S.~R{\"u}diger, and R.~Erban.
\newblock Particle-based multiscale modeling of intracellular calcium dynamics.
\newblock {\em Multiscale. Model. Sim.}, 14(3):997--1016, 2015.

\bibitem{duncan2016hfs}
A.~Duncan, R.~Erban, and K.~Zygalakis.
\newblock Hybrid framework for the simulation of stochastic chemical kinetics.
\newblock {\em J. Comput. Phys.}, 326:398--419, 2016.

\bibitem{elf2004ssb}
J.~Elf and M.~Ehrenberg.
\newblock Spontaneous separation of bi-stable biochemical systems into spatial
  domains of opposite phases.
\newblock {\em Syst. Biol.}, 1(2):230--236, 2004.

\bibitem{engblom2009ssr}
S.~Engblom, L.~Ferm, A.~Hellander, and P.~L\"{o}tstedt.
\newblock Simulation of stochastic reaction-diffusion processes on unstructured
  meshes.
\newblock {\em SIAM J. Sci. Comput.}, 31(3):1774--1797, 2009.

\bibitem{Erban2007rbc}
R.~Erban and S.~Chapman.
\newblock Reactive boundary conditions for stochastic simulations of
  reaction{--}diffusion processes.
\newblock {\em Phys. Biol.}, 4(1):16--28, 2007.

\bibitem{erban2009smr}
R.~Erban and S.~Chapman.
\newblock Stochastic modelling of reaction--diffusion processes: algorithms for
  bimolecular reactions.
\newblock {\em Phys. Biol.}, 6(4):1--18, 2009.

\bibitem{erban2007pgs}
R.~Erban, S.~Chapman, and P.~Maini.
\newblock A practical guide to stochastic simulations of reaction-diffusion
  processes.
\newblock {\em arXiv preprint arXiv:0704.1908}, 2007.

\bibitem{erban2014msr}
R.~Erban, M.~Flegg, and G.~Papoian.
\newblock Multiscale stochastic reaction--diffusion modeling: application to
  actin dynamics in filopodia.
\newblock {\em Bull. Math. Biol.}, 76(4):799--818, 2014.

\bibitem{eymard2000fvm}
R.~Eymard, T.~Gallou{\"e}t, and R.~Herbin.
\newblock Finite volume methods.
\newblock {\em Handbook of numerical analysis}, 7:713--1018, 2000.

\bibitem{ferm2010aas}
L.~Ferm, A.~Hellander, and P.~L{\"{o}}tstedt.
\newblock An adaptive algorithm for simulation of stochastic reaction-diffusion
  processes.
\newblock {\em J. Comput. Phys.}, 229(2):343--360, 2010.

\bibitem{flegg2012trm}
M.~Flegg, S.~Chapman, and R.~Erban.
\newblock The two-regime method for optimizing stochastic reaction--diffusion
  simulations.
\newblock {\em J. Roy. Soc. Interface}, 9(70):859--868, 2012.

\bibitem{flegg2014atr}
M.~Flegg, S.~Chapman, L.~Zheng, and R.~Erban.
\newblock Analysis of the two-regime method on square meshes.
\newblock {\em (SIAM) J. Sci. Comput.}, 36(3):B561--B588, 2014.

\bibitem{flegg2015cmc}
M.~Flegg, S.~Hellander, and R.~Erban.
\newblock Convergence of methods for coupling of microscopic and mesoscopic
  reaction-diffusion simulations.
\newblock {\em J. Comput. Phys.}, 289(C):1--17, 2015.

\bibitem{flegg2013dsn}
M.~Flegg, S.~R{\"u}diger, and R.~Erban.
\newblock Diffusive spatio-temporal noise in a first-passage time model for
  intracellular calcium release.
\newblock {\em J. Chem. Phys.}, 138(15):154103, 2013.

\bibitem{flekkoy2001cpf}
E.~Flekk{\o}y, J.~Feder, and G.~Wagner.
\newblock Coupling particles and fields in a diffusive hybrid model.
\newblock {\em Phys. Rev. E}, 64(6):066302, 2001.

\bibitem{franz2012mrd}
B.~Franz, M.~Flegg, S.~Chapman, and R.~Erban.
\newblock {Multiscale reaction-diffusion algorithms: PDE-assisted Brownian
  dynamics}.
\newblock {\em {SIAM} J. Appl. Math.}, 73(3):1224--1247, 2013.

\bibitem{geyer2004ibd}
T.~Geyer, C.~Gorba, and V.~Helms.
\newblock Interfacing brownian dynamics simulations.
\newblock {\em J. Chem. Phys.}, 120(10):4573--4580, 2004.

\bibitem{gillespie1977ess}
D.~Gillespie.
\newblock Exact stochastic simulation of coupled chemical reactions.
\newblock {\em J. Phys. Chem.}, 81(25):2340--2361, 1977.

\bibitem{gillespie2009dls}
D.~Gillespie.
\newblock Deterministic limit of stochastic chemical kinetics.
\newblock {\em J. Phys. Chem. B}, 113(6):1640--1644, 2009.

\bibitem{gorba2004bds}
C.~Gorba, T.~Geyer, and V.~Helms.
\newblock Brownian dynamics simulations of simplified cytochrome c molecules in
  the presence of a charged surface.
\newblock {\em J. Chem. Phys.}, 121(1):457--464, 2004.

\bibitem{harrison2016hac}
J.~Harrison and C.~Yates.
\newblock {A hybrid algorithm for coupling PDE and compartment-based dynamics}.
\newblock {\em J. Roy. Soc. Interface}, 13(122):20160335, 2016.

\bibitem{hellander2012cmm}
A.~Hellander, S.~Hellander, and P.~Lotstedt.
\newblock Coupled mesoscopic and microscopic simulation of stochastic
  reaction-diffusion processes in mixed dimensions.
\newblock {\em Multiscale. Model. Sim.}, 10(2):585--611, 2012.

\bibitem{hellander2015rrm}
S.~Hellander, A.~Hellander, and L.~Petzold.
\newblock Reaction rates for mesoscopic reaction-diffusion kinetics.
\newblock {\em Phys. Rev. E}, 91(2):023312, 2015.

\bibitem{hillen2008ugp}
T.~Hillen and K.~Painter.
\newblock {A user's guide to PDE models for chemotaxis}.
\newblock {\em J. Math. Biol.}, 58(1):183--217, 2009.

\bibitem{isaacson2008rbr}
S.~Isaacson.
\newblock Relationship between the reaction--diffusion master equation and
  particle tracking models.
\newblock {\em J. Phys. A.-Math. Theor.}, 41(6):065003, 2008.

\bibitem{isaacson2013crd}
S.~Isaacson.
\newblock A convergent reaction-diffusion master equation.
\newblock {\em J. Chem. Phys.}, 139(5):054101, 2013.

\bibitem{keller1970ism}
E.~Keller and L.~Segel.
\newblock {Initiation of slime mold aggregation viewed as an instability.}
\newblock {\em J. Theor. Biol.}, 26(3):399--415, 1970.

\bibitem{keller1971mfc}
E.~Keller and L.~Segel.
\newblock {Model for chemotaxis}.
\newblock {\em J. Theor. Biol.}, 30(2):225--234, 1971.

\bibitem{keller1971tbc}
E.~Keller and L.~Segel.
\newblock Traveling bands of chemotactic bacteria: a theoretical analysis.
\newblock {\em J. Theor. Biol.}, 30(2):235--248, 1971.

\bibitem{khan2011scd}
S.~Khan, Y.~Zou, A.~Amjad, A.~Gardezi, C.~Smith, C.~Winters, and T.~Reese.
\newblock {Sequestration of CaMKII in dendritic spines \textit{in silico}}.
\newblock {\em J. Comput. Neurosci.}, 31(3):581--594, 2011.

\bibitem{king2017ppf}
J.~King and R.~O'Dea.
\newblock Pushed and pulled fronts in a discrete reaction--diffusion equation.
\newblock {\em J. Eng. Math.}, 102(1):89--116, 2017.

\bibitem{klann2012hsg}
M.~Klann, A.~Ganguly, and H.~Koeppl.
\newblock Hybrid spatial gillespie and particle tracking simulation.
\newblock {\em Bioinformatics}, 28(18):i549--i555, 2012.

\bibitem{kurtz1972rbs}
T.~Kurtz.
\newblock {The relationship between stochastic and deterministic models for
  chemical reactions}.
\newblock {\em J. Chem. Phys.}, 57(7):2976--2978, 1972.

\bibitem{lipkova2011abd}
J.~Lipkov{\'a}, K.~Zygalakis, S.~Chapman, and R.~Erban.
\newblock Analysis of brownian dynamics simulations of reversible bimolecular
  reactions.
\newblock {\em {SIAM} J. Appl. Math.}, 71(3):714--730, 2011.

\bibitem{lo2019hsm}
W.-C. Lo and S.~Mao.
\newblock A hybrid stochastic method with adaptive time step control for
  reaction--diffusion systems.
\newblock {\em J. Comput. Phys.}, 379:392--402, 2019.

\bibitem{lo2016hcd}
W.-C. Lo, L.~Zheng, and Q.~Nie.
\newblock A hybrid continuous-discrete method for stochastic
  reaction--diffusion processes.
\newblock {\em R. Soc. Open Sci.}, 3(9):160485, 2016.

\bibitem{moro2004hms}
E.~Moro.
\newblock Hybrid method for simulating front propagation in reaction-diffusion
  systems.
\newblock {\em Phys. Rev. E}, 69(6):060101, 2004.

\bibitem{mort2016rdm}
R.~Mort, R.~Ross, K.~Hainey, O.~Harrison, M.~Keighren, G.~Landini, R.~Baker,
  K.~Painter, I.~Jackson, and C.~Yates.
\newblock Reconciling diverse mammalian pigmentation patterns with a
  fundamental mathematical model.
\newblock {\em Nat. Commun.}, 7:10288, 2016.

\bibitem{morton2005nsp}
K.~Morton and D.~Mayers.
\newblock {\em {Numerical Solution of Partial Differential Equations}}.
\newblock Cambridge University Press, 2005.

\bibitem{othmer1988mod}
H.~Othmer, S.~Dunbar, and W.~Alt.
\newblock Models of dispersal in biological systems.
\newblock {\em J. Math. Biol.}, 26(3):263--298, 1988.

\bibitem{othmer1997abc}
H.~Othmer and A.~Stevens.
\newblock Aggregation, blowup, and collapse: the {A}{B}{C}'s of taxis in
  reinforced random walks.
\newblock {\em {SIAM} J. Appl. Math.}, 57(4):1044--1081, 1997.

\bibitem{painter2002vfq}
K.~Painter and T.~Hillen.
\newblock {Volume-filling and quorum-sensing in models for chemosensitive
  movement}.
\newblock {\em Can. Appl. Math. Quart.}, 10(4):501--543, 2002.

\bibitem{painter1999sfj}
K.~Painter, P.~Maini, and H.~Othmer.
\newblock {Stripe formation in juvenile \textit{Pomacanthus} explained by a
  generalized Turing mechanism with chemotaxis}.
\newblock {\em Proc. Natl. Acad. Sci. USA}, 96(10):5549--5554, 1999.

\bibitem{plapp2000mrw}
M.~Plapp and A.~Karma.
\newblock Multiscale random-walk algorithm for simulating interfacial pattern
  formation.
\newblock {\em Phys. Rev. Lett.}, 84(8):1740, 2000.

\bibitem{redner2001gfp}
S.~Redner.
\newblock {\em {A Guide to First-Passage Processes}}.
\newblock Cambridge University Press, 2001.

\bibitem{risken1989fokker}
H.~Risken.
\newblock {\em The Fokker-Planck Equation. Methods of Solution and
  Applications}.
\newblock 1989.

\bibitem{robinson2017pbm}
M.~Robinson and M.~Bruna.
\newblock Particle-based and meshless methods with aboria.
\newblock {\em SoftwareX}, 6:172--178, 2017.

\bibitem{robinson2014atr}
M.~Robinson, M.~Flegg, and R.~Erban.
\newblock Adaptive two-regime method: application to front propagation.
\newblock {\em J. Chem. Phys.}, 140(12):124109, 2014.

\bibitem{rossinelli2008ash}
D.~Rossinelli, B.~Bayati, and P.~Koumoutsakos.
\newblock Accelerated stochastic and hybrid methods for spatial simulations of
  reaction--diffusion systems.
\newblock {\em Chem. Phys. Lett.}, 451(1):136--140, 2008.

\bibitem{schulze2003ckm}
T.~Schulze, P.~Smereka, and W.~E.
\newblock {Coupling kinetic Monte-Carlo and continuum models with application
  to epitaxial growth}.
\newblock {\em J. Comput. Phys.}, 189(1):197--211, 2003.

\bibitem{sherratt2005avs}
J.~Sherratt.
\newblock An analysis of vegetation stripe formation in semi-arid landscapes.
\newblock {\em J. Math. Biol.}, 51(2):183--197, 2005.

\bibitem{smith2018seh}
C.~Smith and C.~Yates.
\newblock Spatially extended hybrid methods: a review.
\newblock {\em J. Roy. Soc. Interface}, 15(139), 2018.

\bibitem{smith2018arm}
C.~Smith and C.~Yates.
\newblock {The auxiliary region method: A hybrid method for coupling a PDE to
  Brownian-based dynamics for reaction-diffusion systems}.
\newblock {\em R. Soc. Open Sci.}, 5(8):180920, 2018.

\bibitem{smith1985nsp}
G.~Smith.
\newblock {\em Numerical solution of partial differential equations: finite
  difference methods}.
\newblock {Oxford University Press}, 1985.

\bibitem{smoluchowski1917vem}
M.~Smoluchowski.
\newblock {Versuch einer mathematischen Theorie der Koagulationskinetik
  kolloider L{\"o}sungen}.
\newblock {\em Z. Phys. Chem.}, 92(129-168):9, 1917.

\bibitem{sokolowski2019ead}
T.~Sokolowski, J.~Paijmans, L.~Bossen, T.~Miedema, M.~Wehrens, N.~Becker,
  K.~Kaizu, K.~Takahashi, M.~Dogterom, and P.~Ten~Wolde.
\newblock {eGFRD in all dimensions}.
\newblock {\em J. Chem. Phys.}, 150(5):054108, 2019.

\bibitem{spill2015ham}
F.~Spill, P.~Guerrero, T.~Alarcon, P.~Maini, and H.~Byrne.
\newblock Hybrid approaches for multiple-species stochastic reaction--diffusion
  models.
\newblock {\em J. Comput. Phys.}, 299:429--445, 2015.

\bibitem{strehl2015hss}
R.~Strehl and S.~Ilie.
\newblock Hybrid stochastic simulation of reaction-diffusion systems with slow
  and fast dynamics.
\newblock {\em J. Chem. Phys.}, 143(23):234108, 2015.

\bibitem{taylor2015dab}
P.~Taylor, R.~Baker, and C.~Yates.
\newblock Deriving appropriate boundary conditions, and accelerating
  position-jump simulations, of diffusion using non-local jumping.
\newblock {\em Phys. Biol.}, 12(1):016006, 2015.

\bibitem{turing1952cbm}
A.~Turing.
\newblock The chemical basis of morphogenesis.
\newblock {\em Phil. Trans. R. Soc. B.}, 237(641):37--72, 1952.

\bibitem{van1988dim}
N.~van Kampen.
\newblock Diffusion in inhomogeneous media.
\newblock {\em J. Phys. Chem. Solids.}, 49(6):673--677, 1988.

\bibitem{van2007spp}
N.~van Kampen.
\newblock {\em {Stochastic processes in physics and chemistry}}.
\newblock North-Holland, 2007.

\bibitem{van2005gfr}
J.~van Zon and P.~ten Wolde.
\newblock Green's-function reaction dynamics: a particle-based approach for
  simulating biochemical networks in time and space.
\newblock {\em J. Chem. Phys.}, 123(23):234910, 2005.

\bibitem{volpert2009rdw}
V.~Volpert and S.~Petrovskii.
\newblock Reaction--diffusion waves in biology.
\newblock {\em Phys. Life Rev.}, 6(4):267--310, 2009.

\bibitem{winkelmann2017hmc}
S.~Winkelmann and C.~Sch{\"u}tte.
\newblock Hybrid models for chemical reaction networks: Multiscale theory and
  application to gene regulatory systems.
\newblock {\em J. Chem. Phys.}, 147(11):114115, 2017.

\bibitem{yates2013ivd}
C.~Yates and R.~Baker.
\newblock {Importance of the Voronoi domain partition for position-jump
  reaction-diffusion processes on non-uniform rectilinear lattices}.
\newblock {\em Phys. Rev. E}, 88(5):054701, 2013.

\bibitem{yates2012gfm}
C.~Yates, R.~Baker, R.~Erban, and P.~Maini.
\newblock Going from microscopic to macroscopic on non-uniform growing domains.
\newblock {\em Phys. Rev. E}, 86(2):021921, 2012.

\bibitem{yates2015pcm}
C.~Yates and M.~Flegg.
\newblock The pseudo-compartment method for coupling partial differential
  equation and compartment-based models of diffusion.
\newblock {\em J. Roy. Soc. Interface}, 12(106):20150141, 2015.

\end{thebibliography}
\end{document}